\documentclass[twocolumn]{aastex701}

\usepackage{lineno}
\usepackage{amssymb}
\usepackage{amsmath}
\usepackage{booktabs}
\usepackage{hyperref}
\usepackage{multirow}
\usepackage{times}
\PassOptionsToPackage{hyphens}{url}\usepackage{hyperref}
\usepackage[hyphenbreaks]{breakurl}

\newcommand{\degg}{\hbox{$^\circ$}}

\newcommand{\cgs}{ergs\,cm$^{-2}$\,s$^{-1}$}
\newcommand{\pg}{PG\,1211+143}
\newcommand{\xmm}{{\it XMM-Newton}}
\newcommand{\xrism}{{\it XRISM}}
\newcommand{\nustar}{{\it NuSTAR}}
\newcommand{\swift}{{\it Swift}}

\newcommand{\ls}
{\mathrel{\hbox{\rlap{\hbox{\lower4pt\hbox{$\sim$}}}\hbox{$<$}}}}
\newcommand{\gs}
{\mathrel{\hbox{\rlap{\hbox{\lower4pt\hbox{$\sim$}}}\hbox{$>$}}}}

\begin{document}

\title{Resolving the Multiple Component Outflows in PG 1211+143: II. The Soft X-ray View of the Ultra Fast Outflow.}

\author{James N.\ Reeves} 
\affiliation{Institute for Astrophysics and Computational Sciences, Department of Physics, The Catholic University of America, Washington, DC 20064, USA}
\affiliation{INAF, Osservatorio Astronomico di Brera, Via Bianchi 46, I-23807 Merate (LC), Italy}
\email{james.n.reeves456@gmail.com}

\author{Valentina Braito}
\affiliation{INAF, Osservatorio Astronomico di Brera, Via Brera 20, I-20121 Milano, Italy}
\affiliation{Dipartimento di Fisica, Università di Trento, Via Sommarive 14, Trento 38123, Italy}
\affiliation{Institute for Astrophysics and Computational Sciences, Department of Physics, The Catholic University of America, Washington, DC 20064, USA}
\email{valentina.braito@gmail.com}

\author{Misaki Mizumoto}
\affiliation{Science Education Research Unit, University of Teacher Education Fukuoka, Munakata, Fukuoka 811-4192, Japan}
\email{mizumoto-m@fukuoka-edu.ac.jp}  

\author{Steven B.\ Kraemer}
\affiliation{Institute for Astrophysics and Computational Sciences, Department of Physics, The Catholic University of America, Washington, DC 20064, USA}
\email{kraemer@cua.edu}

\author{Ehud~Behar}
\affiliation{Department of Physics, Technion, Technion City, Haifa 3200003, Israel}
\affiliation{Kavli Institute for Astrophysics and Space Research, Massachusetts Institute of Technology, Cambridge, MA 02139, USA}
\email{behar@physics.technion.ac.il}

\author{Chris Done}
\affiliation{Centre for Extragalactic Astronomy, Department of Physics, Durham University, South Road, Durham, DH1 3LE, UK
}
\affiliation{Kavli Institute for the Physics and Mathematics of the Universe (WPI), University of Tokyo, Kashiwa, Chiba 277-8583, Japan
}
\email{chris.done@durham.ac.uk}

\author{Kouichi Hagino}
\affiliation{Department of Physics, University of Tokyo, Hongo, Bunkyo-ku, Tokyo 113-0033, Japan
}
\email{kouichi.hagino@phys.s.u-tokyo.ac.jp}

\author{Gabriele~A.~Matzeu}
\affiliation{Quasar Science Resources SL for ESA, European Space Astronomy Centre (ESAC), Science Operations Department, 28692, Villanueva de la Cañada, Madrid, Spain
}
\email{Gabriele.Matzeu@ext.esa.int}

\author{Hirofumi Noda}
\affiliation{Astronomical Institute, Tohoku University, Aramaki, Aoba-ku, Sendai, Miyagi 980-8578, Japan
}
\email{hirofumi.noda@astr.tohoku.ac.jp}

\author{Mariko Nomura}
\affiliation{Faculty of Science and Technology, Graduate School of Science and Technology, Hirosaki University, Hirosaki, Aomori 036-8561, Japan
}
\email{nomura@hirosaki-u.ac.jp}

\author{Shoji Ogawa}
\affiliation{Institute of Space and Astronautical Science (ISAS), Japan Aerospace Exploration Agency (JAXA), Sagamihara, Kanagawa 252-5210, Japan
}
\email{ogawa.shohji@jaxa.jp}

\author{Ken~Ohsuga}
\affiliation{Center for Computational Science, University of Tsukuba, Tennodai, Tsukuba, Ibaraki 305-8577, Japan
}
\email{ohsuga@ccs.tsukuba.ac.jp}

\author{Atsushi Tanimoto}
\affiliation{Graduate School of Science and Engineering, Kagoshima University, Kagoshima 890-0065, Japan
}
\email{atsushi.tanimoto@sci.kagoshima-u.ac.jp}

\author{Tracey~J.~Turner}
\affiliation{Eureka Scientific, Inc., 2452 Delmer Street Suite 100, Oakland, CA 94602-3017, USA
}
\email{turnertjane@gmail.com}

\author{Yoshihiro Ueda}
\affiliation{Department of Astronomy, Kyoto University, Kitashirakawa-Oiwake-cho, Sakyo-ku, Kyoto 606-8502, Japan
}
\email{ueda@kusastro.kyoto-u.ac.jp}

\author{Satoshi Yamada}
\affiliation{The Frontier Research Institute for Interdisciplinary Sciences, Tohoku University, Aramaki, Aoba-ku, Sendai, Miyagi 980-8578, Japan
}
\affiliation{Astronomical Institute, Tohoku University, Aramaki, Aoba-ku, Sendai, Miyagi 980-8578, Japan
}
\email{satoshi.yamada@terra.astr.tohoku.ac.jp}

\author{Sreeparna Ganguly}
\affiliation{Dipartimento di Fisica, Università di Trento, Via Sommarive 14, Trento 38123, Italy}
\email{sreeparna.ganguly@unitn.it}

\author{Paolo Somenzi}
\affiliation{Dipartimento di Fisica, Università di Trento, Via Sommarive 14, Trento 38123, Italy}
\email{paolo.somenzi@studenti.unitn.it}

\author{Omer~Reich}
\affiliation{Department of Physics, Technion, Technion City, Haifa 3200003, Israel}
\email{reich@campus.technion.ac.il}

\begin{abstract}

The nearby quasar, \pg, has one of the prototype examples of an ultra fast outflow (UFO), as seen in several past \xmm\ and {\it Chandra} observations.
In December 2024, \pg\ was observed simultaneously with \xrism\ Resolve and \xmm, allowing both the Fe K and soft X-ray outflows to be examined at high resolution 
simultaneously. The Resolve spectrum revealed a forest of Fe K band absorption lines from the UFO \citep{Mizumoto25}, comprising of up to six discrete velocity components ranging from 
$v/c=-0.074$ to $v/c=-0.40$. Here we present the simultaneous \xmm\ RGS (Reflection Grating Spectrometer) spectrum, where three lower ionization counterparts of the 
Fe K velocity zones are observed; at $v/c=-0.074, -0.12$ and $-0.33$. The soft X-ray absorbers tend to be somewhat less ionized than their Fe K counterparts, 
with their opacity mainly arising from Fe L shell lines and highly ionized Oxygen. From comparing the Resolve and RGS absorbers, we show that the outflow 
can be parameterized with a density profile varying with radius as $r^{-5/3}$, while the lower ionization zones likely originate from denser clumps of gas. 
Pure electron scattering appears insufficient to provide enough thrust to power the wind, unless sufficient low ionization 
gas capable of radiative line driving exists outside of the line of sight. Overall, \pg\ provides further evidence for the clumpy nature of accretion disk winds, as was recently revealed in the quasar PDS\,456 with \xrism. 

\end{abstract}

\keywords{\uat{High Energy astrophysics}{739} --- \uat{X-ray active galactic nuclei}{2035}}

\section{Introduction}

The discovery of blueshifted absorbers in the X-ray band has provided strong evidence for the existence of Ultra Fast Outflows (UFOs); see \citet{Chartas02,Chartas03,Pounds03,Reeves03} for some of the first known examples. They primarily manifest themselves via absorption lines from K-shell transitions of He or H-like iron, while their measured blueshifts imply outflow velocities of the order of $\sim0.1c$ or higher \citep{Tombesi10,Gofford13,Matzeu23,Yamada24,Laurenti25}. 
Their high velocities and ionizations suggest they originate much closer to the black hole than the slower, less-ionized, warm absorbers \citep{Tombesi13}.  
UFOs are estimated to carry significant kinetic power, of a few per cent or higher of the bolometric luminosity, $L_{\rm bol}$ \citep{Tombesi13,Gofford15,Nardini15,Gianolli24}.  As such, they likely play an important role in black-hole and host galaxy co-evolution \citep{FM00,Gebhardt00}, providing a channel for the mechanical feedback from the accreting black hole out to the interstellar medium of the host galaxy \citep{King03,King10}. 

PG\,1211+143 is a luminous narrow-line QSO (Quasi Stellar Object), at a distance of 331\,Mpc ($z = 0.0809$; \citealt{Marziani96}).  
It is bright in the X-ray and optical bands, with $L_{\rm X}\approx$10$^{44}$\,erg\,s$^{-1}$ and $L_{\rm bol}\approx10^{46}$\,erg\,s$^{-1}$ \citep{Lobban16}. 
Its black hole mass lies in the range $M_{\rm BH}\approx 0.5-1\times10^{8}\,{\rm M}_{\odot}$ \citep{Kaspi00,Peterson04}, which implies that \pg\ likely accretes at close to Eddington. 
Alongside with the luminous quasar, PDS\,456 \citep{Reeves03,Reeves09,Nardini15,Matzeu17}, PG\,1211+143 is famous for being a `prototype' UFO source, as it provided the first detection of a mildly-relativistic outflow in a non-broad-absorption-line (BAL) AGN \citep{Pounds03}, with an outflow velocity of $v_{\rm out} \approx -0.1c$.  In the original 50\,ks 2001 \xmm\ observation, the absorption lines originated both from the iron K-shell lines, as seen in CCD resolution spectra and through blue-shifted lines of C, N, O and Ne observed at higher resolution with \xmm\ RGS (Reflection Grating Spectrometer) in the soft X-ray band. 

A compelling confirmation of the UFO in \pg\ came from a deep 600\,ks {\it XMM/RGS} exposure in 2014 \citep{Pounds16a,Pounds16b, Reeves18}.  
The iron K absorption was confirmed in the mean spectrum, with outflow velocities ranging from $0.06-0.13c$. 
In particular, the RGS spectrum at soft X-rays showed a plethora of blue-shifted absorption lines from N, O, Ne and L-shell Fe, with a common outflow velocity of $v/c\approx-0.06$, as well as possible higher velocity components 
\citep{Pounds16b,Reeves18,PP25}. Soft X-ray UFO variability, on timescales of weeks, was also found from a time dependent analysis of that \xmm\ dataset \citep{Reeves18}. This implied an origin from a clumpy, two phase medium on the size-scale of the AGN broad line region. 

Independently of the \xmm\ studies, a deep (400\,ks) Chandra HETG (High Energy Transmission Grating) exposure also confirmed the presence of the fast soft X-ray wind \citep{Danehkar18}, where the outflow velocity was consistent with the value found from the \xmm\ RGS analysis.  Simultaneously to the Chandra campaign, the {\it HST} Cosmic Origins Spectrograph (COS) also measured a fast UV counterpart of the wind in \pg, 
originating from a broad Ly$\alpha$ trough and with the same outflow velocity ($\approx -0.06c$) as per the X-ray observations \citep{Kriss18}.

While the soft X-ray UFO in \pg\ has been extensively studied via these deep grating exposures, confirming its detection and physical properties, 
the past Fe K band UFO studies have been limited to CCD resolution spectra, e.g. with EPIC-pn \citep{Struder01} and 
EPIC-MOS \citep{Turner01} onboard \xmm.  
Since the launch of \xrism\ in 2024 \citep{Tashiro25}, the Resolve micro-calorimeter spectrometer is revolutionizing our understanding of the Fe K UFO absorbers and AGN outflows generally, through precision, non-dispersive, measurements of the wind velocities and kinematics. This is made possible through the exquisite resolution (5\,eV FWHM at 6\,keV) of the Resolve array. 
The first observations of the Fe K band UFOs in AGN with \xrism\ reveal an unexpected picture. All of the examples of the UFOs to date appear to be clumpy or highly structured and variable, 
e.g. PDS\,456 \citep{xrism25,Xu25}, IRAS~05189$-$2524 \citep{Noda25}, \pg\ \citep{Mizumoto25}, NGC\,3783 \citep{Mehdipour25,Gu25}, NGC\,4151 \citep{Xiang25}. 
This challenges the previous picture of smooth, wide angle accretion disk winds as the origin of the UFOs \citep{PR09, Nardini15} and may have implications for the wind driving and energetics, as well as the efficiency of AGN feedback on larger kilo-parsec scales \citep{Wagner13,Hopkins16}. In the case of PDS\,456, the mass outflow rate was found to be very high, at least $50\,{\rm M}_{\odot}$\,yr$^{-1}$, which exceeded the Eddington accretion rate \citep{xrism25}.  

\begin{deluxetable*}{lccccc}
\tablecaption{Observation log of the 2024 PG\,1211+143 campaign.}
\tablewidth{0pt}
\tablehead{
\colhead{Instrument} & \colhead{Obs. ID} & \colhead{Start Date} & \colhead{End Date} & \colhead{Exposure$^{a}$} & \colhead{Count Rate$^{b}$}}
\startdata
XMM/RGS\,1+2 & 0953790101 & 2024-12-01 04:04:35 & 2024-12-02 03:09:54 & 76.9 & $0.740\pm0.003$\\
& 0953791101 & 2024-12-02 12:57:26 & 2024-12-02 15:49:58 & 5.6 & $0.596\pm0.011$\\
& 0953791201 & 2024-12-03 05:33:26 & 2024-12-03 09:52:15 & 15.1 & $0.648\pm0.007$\\
& Total & $-$ & $-$ & 97.6 & $0.725\pm0.002$\\
XMM/EPIC-pn & Total & $-$ & $-$ & 72.5 & $9.90\pm0.01$\\ 
NuSTAR FPMA+B & 91001637002 & 2024-12-01 07:26:09 & 2024-12-03 06:36:09 & 83.9 & $0.190\pm0.002$\\
XRISM Resolve & 201065010 & 2024-11-29 16:58:38 & 2024-12-04 18:14:23 & 226.8 & $0.0797\pm0.0006$\\
Swift/XRT & 904070--904102 & 2024-11-13 06:34:57 & 2024-12-20 02:04:56 & $33\times1.5$ & $0.19-0.98$
\enddata
\tablenotetext{a}{Net exposure time in units of ks.}
\tablenotetext{b}{Observed background-subtracted count rate in units of ct\,s$^{-1}$.}
\label{tab:obs_log}
\end{deluxetable*}

\pg\ was observed with \xrism\ in December 2024 in order to obtain the first high resolution spectrum of its iron K band UFO. The observation was coordinated with TOOs (Target of Opportunities) from \xmm\ in order to obtain simultaneous RGS spectroscopy of the soft X-ray wind, with \nustar\ to measure the high energy continuum, while \swift\ completed the picture with daily UV and X-ray monitoring throughout the campaign. 
The detailed spectra of the Fe K band UFO with \xrism\ Resolve was reported by \citep{Mizumoto25}; hereafter paper I. The Fe K UFO was resolved into a forest of narrow $\sigma\ls 1000$\,km\,s$^{-1}$ 
absorption lines, arising from up to six different velocity components. These components (zones 1--6 in paper I) have outflow velocities of $v/c=-0.074, -0.120, -0.121, -0.287, -0.357, -0.405$. 
The three slow components are consistent the outflow velocities reported in the earlier \xmm\  spectra \citep{Pounds03,Pounds16a, Pounds16b}, 
The \xrism\ observations supports the emerging picture of clumpy or stratified accretion disk winds in AGN; e.g. PDS\,456 \citep{xrism25}, IRAS~05189$-$2524 \citep{Noda25}. 

In this paper, we present the simultaneous RGS spectrum of \pg\ in 2024, which reveals the soft X-ray counterparts of the Fe K wind components seen with \xrism. 
The unique combination of RGS and Resolve makes it possible to study both the Fe K and soft X-ray UFOs at high resolution, thereby providing a complete picture of the overall wind properties. 
The 2024 campaign is described in Section~2, the broad-band X-ray spectrum is described in Section~3 and the soft X-ray UFO components with RGS are modeled in Section~4. 
Section~5 presents a joint RGS and Resolve analysis, comparing the different wind components in velocity and ionization, while the implications are discussed in Section~6.

In the subsequent analysis, outflow velocities are given in the rest-frame of PG\,1211+143 at $z=0.0809$, 
accounting for relativistic Doppler shifts along the line of sight, where negative values of $v/c$ correspond to blue-shifted velocities. 
The following prescription is used:-
\begin{equation}
\frac{v_{\rm out}}{c} = \frac{(1 + z_{\rm obs})^2 - (1 + z_{\rm QSO})^2}{(1 + z_{\rm obs})^2 + (1 + z_{\rm QSO})^2}
\end{equation}

\noindent where $z_{\rm obs}$ is the redshift of an absorption system in the observed frame and 
$z_{\rm QSO}$ is the QSO redshift ($z=0.0809$).
Errors are quoted at 90\% confidence for one parameter of interest ($\Delta \chi^{2}$ or $\Delta C = 2.7$).  All subsequent fits are performed using the \textsc{xspec} spectral fitting package \citep{Arnaud96}.
$\chi^2$ statistics are used for the EPIC-pn and {\it NuSTAR} data and the C-statistic \citep{Cash79} is used for {\it Resolve} and RGS. 

\begin{figure*}
\begin{center}
\rotatebox{-90}{\includegraphics[height=8cm]{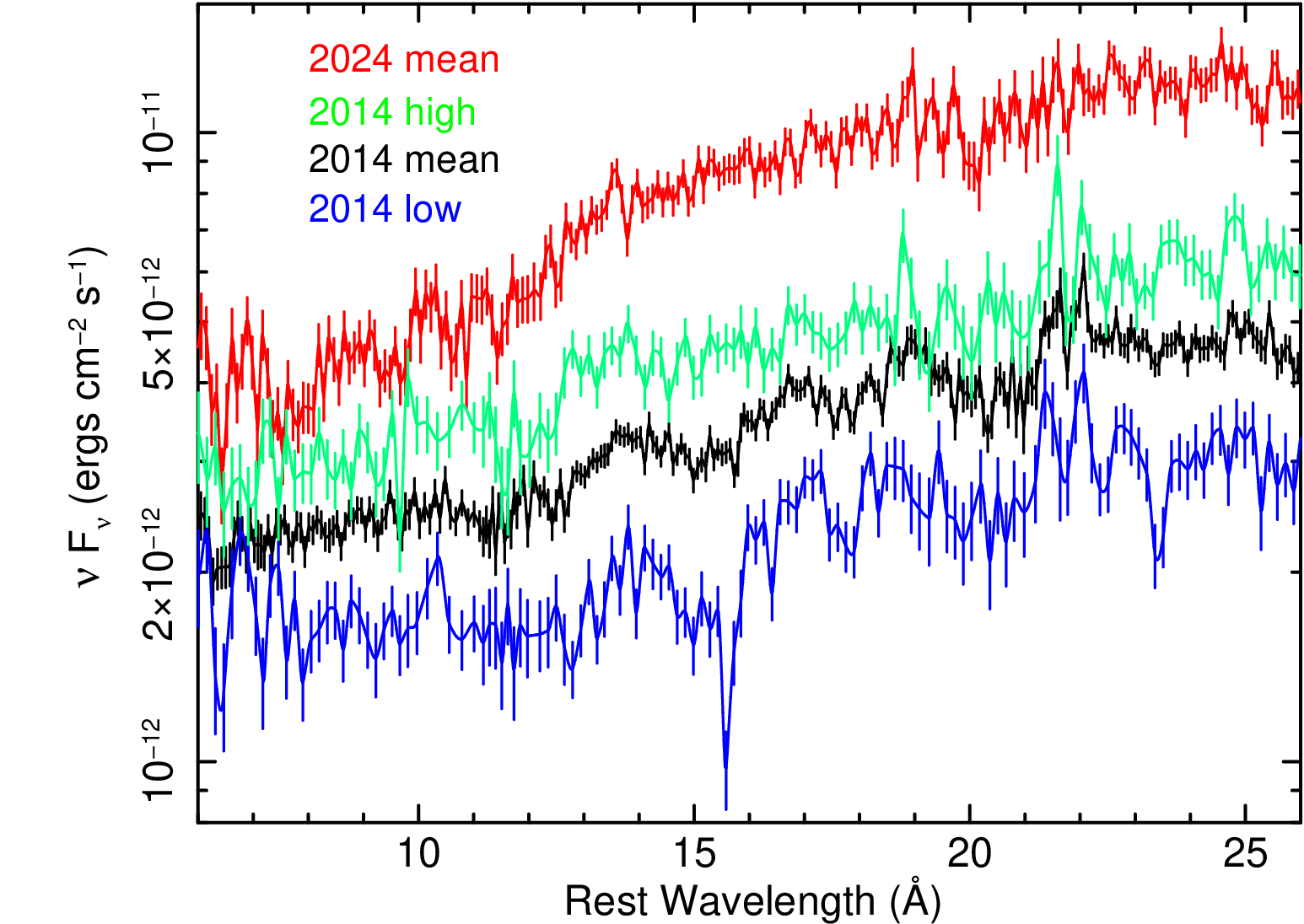}}
\rotatebox{-90}{\includegraphics[height=8cm]{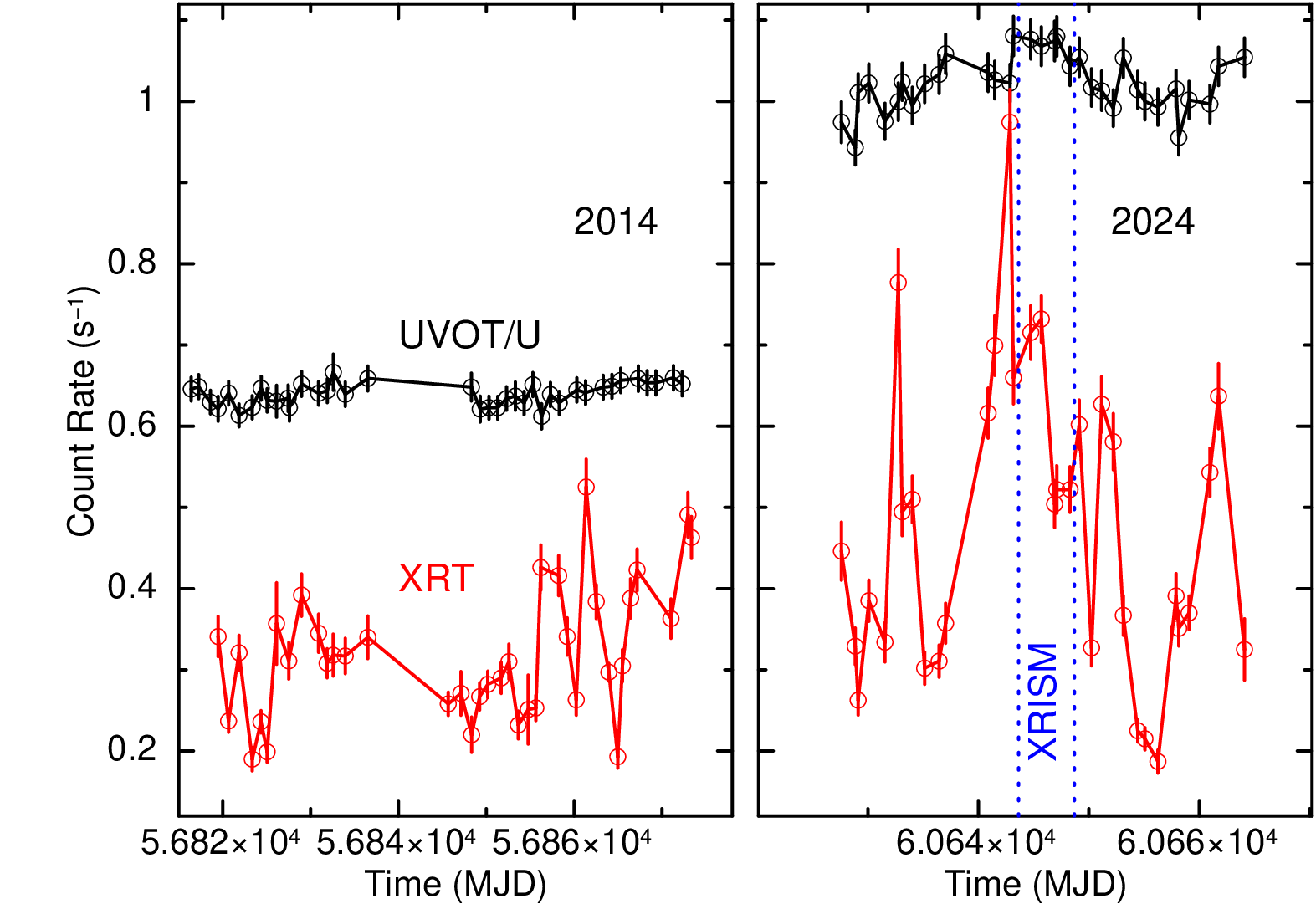}}
\end{center}
\caption{The X-ray variability of \pg. The left panel shows the 2024 RGS spectrum of \pg\ (red points), which is compared to the previous 2014 epoch \citep{Reeves18}. Here, the mean 2014 spectrum from 7 \xmm\ orbits is shown in black, 
the 2014 low flux spectrum (revolution 2659) is in blue and the 2014 high flux spectrum (revolution 2664) is in green. The 2024 observation is brighter than all of the previous epochs. The right panels show the Swift XRT (0.3--10\,keV, red) and Swift UVOT (U band, black) light-curves comparing the 2014 versus 2024 campaigns. Dotted vertical blue lines corresponds to the interval of the 2024 {\it XRISM} observation. Overall, \pg\ is brighter in both the UV and X-rays in 2024, where the \xrism\ and \xmm\ observations caught the AGN near to a strong flare. Note the U data points are divided by 100 to fit the scale of the plot.}
\label{fig:lightcurve}
\end{figure*}

\section{The 2024 Campaign on \pg.}

The coordinated observations of PG\,1211+143 were performed in December 2024, with \xrism, \xmm, \nustar\ and \swift; see the observation log in Table~\ref{tab:obs_log} for a summary. The observations were co-ordinated around the longer \xrism\ observation, 
as is presented in paper I, which overlaps both the \xmm\ and \nustar\ exposures. The \xmm\ observation, with the RGS \citep{denHerder01} as prime instrument, was split into three sequences and over two satellites orbits as per Table~\ref{tab:obs_log}, but is dominated by the first and longest sequence from the first orbit. 

The RGS spectra were extracted using the latest version of the \xmm\ Scientific Analysis Software (\textsc{sas}\footnote{\url{http://xmm.esac.esa.int/sas/}}) package and with the latest calibration files at the time of the observation. 
The \textsc{rgsproc}\footnote{\url{http://xmm-tools.cosmos.esa.int/external/sas/current/doc/rgsproc/}} script was used to extract first-order dispersed spectra from each of the two RGS modules.  Observations were screened for periods of high background by examining light-curves from the CCD closest to the optical axis of the telescope, resulting in a total net exposure of 97.6\,ks. 
As there was no spectral variability between the three sequences and only very modest ($\sim20$\%) changes in flux, the data from all three were combined. RGS\,1 and RGS\,2 spectra were also found to be consistent within errors 
and subsequently were combined using the \textsc{sas} task \textsc{rgscombine} into a single mean RGS spectrum for the 2024 observation. The spectrum was binned to 50\,m\AA\ bins, which slightly over-samples the FWHM resolution of the RGS array of 60--80\,m\AA\ \citep{denHerder01}. 

The mean RGS\,1+2 (combined) count rate was $0.725\pm0.002$\,ct\,s$^{-1}$. This is substantially higher than the count rate range of $0.17-0.36$\,ct\,s$^{-1}$ observed from the previous 2014 \xmm\ RGS observations \citep{Lobban16,Reeves18}, where this earlier campaign encompassed seven satellite orbits (\xmm\ revolutions 2652, 2659, 2661, 2663, 2664, 2666, 2670). Figure~\ref{fig:lightcurve} (left panel) shows the 2024 fluxed RGS spectrum versus the mean, lowest (revolution 2659) and 
highest flux (revolution 2664) spectra from the 2014 campaign. The 2024 spectrum is brighter even compared to the highest flux 2014 spectrum; for example over the full 0.3--2.0\,keV soft X-ray band, the 2024 flux is 
$1.5\times10^{-11}$\,\cgs, whereas the 2014 observations covered a flux range of $4.0-8.0\times10^{-12}$\,\cgs. 

Long term spectral variability is also observed between the different RGS epochs, where the 2024 spectrum is softer versus the 2014 ones. Furthermore, no absorption is present in the 15--17\,\AA\ band, which is coincident with the expected wavelengths of the Fe M-shell Unresolved Transition Array or UTA \citep{Sako01,Behar01}. This suggests that any low ionization absorption diminishes with increasing soft X-ray flux and the 2024 observations caught \pg\ during a relatively bare state, in contrast to the lowest flux 2014 spectrum. 

Further context on the long term variability of \pg\ can be obtained from the \swift\ monitoring observations. During 2024, $33\times1.5$\,ks \swift\ XRT (X-ray Telescope; \citealt{Burrows05}) snapshots were obtained, on an approximately daily basis from 2024/11/13 -- 2024/12/20 and encompassing the \xrism\ campaign; see Table~\ref{tab:obs_log}. A similar \swift\ monitoring campaign was also performed in 2014, coincident with these seven earlier \xmm\ sequences. Both the 2014 and 2024 monitoring observations 
were performed with the U, UVW1 and UVW2 photometric filters onboard the \swift\ UVOT (Ultra Violet and Optical Telescope; \citealt{Roming05}). 

Figure~\ref{fig:lightcurve} (see the two righthand panels) shows the Swift XRT and UVOT (U band) light-curves in 2014 vs 2024, where the XRT light-curve was extracted over the 0.3--10\,keV band\footnote{The Swift XRT lightcurves were processed from the UK Swift Science Data Centre at www.swift.ac.uk.}. Overall, \pg\ was brighter in the X-ray band in 2024, whereby the average XRT count rate increased by about 50\%, from $0.32\pm0.01$ in 2014 to $0.48\pm0.03$\,ct\,s$^{-1}$ in 2024. 
The long term increase in X-ray flux was also accompanied by an increase in the U band flux by a similar 50\% factor, with 
the same behavior also seen in the shorter wavelength UVW1 and UVW2 light-curves. On shorter timescales of days, the \xrism\ and \xmm\ observations occurred just after the peak in the XRT light curve due to a strong X-ray flare. 
Thus the 2024 observations have captured an historically bright state of \pg, with the increase in UV flux possibly corresponding to an increase in the overall accretion rate since 2014.   

\begin{figure*}
\begin{center}
\rotatebox{-90}{\includegraphics[height=8.5cm]{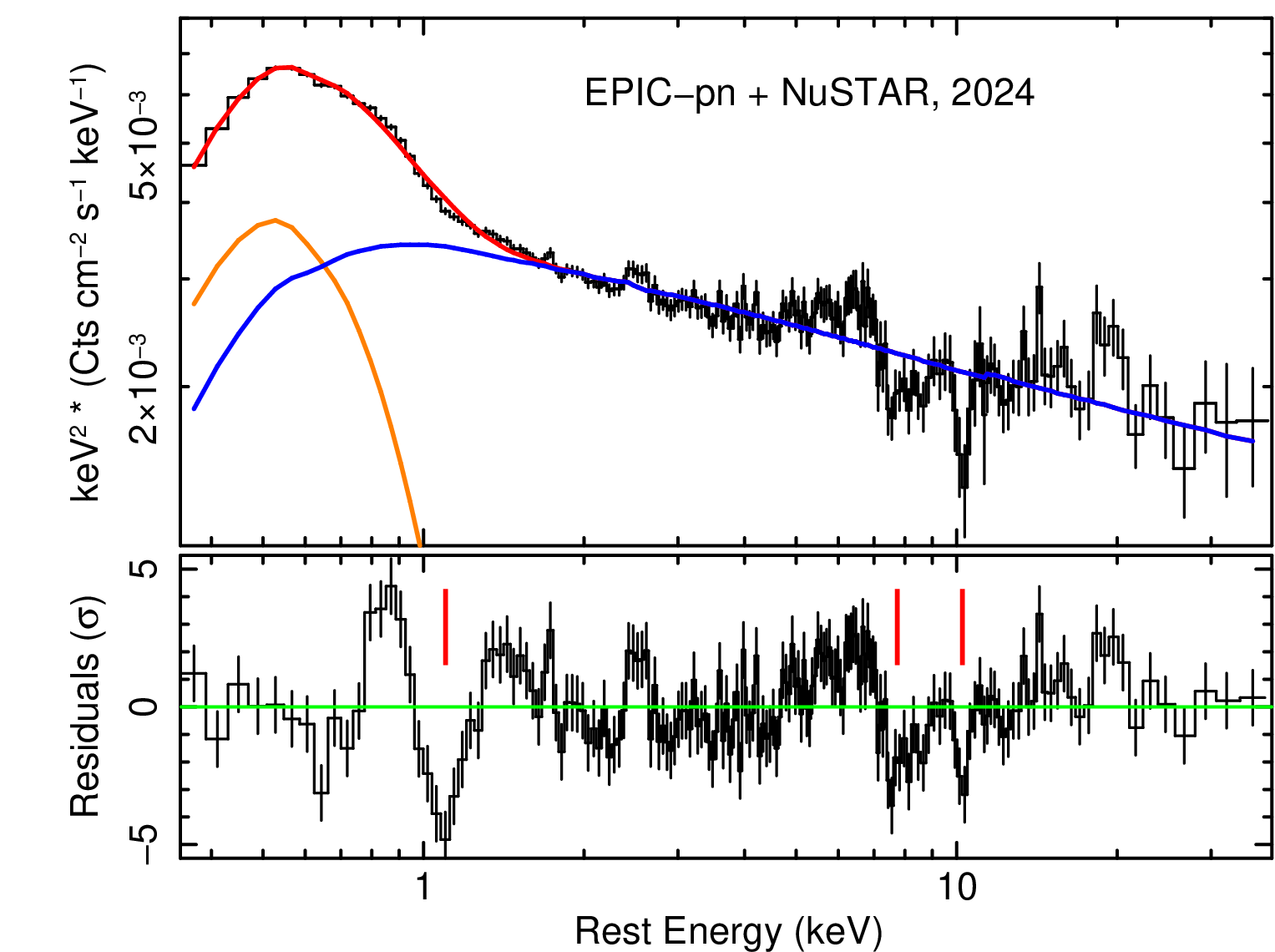}}
\rotatebox{-90}{\includegraphics[height=8.5cm]{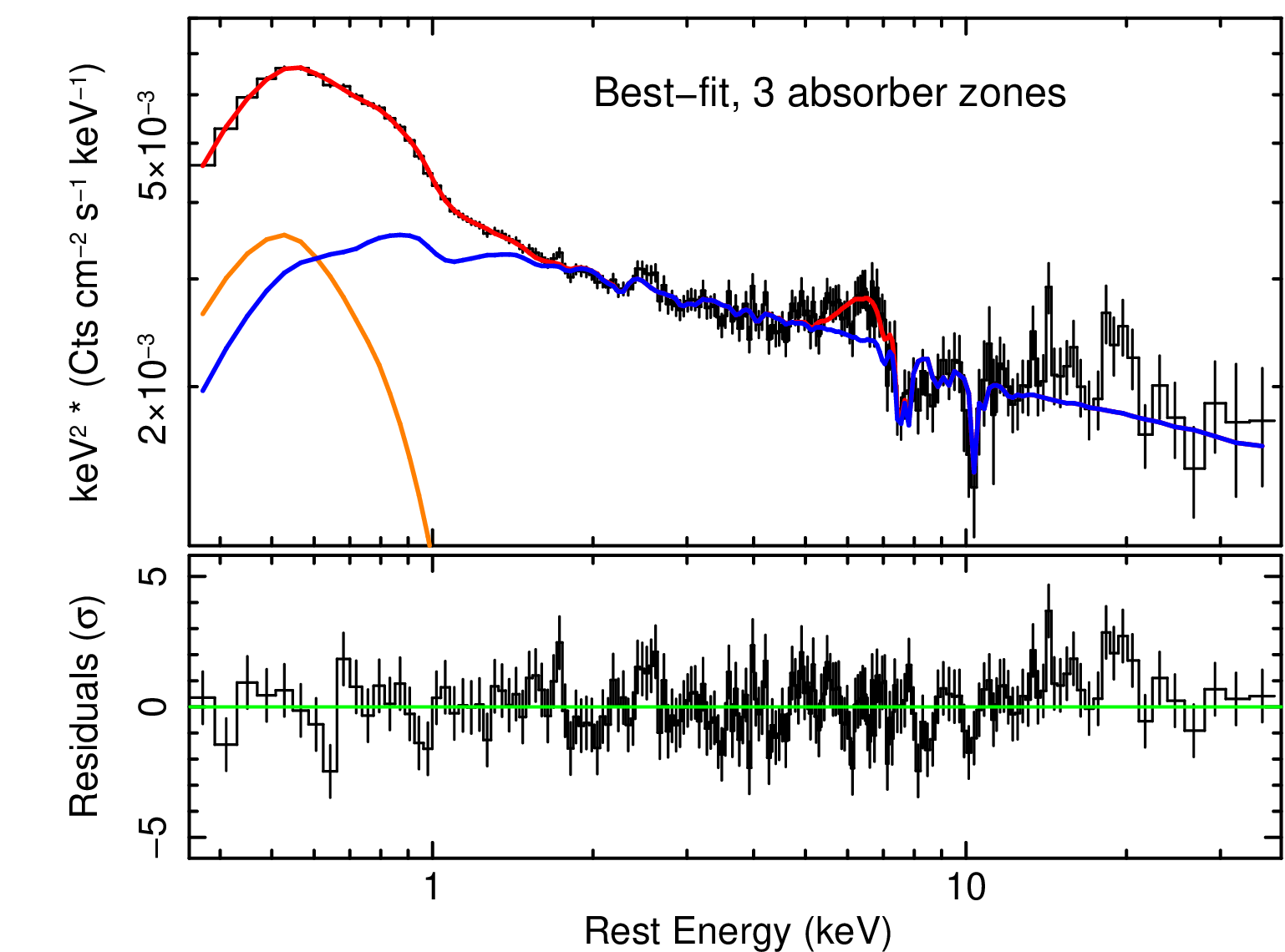}}
\rotatebox{-90}{\includegraphics[height=14cm]{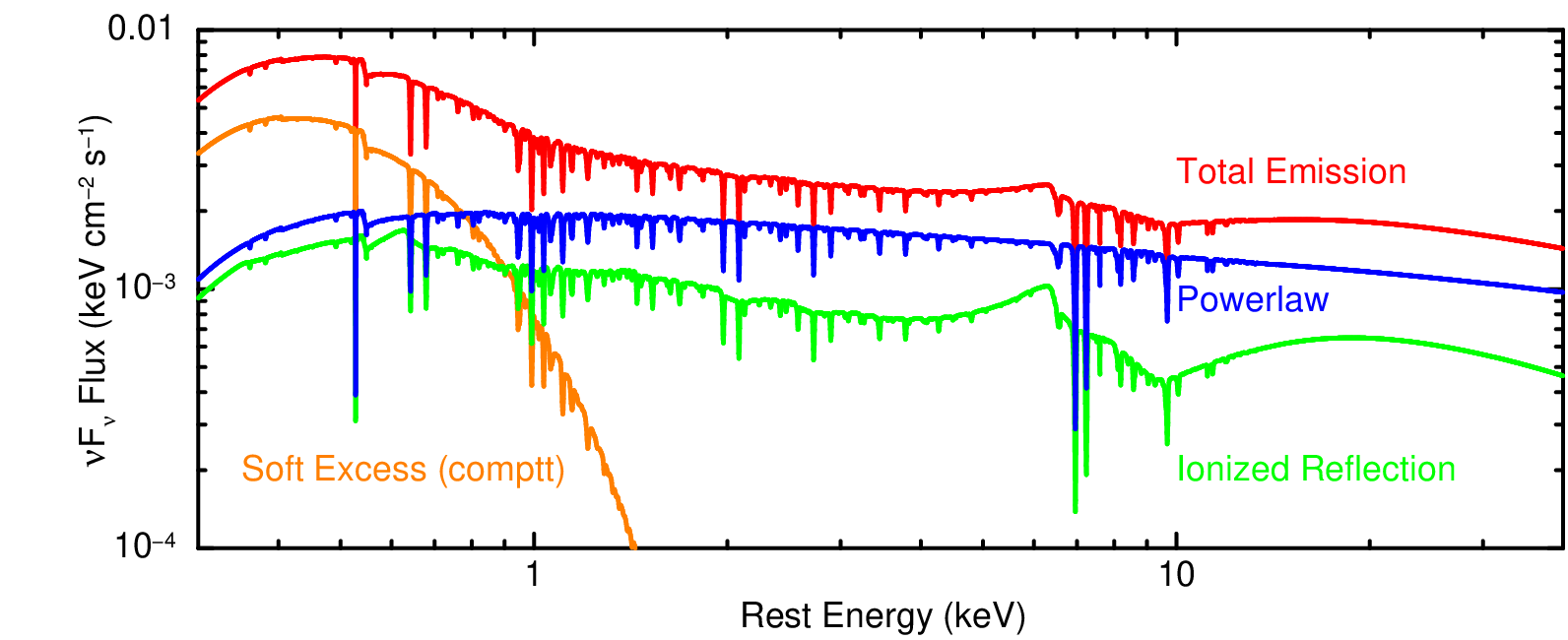}}
\end{center}
\caption{2024 \xmm\ EPIC-pn and {\it NuSTAR} spectrum of \pg, fitted from 0.3--40\,keV in the AGN rest frame. The upper panels show the data points (black) fitted with a simple continuum model, consisting of a $\Gamma\approx2.2$ power-law (blue line) and a soft X-ray excess represented by a warm Comptonization component (orange), while the total emission is in red. The lower panel shows the $\sigma$ residuals of the data about this model, which show strong emission and absorption features. The energies of the UFO absorber troughs are marked by solid red lines. The upper right panels plot the best fit model, modified by the addition of three UFO absorbers, as well as a broad iron K$\alpha$ line. The UFO absorbers correspond to outflow velocities of $-0.07c$, $-0.12c$ and $-0.40c$ and are parameterized in Table~2. The lower panel shows the best-fit model and continuum components, after the addition of a highly ionized 
reflector.}
\label{fig:pn}
\end{figure*}

\section{The Broad Band X-ray Spectrum}

The aim of this paper is to compare any soft X-ray outflow components with the 6 possible UFO absorbers observed in {\it Resolve}, in order to fully characterize the fast wind in \pg. 
The first step is to model the broad band continuum seen in the EPIC-pn and {\it NuSTAR} spectrum. 
The 2024 spectra from EPIC-pn and \nustar\ were used, where their reduction is described in paper 1 and their exposures and count rates are summarized in Table~1. The pn and \nustar\ spectra were analyzed over the 0.3--10\,keV and 3--35\,keV observed frame bands respectively and were binned to 100 counts per bin. The separate \nustar\ FPMA+B spectra were consistent and were subsequently combined into a single 
spectrum, while data above 40\,keV was not utilized as the background dominates at high energies due to the steep hard X-ray spectrum of \pg. 
The update to the high energy effective area of the EPIC-pn was applied, via the \textsc{arf} response file, which produces a better cross-calibration agreement between EPIC-pn and \nustar\ over the 
overlapping 3--10\,keV band.\footnote{See https://xmmweb.esac.esa.int/docs/documents/CAL-TN-0230-1-3.pdf for details.}

The 2024 Spectral Energy Distribution (SED) of \pg\ is described in paper I; see the corresponding Figure~2. A prominent soft X-ray excess is observed below 1\,keV, which forms part of the UV to soft X-ray bump, while at high energies a steep power-law, of photon index $\Gamma\approx2.2$ is observed up to 40\,keV. Thus a two component continuum was adopted to model the simultaneous pn and \nustar\ spectrum, consisting of a high energy 
power-law and a Comptonized disk spectrum for the soft excess described by the \textsc{comptt} model \citep{Titarchuk94}. For the latter, the temperature of the input seed photons was set to $kT=20$\,eV, 
which is consistent with the expected peak of the SED in the UV.  
Both continuum components were absorbed by a neutral Galactic absorber, via the \textsc{xspec} \textsc{tbabs} model \citep{Wilms00}, 
This was set to the expected column density of $N_{\rm H}=3\times10^{20}$\,cm$^{-2}$ \citep{Kalberla05}. 
A multiplicative cross normalization factor was included between the pn and {\it NuSTAR} spectra. 

\begin{deluxetable*}{lcccccc}
\tabletypesize{\footnotesize}
\tablecaption{XMM-Newton and NuSTAR Spectral fits to PG\,1211+143..}
\tablewidth{0pt}
\tablehead{
\colhead{} & \colhead{pn/NuSTAR} & \colhead{} & \colhead{RGS}}
\startdata
Continuum:-\\							
$\Gamma$ & $2.25\pm0.02$ & & $2.25^f$ \\
$N_{\rm PL} \times10^{-3}$$^a$ & $3.4\pm0.1$ & & $3.0\pm0.2$ \\
$kT$\,(eV)	& $135\pm4$ & & $147\pm8$ \\
$\tau$ & $28\pm2$ & & $26\pm3$ \\
Flux (0.3--2\,keV)$^b$ & $1.55$ & & $1.47$\\
Flux (2--10\,keV)$^b$ & $0.60$\\
\hline
Emission Lines:-\\
Identification & Fe K$\alpha$ & & N\,\textsc{vii} & O\,\textsc{viii} & Ne\,\textsc{ix}\\
$E\,({\rm keV})$ & $6.34\pm0.17$ & & $0.504\pm0.003$ & $0.6549\pm0.0007$ & $0.920\pm0.003$\\
$\lambda$\,(\AA) & & & $24.58\pm0.15$ & $18.93\pm0.02$ & $13.48\pm0.04$\\
$\sigma\,({\rm eV})$ & $540^{+240}_{-190}$ & & $3.4^{+2.5}_{-1.6}$ & $<1.2$ & $5.6\pm1.8$\\
FWHM (km\,s$^{-1}$)$^f$ & & & $4700^{+3600}_{-2200}$ & $<1300$ & $4300^{+2200}_{-1400}$\\
$v_{\rm out}$\,(km\,s$^{-1}$)$^f$ & & & $-2400\pm1800$ & $-630\pm310$ & $-4800\pm900$$^{j}$ \\ 
$N_{\rm Gauss}\times10^{-5}$$^c$ & $1.5\pm0.4$ & & $4.8\pm2.6$ & $2.6\pm1.3$ & $4.9\pm1.7$\\	
${\rm EW}\,({\rm eV})$ & $207\pm55$ & & $1.0\pm0.5$ & $1.2\pm0.6$ & $5.7\pm1.8$\\
$\Delta$ C / $\Delta \nu$$^d$ & $-150.7/3$ & & $-9.6/3$ & $-11.7/2$ & $-31.2/3$\\
$P_{\rm AIC}$$^i$ & $3.8\times10^{-32}$ & & $0.061$ & $0.021$ & $3.4\times10^{-6}$\\	
\hline
Fe K Absorption:-\\							
Zone	 & Fe K Absorber~1 & Fe K Absorber~2\\
$N_{\rm H}\times10^{22}$\,cm$^{-2}$ & $13.2^{+6.1}_{-3.9}$ & $6.0^{+2.0}_{-2.6}$\\     
$\log \xi$$^e$ & $4.86^{+0.20}_{-0.17}$ & $4.65^{+0.24}_{-0.21}$ \\ 
$v/c$$^g$ & $-0.117\pm0.007$ & $-0.41\pm0.01$\\    
$\Delta\chi^2$ / $\Delta \nu$$^d$ & $-51.4/3$ & $-18.5/3$\\ 
$P_{\rm AIC}$$^i$ & $1.4\times10^{-10}$ & $1.9\times10^{-3}$\\
\hline
Soft Absorption:-\\							
Zone	 & pn soft & & RGS Zone\,0 & RGS Zone\,1b & RGS Zone 3b & RGS--fast\\
$N_{\rm H}\times10^{22}$\,cm$^{-2}$ & $0.80^{+0.35}_{-0.25}$ & & $(4.6\pm1.7)\times10^{-3}$ & $0.11\pm0.04$ & $0.81^{+0.34}_{-0.31}$ & $3.6^{+1.6}_{-1.5}$\\
$\log \xi$$^e$ & $3.91\pm0.09$ & & $0.85\pm0.31$ & $3.64\pm0.16$ & $4.16^{+0.14}_{-0.18}$ & $4.35^{+0.29}_{-0.13}$\\
$v/c$$^g$ & $-0.071\pm0.012$ & & $-0.078\pm0.001$$^h$ & $-0.0745\pm0.0017$ & $-0.120\pm0.002$ & $-0.331\pm0.002$\\
$\Delta$ C / $\Delta \nu$$^d$ & $-141.6/3$ & & $-22.4/3$ & $-22.1/3$ & $-29.5/3$ & $-23.1/3$\\
$P_{\rm AIC}$$^i$ & $3.6\times10^{-30}$ & & $2.7\times10^{-4}$ & $3.2\times10^{-4}$ & $7.9\times10^{-6}$ & $1.9\times10^{-4}$\\ 
\hline
Fit Statistic & $912.6/861$	 & & $551.9/602$		
\enddata
\tablenotetext{a}{Power-law flux in units of $\times10^{-3}$\,photons\,cm$^{-2}$\,s$^{-1}$\,keV$^{-1}$ at 1\,keV.}
\tablenotetext{b}{0.3--2\,keV and 2--10\,keV flux in units of $\times10^{-11}$\,ergs\,cm$^{-2}$\,s$^{-1}$.}
\tablenotetext{c}{Line flux in units of $\times10^{-5}$\,photons\,cm$^{-2}$\,s$^{-1}$.}
\tablenotetext{d}{Improvement in $\Delta C$ (or $\Delta\chi^2$ for pn) upon adding the line to the model. $\Delta\nu$ is the change in degrees of freedom.}
\tablenotetext{e}{Units of Ionization ($\xi$) are ergs\,cm\,s$^{-1}$. }
\tablenotetext{f}{Kinematics of RGS emission lines. These are the FWHM velocity width and the outflow velocity, in units of km\,s$^{-1}$.}
\tablenotetext{g}{The redshift is fitted as a proxy for the outflow velocity. This translates into an outflow velocity as per equation~1 in the text.}
\tablenotetext{h}{The best-fit redshift of zone\,0 is $z=(0.0\pm1.2)\times10^{-3}$ and is consistent with our Galaxy at $z=0$.}
\tablenotetext{i}{Null hypothesis probability derived from the AIC statistic.}
\tablenotetext{j}{The blue-shift of the Ne\,\textsc{ix} line depends on its identification. The maximum value of 4800\,km\,s$^{-1}$ is for an identification with the forbidden line at 13.70\,\AA. 
The velocity is lower compared to the inter-combination lines (13.55\,\AA) and is not required for the resonance line (13.70\,\AA).}
\label{tab:emission-lines}
\end{deluxetable*}

The resulting continuum parameters are listed in Table~2 and the fitted spectrum is shown in Figure~2. The electron temperature of the soft X-ray excess is $kT=135\pm4$\,eV and is optically thick ($\tau>>1$), while the power-law is steep with $\Gamma=2.25\pm0.02$. The cross normalization between \nustar\ and EPIC-pn is $1.06\pm0.02$. While this model provides a good description of the broad-band X-ray continuum between 
0.3--40\,keV, strong absorption and emission residuals are present in the data; see Figure~2 (left panel). The resulting fit statistic is also unacceptable, with $\chi^2/\nu = 1274.8/873$. 
As per paper I, a broadened P-Cygni like Fe K emission and absorption feature is observed either side of 7\,keV, while a secondary absorption trough is also seen just above 10\,keV in the QSO rest frame. 
Most notable at soft X-rays is a strong absorption trough located just above 1\,keV, which coincides with the Fe L-shell absorption band.

The absorption features were modeled with \textsc{xstar} \citep{Kallman04}, using the same multiplicative tables described in paper I and which adopts the SED of \pg\ from the 2024 campaign as the photo-ionizing continuum. 
The turbulence velocity (velocity width) of the absorbers was fixed to 1000\,km\,s$^{-1}$ as this could not be determined in the CCD resolution spectrum.  
Three absorption components were required to model the absorption structure in the broad band spectrum, together with a broad emission component at Fe K$\alpha$\ described by a Gaussian profile. The properties of the absorbers are listed in Table~2; one models the soft band (pn soft absorber in Table~2) and the other two the absorption troughs in the Fe K band (Fe K absorbers 1 and 2). 
The broad Fe K$\alpha$ emission is consistent with what was found in paper I.

The outflow velocities of the two Fe K absorbers are $v/c=-0.117\pm0.007c$ and $v/c=-0.41\pm0.01$; the former closely corresponds to the zone 3 absorption trough observed in the {\it Resolve} spectrum ($v/c=-0.120$), while the velocity of the second absorber above 10\,keV is consistent with the highest velocity zone\,6 in {\it Resolve} 
($v/c=-0.405$) and provides independent evidence of this fast zone. The soft absorber has a lower velocity $v/c=-0.071\pm0.012$ and is somewhat less ionized, its velocity is consistent with the zone\,1 absorber observed in {\it Resolve} (with $v/c=-0.074$). The latter may be produced by a blend of Fe L absorption lines, which when convolved with the CCD response function of EPIC-pn, appear as a broadened absorption trough. 
All three zones are formally significant according to the Akaike Information Criteria \citep{Akaike74}, as is listed in Table~2.
The fit statistic is then acceptable, where $\chi^2/\nu = 912.6/861$ and no significant residuals in emission or absorption are present in the data (Figure~2, right panel).

\subsection{Reflection Models}

As a final step, the broad iron emission line was replaced by the \textsc{relxill} relativistic reflection model \citep{Garcia14} to model any disk reflection. As is seen in Figure~2, the \nustar\ data above 10\,keV does not show a strong hard X-ray excess associated with the Compton hump. As a result, it is difficult to model the data with a neutral or low ionization reflector which then over predicts the hard X-ray flux. 
However a highly ionized reflector can model the spectrum, as essentially the scattering medium is more reflective, with less photoelectric absorption at lower energies and this lack of contrast with the continuum produces a weaker Compton hump. 

The final model fit, showing the different continuum components including the ionized reflection, is shown in the lower panel of Figure~2. The reflector has a high ionization parameter, with $\log\xi = 3.35\pm0.12$ and a has a flux ratio of $R=0.58^{+0.17}_{-0.12}$ compared to the input power-law. 
The inclination of the reflector was found to be $\theta=33\pm4\degg$ with an inner radius of $R_{\rm in}=8^{+8}_{-4}$\,R$_{\rm g}$ (where $R_{\rm g}$ is the gravitational radius). The power-law disk emissivity
was fixed to $q=3$ and as the spin could not be determined it was fixed to the stationary value ($a=0$), while Solar abundances are assumed. 
The warm Comptonization (\textsc{comptt}) component still dominates the soft X-ray band below 1 keV, with parameters that are consistent with what was found above 
($kT=140\pm5$\,eV), while the ionized reflector is featureless at low energies. The resulting power-law photon index is similar to the previous value, of $\Gamma=2.18\pm0.12$, while the overall fit statistical is acceptable with $\chi^2/\nu = 879.5/860$. The addition of the reflection component has no effect on the absorber parameters.

Overall, the reflection could originate from a highly ionized inner disk, but given its high ionization it may also arise from scattering off the surface of the wind, e.g. \citet{Sim08,Sim10}. The physical origins of the reflected emission will be considered in future works. The modeling tends to favor a warm corona \citep{Rozanska15,Petrucci18,Petrucci20} for the origin of the soft X-ray excess in \pg\ and is physically distinct from the reflection component.

\section{RGS Spectral Analysis}

\subsection{Initial Identification of the Absorption Lines}\label{sec:emission}

The above baseline continuum model, without the ionized absorbers, was then applied to the 2024 RGS spectrum. Note the power-law photon index was fixed at the above best fit value ($\Gamma=2.25$) as the RGS is less sensitive to its exact slope above 2\,keV. The residuals of the RGS data to this continuum are plotted in Figure\,3, showing the whole spectrum and then as zoom in portions around the Fe-L and O K-shell bands. The rest-frame band between 7--13\,\AA\ shows that the single broad absorption trough in the lower resolution EPIC-pn spectrum appears to be split into several narrower absorption lines. This occurs in the region of the spectrum where there are a multitude of Fe L-shell absorption lines from Fe\,\textsc{xix} up to Fe\,\textsc{xxiv}.   
Narrow absorption troughs are also observed between 16--20\,\AA\ and can correspond to the blueshifted absorption lines of O\,\textsc{vii} and O\,\textsc{viii}. At longer wavelengths above 21\,\AA, no strong residuals are observed and the RGS spectrum is well accounted for by the continuum model. 

\begin{figure*}
\begin{center}
\rotatebox{-90}{\includegraphics[height=16cm]{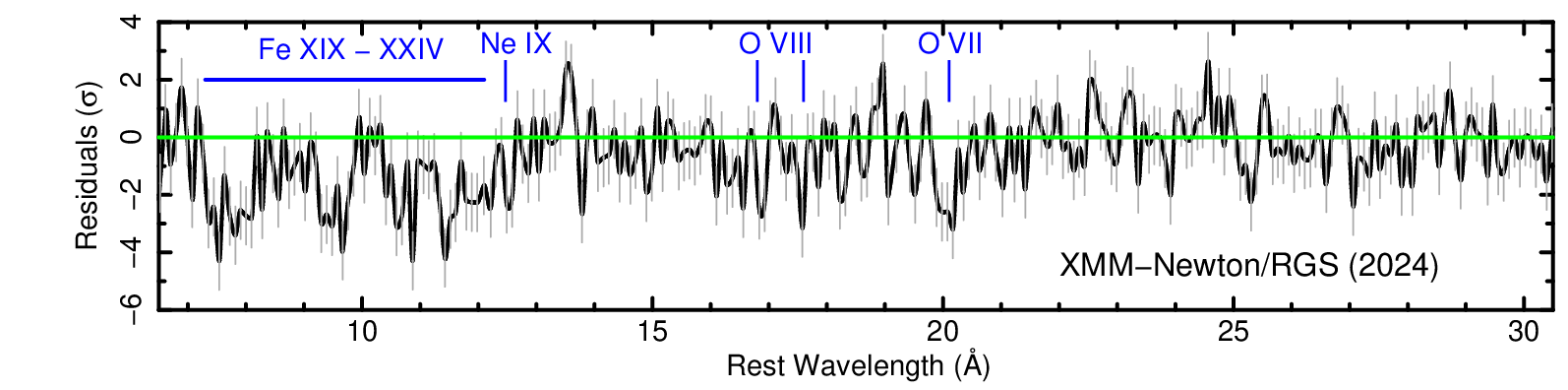}}
\rotatebox{-90}{\includegraphics[height=8.5cm]{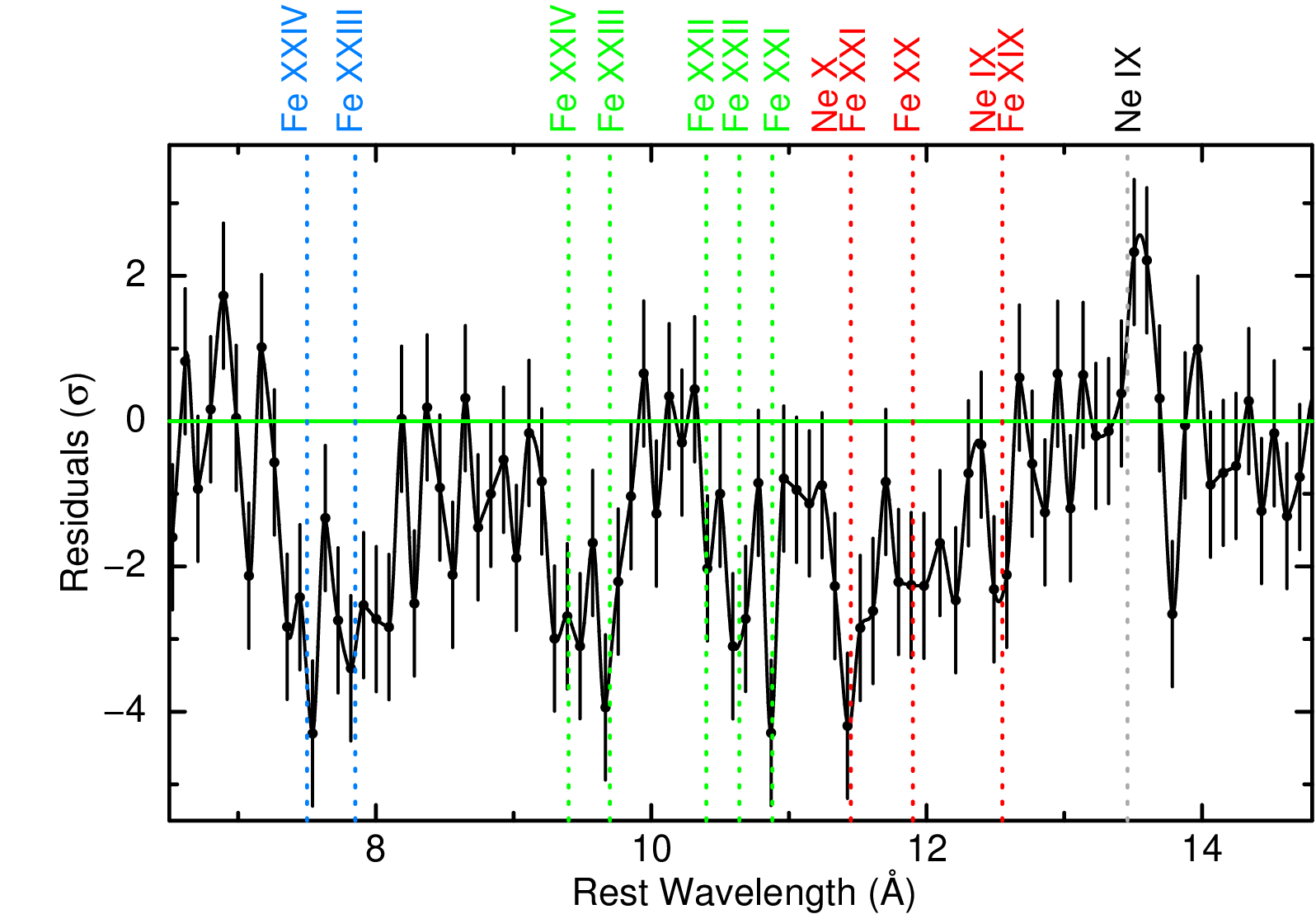}}
\rotatebox{-90}{\includegraphics[height=8.5cm]{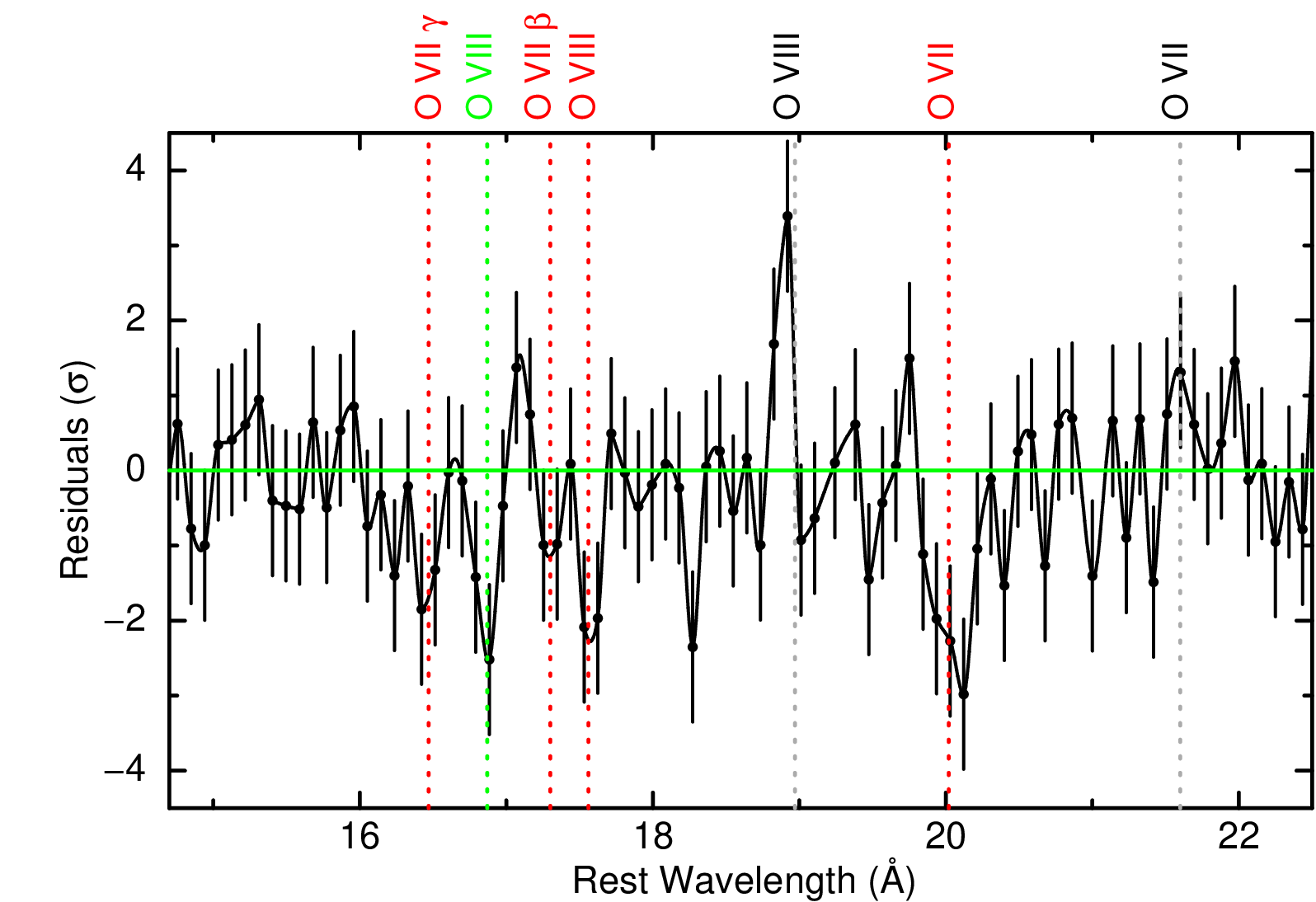}}
\end{center}
\caption{The 2024 RGS spectrum of \pg, plotted as residuals (in $\sigma$) to the power-law plus \textsc{comptt} continuum model described in the text. 
The upper panel shows the broad $7-30$\,\AA\ range, with the strongest absorption features occurring in the short wavelength portion of the spectrum, e.g. due to Fe L-shell ions. 
The lower panels shows zoom-in portions around the Fe L band (left) and the O K-shell band (right). The likely identification of the blue-shifted absorption troughs are marked, as per Table~3 and are color coded 
according to their outflow velocity. Here, red corresponds to $-0.074c$, green to $-0.120c$ and blue to $-0.33c$. The black labels correspond to the expected positions of the O\,\textsc{vii}, O\,\textsc{viii} and
Ne\,\textsc{ix} resonance emission lines in the AGN rest frame, with no velocity shift.} 
\label{fig:rgs}
\end{figure*}

Table~\ref{tab:absorption-lines} lists the centroid rest wavelengths of the most prominent absorption lines observed in the spectrum. With the exception of the weak higher order lines of O\,\textsc{vii}, these deviate from the continuum at $3\sigma$ significance or higher. Their most likely atomic identification and corresponding laboratory frame wavelength are also listed. As has been seen in previous RGS analysis of \pg\ \citep{Pounds03,Pounds16b, Reeves18}, all of the lines appear to be systematically blue-shifted with respect to their expected laboratory wavelengths. In the 2024 data, they appear to be split into 3 different outflow velocity groups, $v/c=-0.075$, $v/c=-0.12$ and $v/c=-0.33$, which is indicative of multiple velocity components. These velocity groups are color coded in Figure~3 (lower panels), along with their prospective identifications. 
For example, the two narrow troughs near 17.6\,\AA\ and 16.9\,\AA\ may result from two velocity components of O\,\textsc{viii} Ly$\alpha$ ($1s\rightarrow 2p$) at $v/c=-0.075$ and $v/c=-0.12$ respectively. In contrast, the lower ionization O\,\textsc{vii} He$\alpha$ absorption near 20\,\AA\ appears to show only the lower velocity component. 
Over the Fe L shell  band between $9-13$\,\AA, the multitude of troughs which appear to result from Fe\,\textsc{xix} to Fe\,\textsc{xxiv} blue-shifted by either $v/c=-0.075$ or $v/c=-0.12$, with a tendency for the higher charge states to be associated with the higher velocity component. The shortest wavelength trough, at $7.5-8.0$\,\AA\ in the AGN rest frame, could originate from a blend of the highest ionization iron L lines of Fe\,\textsc{xxiii} (Be-like) and Fe\,\textsc{xxiv} (Li-like) at $v/c=-0.33$. 

There does not appear to be any strong absorption line residuals at or close to their expected wavelength in the AGN rest frame, i.e. with gas with low or zero outflow velocity. For example, near the expected wavelength of 
O\,\textsc{viii} $1s\rightarrow2p$ (Ly-$\alpha$) at 18.97\,\AA, there is no absorption but weak narrow emission. Likewise, there is emission but no absorption near the Ne\,\textsc{ix} $1s\rightarrow2p$ resonance line at 13.45\,\AA. There is neither emission nor absorption at the position of the O\,\textsc{vii} resonance line at 21.6\,\AA.

\begin{deluxetable}{lccc}
\tabletypesize{\small}
\tablecaption{Blueshifted absorption lines identified in the 2024 RGS spectrum of PG\,1211+143.}
\tablewidth{0pt}
\tablehead{
\colhead{Rest $\lambda$ (\AA)$^{a}$} & \colhead{Line ID$^{b}$} & \colhead{Lab $\lambda$ (\AA)$^{b}$} & \colhead{v/c$^{c}$}}
\startdata
20.02 & O\,\textsc{vii} $1s\rightarrow2p$ & 21.60 & $-0.076\pm0.002$\\
17.60 & O\,\textsc{viii} $1s\rightarrow2p$ & 18.97 & $-0.075\pm0.003$\\
17.29 & O\,\textsc{vii} $1s\rightarrow3p$ & 18.63 & $-0.075\pm0.003$\\
16.87 & O\,\textsc{viii} $1s\rightarrow2p$ &18.97 & $-0.117\pm0.003$\\
16.46 & O\,\textsc{vii} $1s\rightarrow4p$ & 17.77 & $-0.076\pm0.003$\\
12.55$^d$ & Fe\,\textsc{xix} $2p\rightarrow3d$ & 13.52 & $-0.074\pm0.004$\\
11.90 & Fe\,\textsc{xx} $2p\rightarrow3d$ & 12.82 & $-0.074\pm0.004$\\
11.42$^d$ & Fe\,\textsc{xxi} $2p\rightarrow3d$ & 12.28 & $-0.072\pm0.004$\\
10.87$^d$ & Fe\,\textsc{xxi} $2p\rightarrow3d$ &12.28 & $-0.121\pm0.004$\\
10.59$^m$ & Fe\,\textsc{xxii} $2p\rightarrow3d$ & 11.92 & $-0.118\pm0.004$\\
10.41 & Fe\,\textsc{xxii} $2p\rightarrow3d$ & 11.77 & $-0.122\pm0.004$\\
9.70 & Fe\,\textsc{xxiii} $2s\rightarrow3p$ & 10.98 & $-0.123\pm0.005$\\
9.40	& Fe\,\textsc{xxiv} $2s\rightarrow3p$ & 10.62 & $-0.121\pm0.005$\\
7.80	& Fe\,\textsc{xxiii} $2s\rightarrow3p$ & 10.98 & $-0.329\pm0.005$\\
7.52	& Fe\,\textsc{xxiv} $2s\rightarrow3p$ & 10.62 & $-0.332\pm0.005$
\enddata
\tablenotetext{a}{Centroid rest wavelength of an absorption line in \AA, with typical uncertainty of $\Delta\lambda=0.05$\AA.}
\tablenotetext{b}{Probable line ID and known laboratory wavelength from www.nist.gov.}
\tablenotetext{c}{Derived 
blue-shift of absorption line, in units of $c$.} 
\tablenotetext{d}{The Fe\,\textsc{xix} and Fe\,\textsc{xxi} lines may also contain a contribution from Ne\,\textsc{ix} (at 13.45\,\AA) and Ne\,\textsc{x} (at 12.14\,\AA) respectively.} 
\tablenotetext{m}{ The line identification is uncertain. It may originate from the decay down to either the metastable level of $2s^{2}2p\,^{2}P_{3/2}$ at 11.92\,\AA, or the ground state at 11.77\,\AA. For the latter, the velocity is lower with $v/c=-0.101\pm0.004$.} 
\label{tab:absorption-lines}
\end{deluxetable}

We also fitted some of the absorption lines with Gaussian velocity profiles, using the 
$1s\rightarrow 2p$ lines from O\,\textsc{vii} and O\,\textsc{viii} (2 components), which are relatively isolated line profiles. 
For these three lines we obtain a mean width of $\sigma=1380^{+730}_{-400}$\,km\,s$^{-1}$. 
This is broader than some of the Fe K velocity components observed in the {\it Resolve} spectrum of \pg, which can exhibit line widths in the range  $\sigma=100-1300$\,km\,s$^{-1}$; e.g. see paper I.  
However, the spectral resolution of RGS is lower than {\it Resolve}, e.g. 60\,m\AA\ at 19\,\AA\ at O\,\textsc{viii} corresponds $\sim 1000$\,km\,s$^{-1}$ (FWHM). Thus the RGS will not be as sensitive to measuring the widths of very narrow soft X-ray absorption lines. 

\begin{figure*}
\begin{center}
\rotatebox{-90}{\includegraphics[height=16cm]{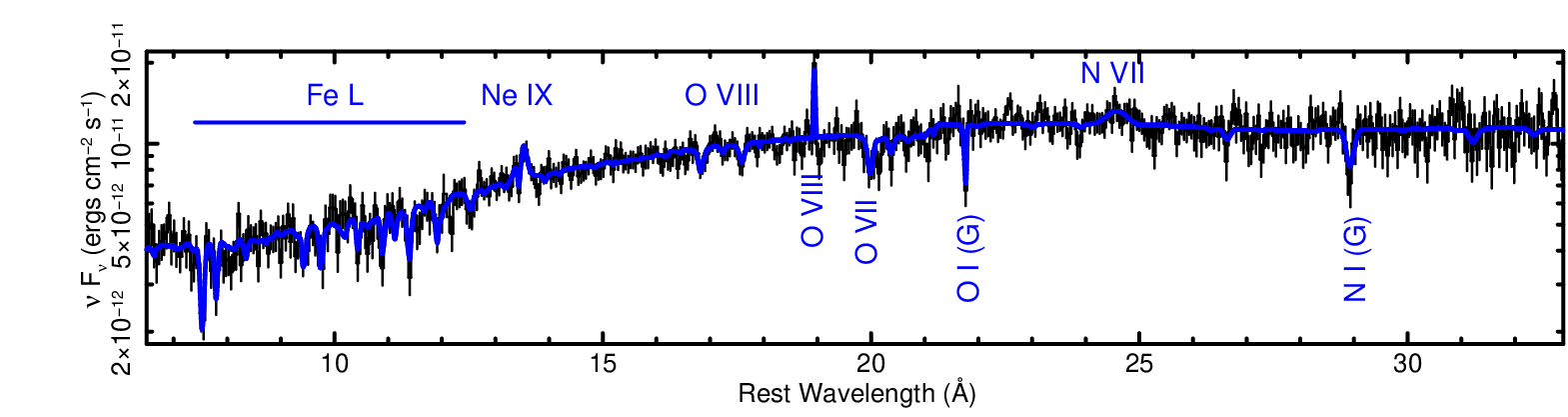}}
\rotatebox{-90}{\includegraphics[height=8.5cm]{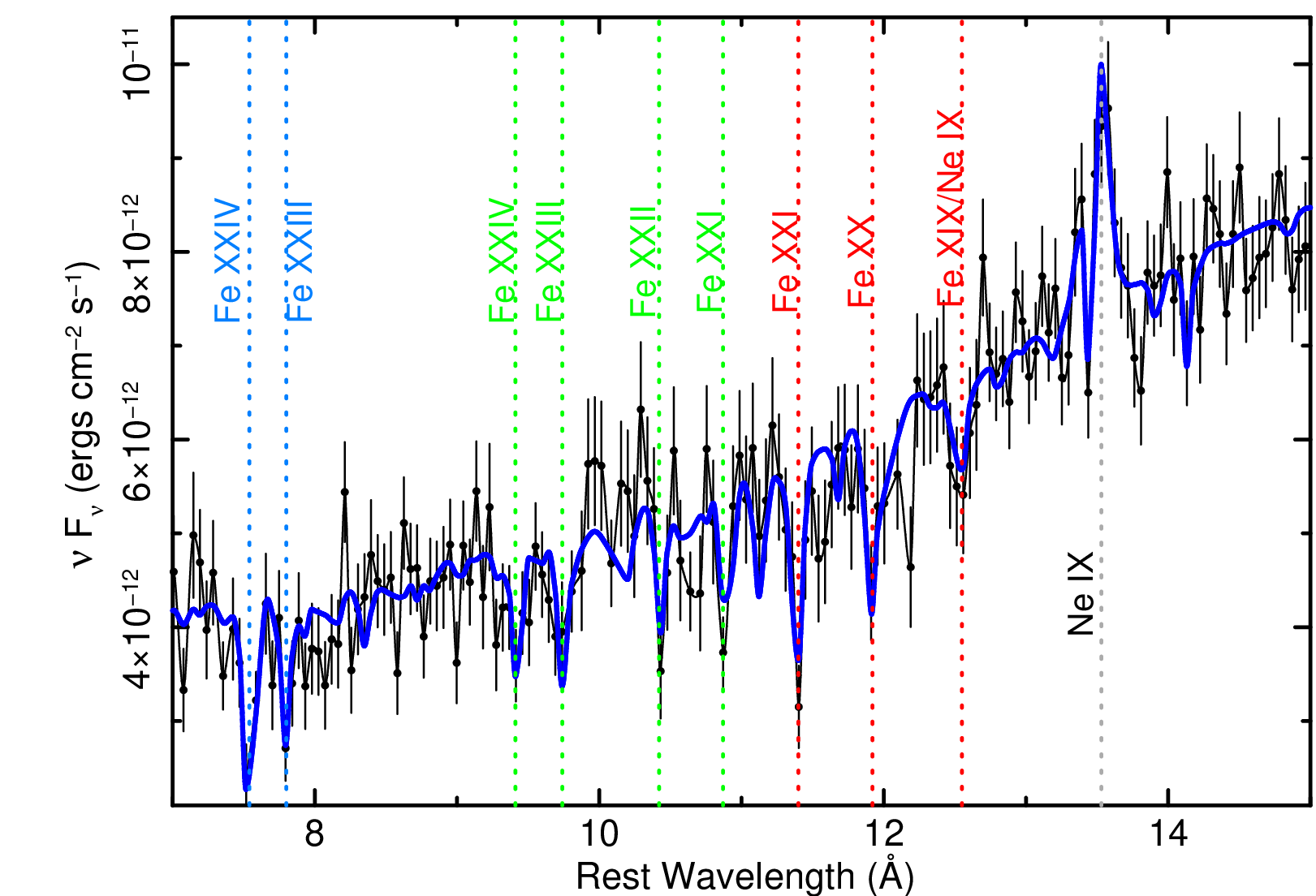}}
\rotatebox{-90}{\includegraphics[height=8.5cm]{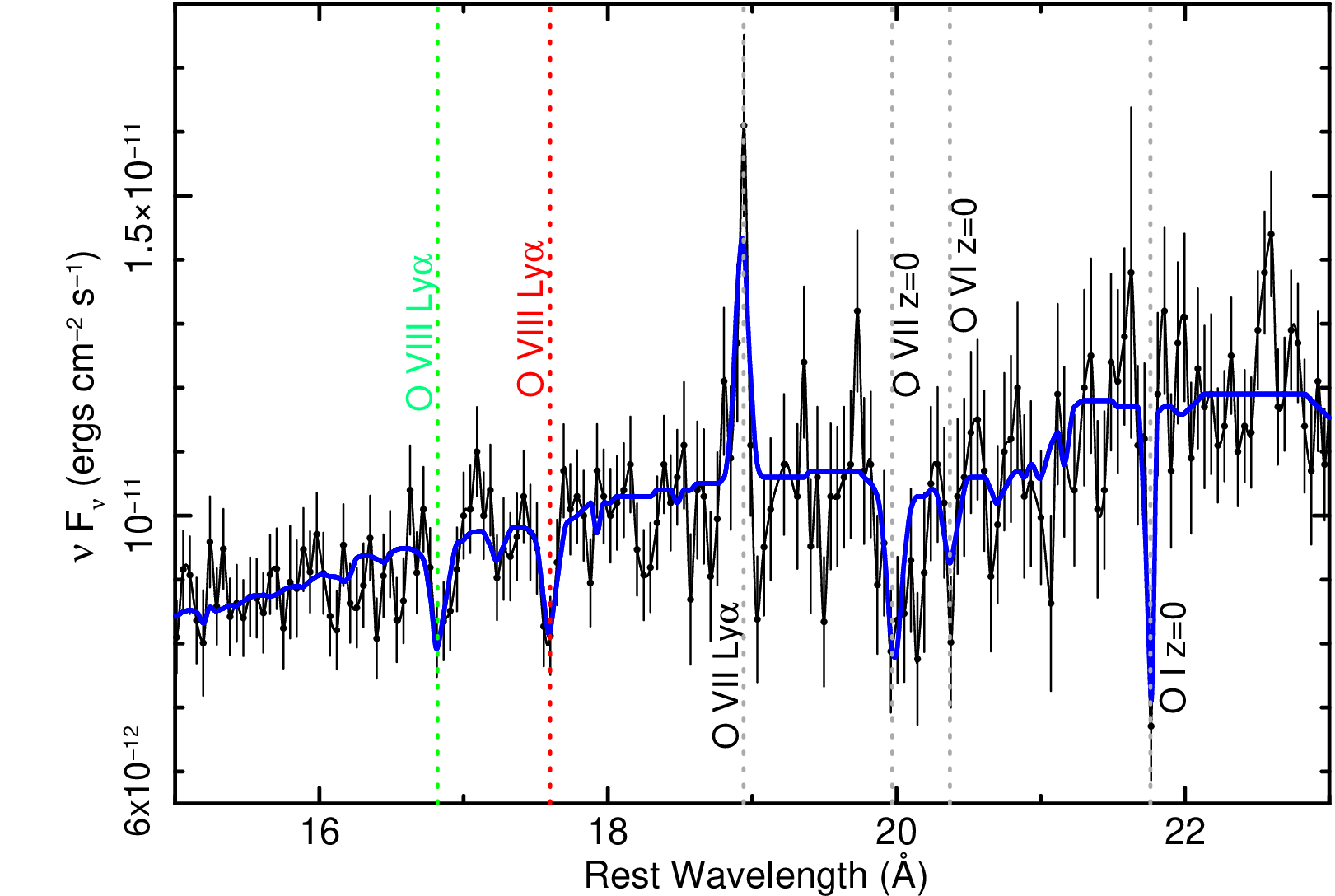}}
\end{center}
\caption{Fluxed RGS spectrum, in $\nu F_{\nu}$ units, with the best fit absorber model overlayed in blue.  
The upper panel shows the fit over the whole band, while the lower panel shows zoom-in portions around the Fe L band (left) and the O K-shell band (right). The likely identification of the blue-shifted absorption troughs are marked, and are color coded 
according to their outflow velocity as per Figure\,3, where the outflow velocities of $-0.074c$ (red), $-0.12c$ (green) and $-0.33c$ (blue) correspond to RGS absorber zones 1, 3 and fast (see Table~3). In addition, neutral and ionized absorption lines from O\,\textsc{i}, N\,\textsc{i} and O\,\textsc{vi-vii} are also marked (at $z=0$). The most prominent emission lines occur from Ne\,\textsc{ix} (13.5\,\AA) and O\,\textsc{viii} (18.9\,\AA) which have little blue-shift.} 
\label{fig:bestfit}
\end{figure*}

\subsection{Soft X-ray emission lines}\label{sec:emission}

Compared to the absorption lines, there is a relative paucity of emission lines in the 2024 RGS spectrum against the bright continuum (see Figure~3).  The emission lines were also more prominent against in the 2014 RGS spectra of \pg, against the weaker soft X-ray continuum \citep{Reeves18}. 
Nonetheless, three weak emission lines are present, which correspond to the H-like lines of O\,\textsc{viii} and N\,\textsc{vii} and the He-like emission of Ne\,\textsc{ix}. The O\,\textsc{viii} Ly$\alpha$ line is narrow, with a FWHM velocity width of $<1300$\,km\,s$^{-1}$, which is narrower than the measured optical H$\beta$ FWHM in \pg\ of 
1860\,km\,s$^{-1}$ \citep{BorosonGreen92}, suggesting an origin in distant gas. The Ne\,\textsc{ix} and N\,\textsc{vii} appear to be somewhat broader than this; Table~2 lists the properties of the emission lines. The blue-shifts of the emission lines are much lower than the outflowing gas seen in absorption.

\subsection{Photoionization Modelling} \label{sec:photoionization_modelling}

Next, the RGS spectrum was modeled with multiplicative absorption tables produced by \textsc{xstar}. 
The phenomenological form of the model is simply:-\\

$\textsc{tbabs} \times [\textsc{xstar}\times (\textsc{pow}+ \textsc{comptt}) + \textsc{gauss}_\textsc{emiss}].$ \\

\noindent where both the power-law (\textsc{pow}) and soft X-ray Comptonization (\textsc{comptt}) continuum components are absorbed by the \textsc{xstar} absorbers, while $\textsc{gauss}_\textsc{emiss}$ represents any soft X-ray emission lines parameterized by Gaussian profiles. 
The neutral Galactic ($z=0$) absorption component (\textsc{tbabs}) absorbs all of the emission components. 
The \textsc{xstar} absorption tables as described in paper I were used, which were also extended to lower ionization values (to $\log\xi=0$) to capture any lower ionization soft X-ray absorption. The absorption tables were calculated for both $\times1$ and $\times3$ Solar abundance of Fe, Co and Ni \citep{Asplund21}, with all other elements fixed at Solar values. Turbulence velocities (i.e. line widths) cover the range from $v_{\rm turb}=100-2700$\,km\,s$^{-1}$. 

\subsection{Soft X-ray Absorption Components} \label{sec:absorption_components}

In order to minimize the parameter space of the ionized absorption in the soft X-ray spectrum, scans in C-statistic space of an \textsc{xstar} absorber grid against the continuum model with no absorption were run.  
The scan methodology and results are described in Appendix~A. 

Overall, four absorption zones are required at $>99.9$\% significance according to the AIC test; these are listed in Table~2. These consist of three high ionization zones and one low ionization zone. The measured redshifts of the high ionization ($\log\xi=3.6-4.4$) zones are:- $z=+(3.1\pm1.7) \times 10^{-3}$, $z=-0.042\pm0.002$ and $z=-0.234\pm0.002$. These correspond to outflow 
velocities of $v/c=-0.0745\pm0.0017$, $v/c=-0.120\pm0.002$ and $v/c=-0.331\pm0.002$ respectively. The first two velocity components are consistent in velocity with the zones 1 and 3 in \xrism\ Resolve,  
where $v/c=-0.0735\pm0.0003$ and $v/c=-0.121\pm0.001$. 
These two velocity components are subsequently referred to as zones 1b and 3b and are soft X-ray counterparts of the Fe K UFOs detected in 
paper I. 

The fastest RGS zone, with $v/c=-0.331\pm0.002$, appears intermediate between velocity zones 4 and 5 in Resolve, where $v/c=-0.2875^{+0.0005}_{-0.0003}$ (zone 4) and $v/c=-0.357^{+0.003}_{-0.002}$ (zone 5). 
This may either suggest some variability in the faster absorption components, as the shorter RGS exposure covers only part (about a day) of the longer five day \xrism\ exposure (Table~1), or that it is an additional component. This fast zone is referred to as RGS--fast in the subsequent discussion. Note that the fitted velocities are consistent with those inferred from the identifications of individual absorption troughs in 
Table~3. 

In contrast, the redshift of the low ionization zone at $z=(0.0\pm1.2) \times 10^{-3}$ most likely corresponds to ionized gas along the line of sight through our Galaxy. We refer to this absorber as zone~0 (for $z=0$). 
It has much lower ionization ($\log\xi=0.8\pm0.3$) and column density ($N_{\rm H}=4.6\times10^{19}$\,cm$^{-2}$) than the high ionization absorbers, where its column density is lower even than the neutral Galactic column of $N_{\rm H}=3\times10^{20}$\,cm$^{-2}$. The latter neutral absorber also imprints weak absorption lines from O\,\textsc{i} and N\,\textsc{i} from our Galaxy. 

The subsequent best-fit model, with four ionized absorbers fitted to the 2024 RGS spectrum, plus the addition of three emission lines, is shown in Figure~4,  
illustrating that the model can reproduce the absorption structures seen in \pg. The fit statistic of the best-fit model is $C/\nu=551.9/602$, versus $C/\nu=701.5/622$ for the continuum model only. The turbulence velocity was assumed to be the same for each of the zones, with a best fit value of $v_{\rm turb}=1330^{+530}_{-360}$\,km\,s$^{-1}$, 
which is consistent with the velocities inferred to the individual absorption troughs. Given the more limited velocity resolution of RGS versus Resolve, it is not possible to infer whether there is any dispersion in the velocity widths between the soft X-ray zones and if they are allowed to vary independently, they are consistent within errors with the above value. 

We also note that the fit with a enhanced abundance of Fe, with $\times 3$ Solar, was slightly preferred to one with Solar values, by a factor of $\Delta C=-10.0$. 
As a result, the higher abundance values are adopted in this paper, which we note was also inferred from an earlier analysis of the 2014 RGS spectra \citep{Reeves18}. This results in a slightly better match to the depths of the Fe L-shell absorption troughs as plotted in Figure~4, which are otherwise somewhat underestimated. The parameters of the absorbers are relatively unaffected by the 
Fe abundance. It slightly alters the column density of the zones, where for example for zone 1b the column decreases from $N_{\rm H}=1.9\pm0.6\times10^{21}$\,cm$^{-2}$ ($1\times$~Solar) to 
$N_{\rm H}=1.1\pm0.4\times10^{21}$\,cm$^{-2}$ ($3\times$~Solar) as a result of the increased Fe abundance. Neither the ionization parameters nor the outflow velocities depend on the relative Fe abundance. 

\begin{figure}
\begin{center}
\rotatebox{-90}{\includegraphics[width=6cm]{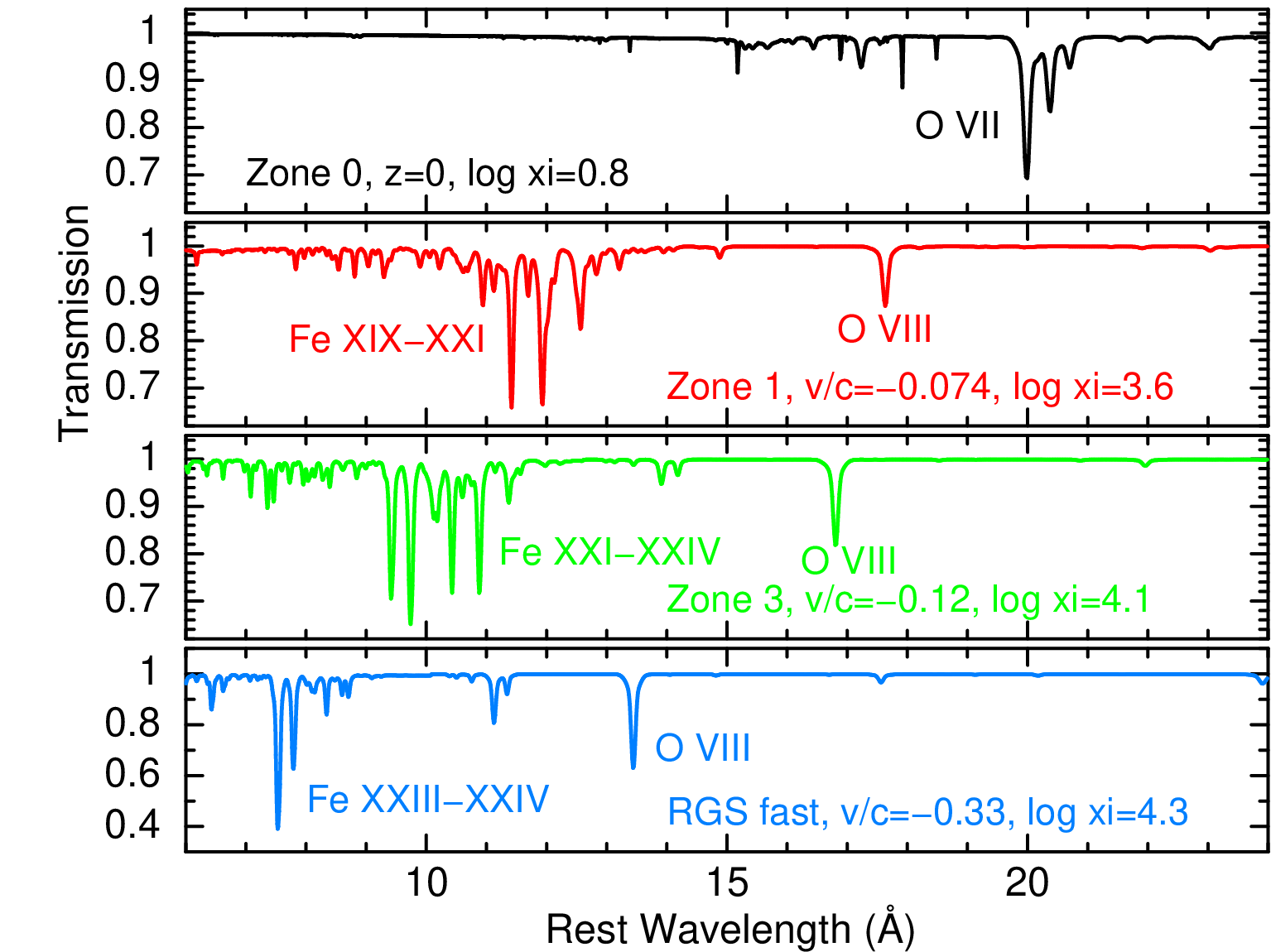}}
\end{center}
\caption{The transmission through the 4 RGS absorber zones; zone 0 (at $z=0$), zone 1 (at $v=-0.074c$), zone 3 (at $v=-0.120c$) and zone RGS--fast ($v=-0.33c$). The $z=0$ zone is likely associated to weak ionized absorption from our Galaxy. The other zones contribute most of their opacity through blue-shifted iron L lines, as marked, as well as from O\,\textsc{viii} 
Ly$\alpha$.} 
\label{fig:absorbers}
\end{figure}

Figure~5 (right) plots the transmission spectra for each of the four ionized absorber zones. Here, the zone~0 absorber contributes little opacity, except for the weak O\,\textsc{vii} $1s\rightarrow 2p$ absorption line at 20.0\,\AA\ in the AGN rest frame. This line corresponds to its expected wavelength at 21.6\,\AA\ in the observed $z=0$ frame, consistent with a Galactic origin. Zone 1b and 3b produces the bulk of the Fe L-shell absorption, but with the tendency of the faster zone 3b to produce the higher ionization ions (Fe\,\textsc{xxi-xxiv}) when compared to the somewhat lower ionization zone 1b (Fe\,\textsc{xiv-xxi}). These two zones also account for the two blue-shifted components of O\,\textsc{viii} at 16.9\,\AA\ and 17.6\,\AA. 
The higher velocity RGS--fast zone (light blue) reproduces the rest frame absorption between $7-8$\,\AA\ in the QSO rest frame, from a combination of highly blue-shifted
Fe\,\textsc{xxiii} and Fe\,\textsc{xxiv} $2s\rightarrow3p$ absorption lines and has negligible opacity elsewhere. 

\begin{deluxetable*}{lccccccc}
\tabletypesize{\footnotesize}
\tablecaption{Summary of Absorption zones fitted to RGS and Resolve.}
\tablewidth{0pt}
\tablehead{
\colhead{Zone} & \colhead{$N_{\rm H, obs}$$^c$} & \colhead{$N_{\rm H, corr}$$^d$} & \colhead{$\log\xi$} & \colhead{$z$} & \colhead{$v/c$} & \colhead{$\Delta C/\Delta\nu$$^a$} & \colhead{Prob (AIC)$^b$}}
\startdata						
0 (RGS) & $(4.6\pm1.7)\times10^{-3}$ & -- & $0.85\pm0.31$ & $(0.4\pm1.2)\times10^{-3}$ & $-$ & $-21.5/3$ & $4.31\times10^{-4}$\\
1 (Resolve) & $0.51^{+0.22}_{-0.20}$ & $0.59^{+0.25}_{-0.23}$ & $4.25^{+0.15}_{-0.10}$ & $(4.1\pm0.4)\times10^{-3}$ & $-0.0736\pm0.0004$\\
1b (RGS) & $0.12^{+0.06}_{-0.04}$ & $0.14^{+0.07}_{-0.05}$ & $3.66\pm0.16$ & $4.1\times10^{-3}$$^t$ & $-0.0736^t$ & $-53.4/5$ & $3.77\times10^{-10}$\\
2 (Resolve) & $1.3\pm0.5$ & $1.6\pm0.6$ & $4.71^t$ & $(-2.58\pm0.23)\times10^{-2}$ & $-0.104\pm0.002$ & $-30/2$ & $2.26\times10^{-6}$\\
3 (Resolve) & $2.3\pm0.6$ & $2.9\pm0.8$ & $4.71\pm0.12$ & $(-4.26\pm0.12)\times10^{-2}$ & $-0.121\pm0.001$\\
3b (RGS)	& $0.74\pm0.29$ & $0.95\pm0.37$ & $4.14^{+0.11}_{-0.18}$ & $-0.0426^t$	& $-0.121^t$ & $-65.1/5$ & $1.08\times10^{-12}$\\
4 (Resolve) & $1.4^{+0.8}_{-0.6}$ & $2.5^{+1.4}_{-1.1}$  & $4.71^t$ & $-0.195\pm0.001$ & $-0.287\pm0.001$ & $-21.6/2$ & $1.51\times10^{-4}$\\
RGS--fast	& $1.7^{+0.9}_{-0.5}$ & $3.4^{+1.8}_{-1.0}$  & $4.26\pm0.08$ & $-0.234\pm0.002$ & $-0.331\pm0.002$ & $-25.9/3$ & $4.77\times10^{-5}$\\
5 (Resolve) & $9^{+11}_{-5}$ & $19^{+23}_{-11}$  & $6.0^t$ & $-0.255\pm0.004$ & $-0.356\pm0.004$ & $-9.8/2$ & $5.5\times10^{-2}$\\
6 (Resolve) & $20^{+13}_{-14}$ & $47^{+30}_{-33}$  & $6.0\pm0.5$ & $-0.296\pm0.002$ & $-0.404\pm0.002$ & $-21.3/2$ & $1.75\times10^{-4}$
\enddata
\tablenotetext{a}{Improvement in fit statistic from adding the velocity component to the model. Note zones 1/1b and 3/3b each correspond to a single velocity component, but with different $\log\xi$.}
\tablenotetext{b}{Null probability of adding the absorber zone, calculated via the AIC test.}
\tablenotetext{c}{Observed column density, in units of $\times10^{22}$\,cm$^{-2}$.}
\tablenotetext{d}{Column density after correction for special relativistic effects, in units of $\times10^{22}$\,cm$^{-2}$.}
\tablenotetext{t}{Denotes parameter is tied. For example, the velocities of zones 1/1b and 3/3b are tied between RGS and Resolve, as well as the ionization parameters of zones 2, 3 and 4. The ionizations between zones 5 and 6 are also tied.}
\label{tab:xstar}
\end{deluxetable*}

\begin{figure*}
\begin{center}
\rotatebox{-90}{\includegraphics[width=6cm]{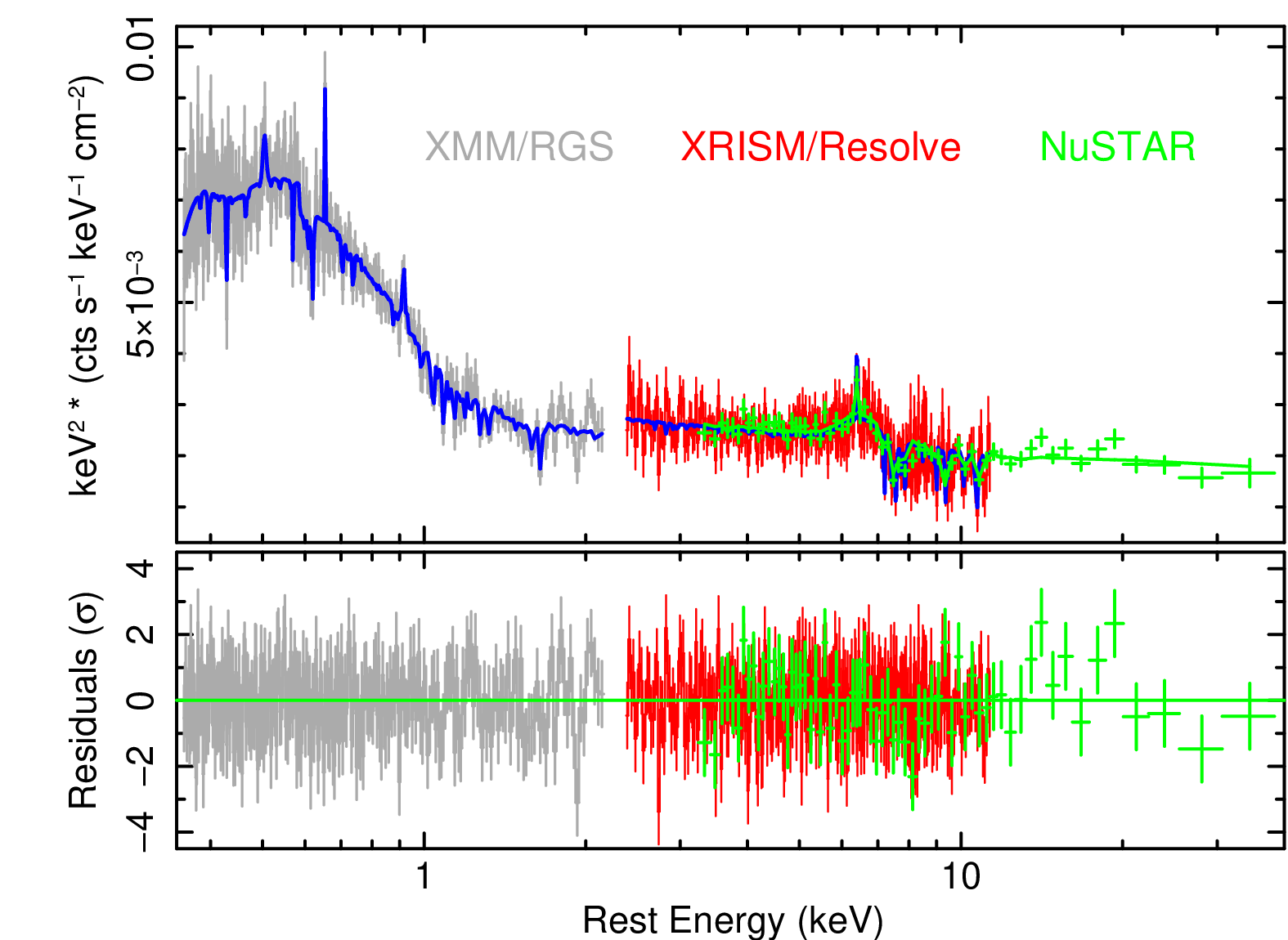}}
\rotatebox{-90}{\includegraphics[width=6cm]{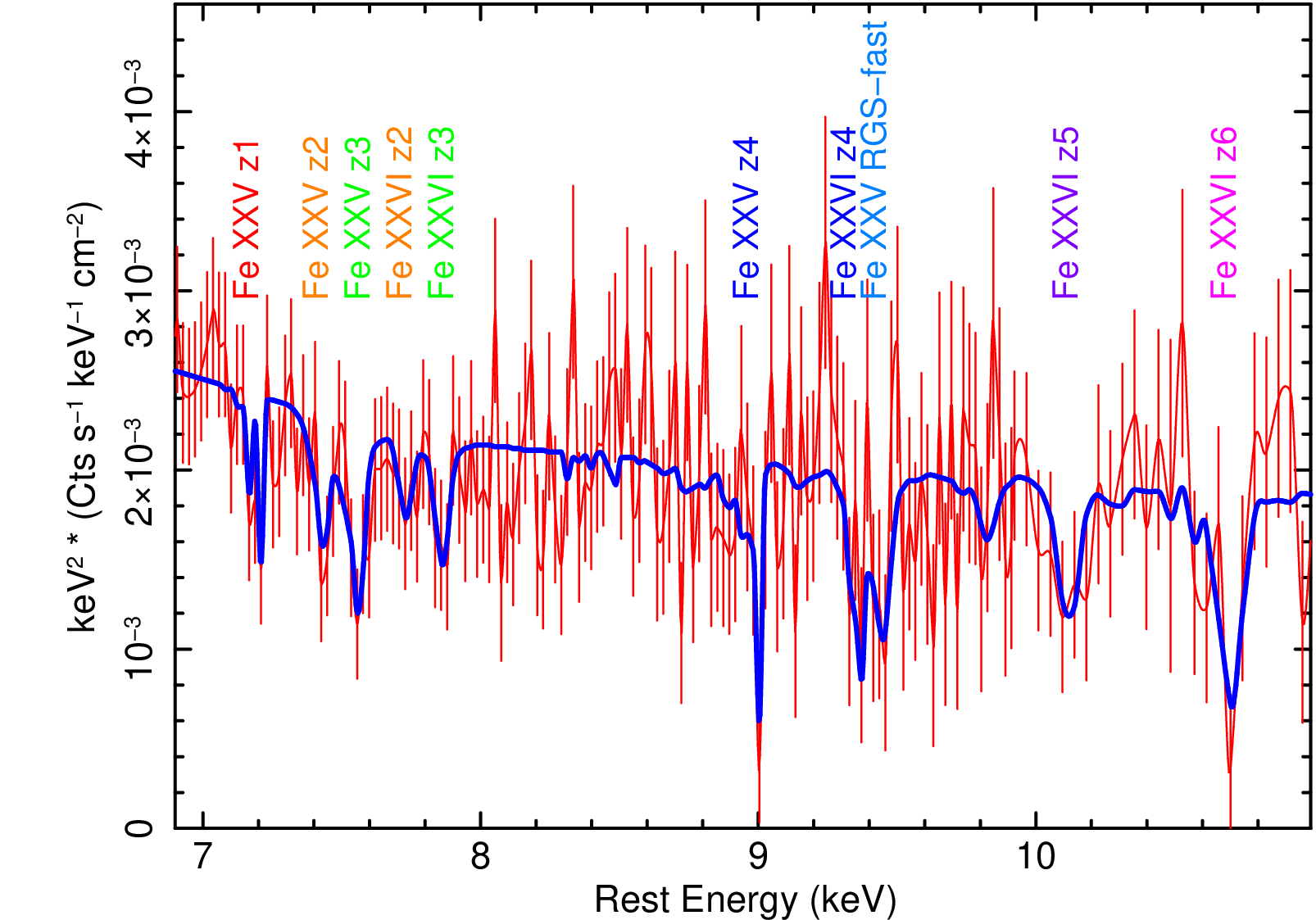}}
\end{center}
\caption{Results of joint spectral fitting between RGS, Resolve and {\it NuSTAR}. The left panels show the broad-band fit, where RGS is in grey, Resolve is in red and {\it NuSTAR} FPMA+B is in green. 
The solid blue line (green for {\it NuSTAR}) shows the best fit model as listed in Table~4, folded through the instrumental responses and allowing for the cross normalization factors. The lower panel shows the residuals (in $\sigma$) versus this model, which reproduces well the absorption troughs. The right panel shows a zoom-in of the Resolve spectrum, from 7--11 keV in the AGN rest frame. The origin of the Fe K absorption features, from zones\,1--6, are marked and are consistent with those described in paper I. Of the soft X-ray zones, only the RGS--fast zone imprints noticeable Fe K absorption upon the Resolve spectrum.}
\label{fig:joint}
\end{figure*}

\section{Comparison between the Soft X-ray and Fe K Absorbers} \label{sec:resolve}

A joint fit was constructed to ensure a consistent comparison between the Fe K absorbers observed in Resolve, versus the soft X-ray absorbers detected in RGS. 
The RGS spectrum was fitted from 0.33--1.8 keV in the observed frame, while Resolve was fitted from 2.2--10\,keV, which ensures a high resolution over a wide energy range. 
The {\it NuSTAR} data was also included up to 35\,keV, which then underpins the hard X-ray continuum, while the cross normalization between instruments was accounted for 
by constant multiplicative factors; these were found to be $C_{\rm RGS/Resolve}=0.91\pm0.02$ and $C_{\rm NuSTAR/Resolve}=1.03\pm0.02$. 
As was shown in Figure~1 of paper I, the individual spectra of \pg\ between \xmm, \nustar\ and \xrism\ are consistent with each other within errors over the 2--10 keV band. 
The above cross-calibration between RGS and Resolve is also in good agreement with the detailed study performed by the \xrism\ science team on NGC\,3783; see \citet{xrism25b} for further details.

The same form of the underlying continuum as described earlier 
was adopted, with values consistent with those in Sections 3 and 4.  The three soft X-ray emission lines were included alongside broad and narrow components of the Fe K emission lines, 
which were shown in detail in paper I from the Resolve spectrum. 

\begin{figure}
\begin{center}
\rotatebox{-90}{\includegraphics[width=6cm]{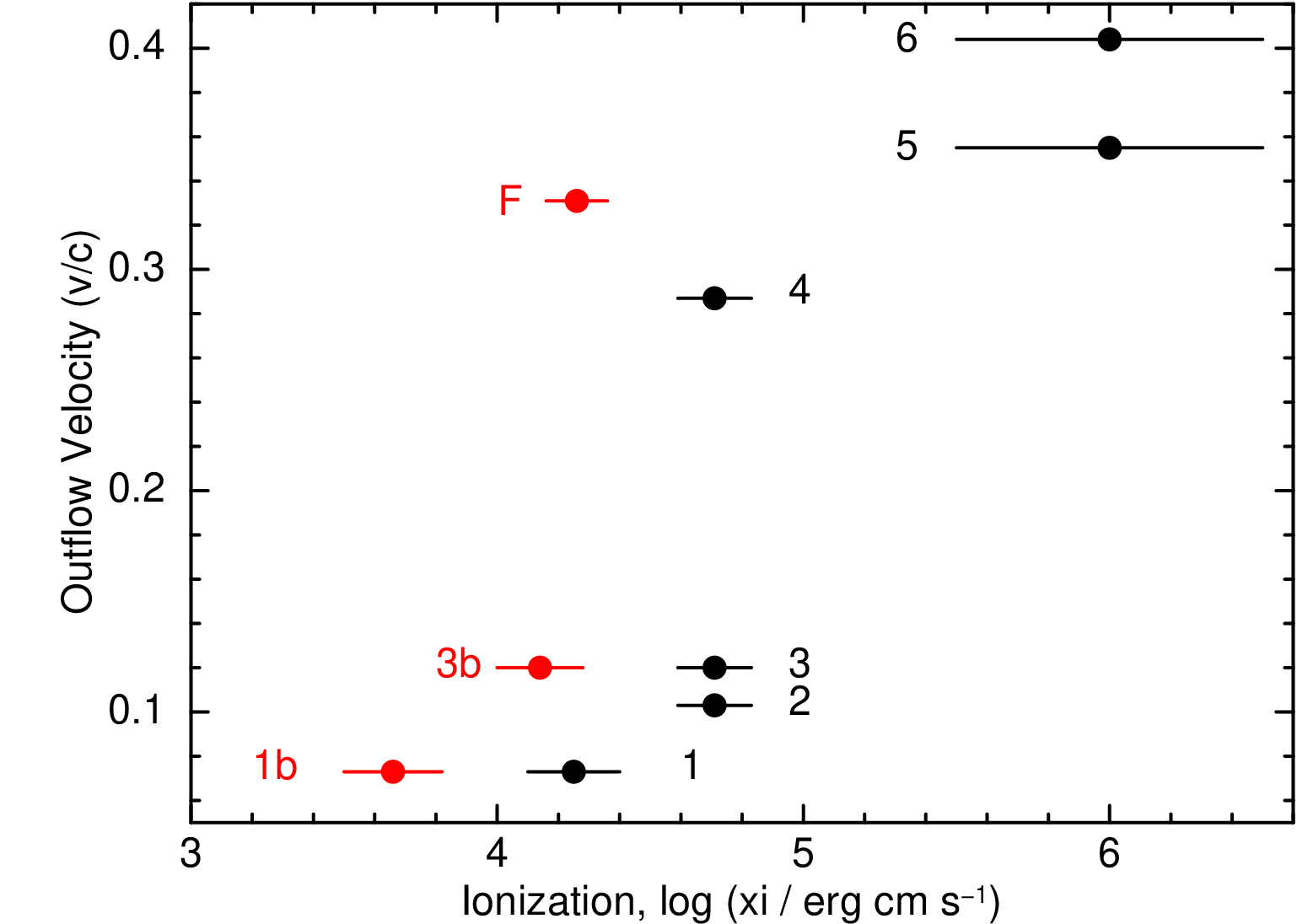}}
\end{center}
\caption{A comparison between the outflow velocities and ionization parameters for the absorber zones measured in RGS versus Resolve (zones\,1--6), where the RGS zones are marked in red, Resolve in black. The velocities of zones 1b and 3b in RGS are consistent with zones 1 and 3 obtained with Resolve at Fe K, but are at somewhat lower ionization. The RGS--fast zone (F) appears intermediate in velocity between Resolve zones 4 and 5, but is similar in ionization to zone 4. Zones 2, 5 and 6 in Resolve have no lower ionization counterparts and are likely too highly ionized to produce significant soft X-ray absorption.}
\label{fig:comparison}
\end{figure}

The model then adopts the six velocity separated absorption zones measured by Resolve, where the turbulence velocities were fixed to the values obtained from the 
detailed Fe K-shell band modeling in paper I, as shown in their Table~3. 
The four soft X-ray absorption zones were also accounted for, where the outflow velocity of zone 1b (RGS) was assumed to be equal to that obtained from Resolve (zone 1) and likewise for zones 3 and 3b, as these are also kinematically consistent components. However the column densities and ionizations between zones 1/1b and 3/3b were allowed to vary independently. 
Thus in total, excluding the $z=0$ absorption, 7 possible outflowing zones with kinematically distinct velocities are applied to both the RGS and Resolve spectra:- zones 1/1b, 2, 3/3b, 4, 5, 6 and RGS--fast.

The joint fit between RGS and Resolve is shown in Figure~6 and can reproduce well both the continuum shape and all of the absorption and emission features observed both at soft X-rays and at Fe K. 
The final value of the underlying photon index is $\Gamma=2.17\pm0.01$. Table~4 lists the detailed parameters of each of the absorption components.
The properties of the Resolve Fe K band absorbers are consistent with those in paper I, with the exception of their column densities which tend to be slightly lower in this paper. This is due to adopting a $\times3$ Solar abundance of Fe in the joint RGS and Resolve fit. The higher relative abundance of Fe compared to other elements appears to be justified, as then  
the resulting fit statistic is worse by $\Delta C$=33.5 if a Solar Fe abundance is used, confirming the result found in the RGS alone.

\begin{figure*}
\begin{center}
\rotatebox{-90}{\includegraphics[height=8.5cm]{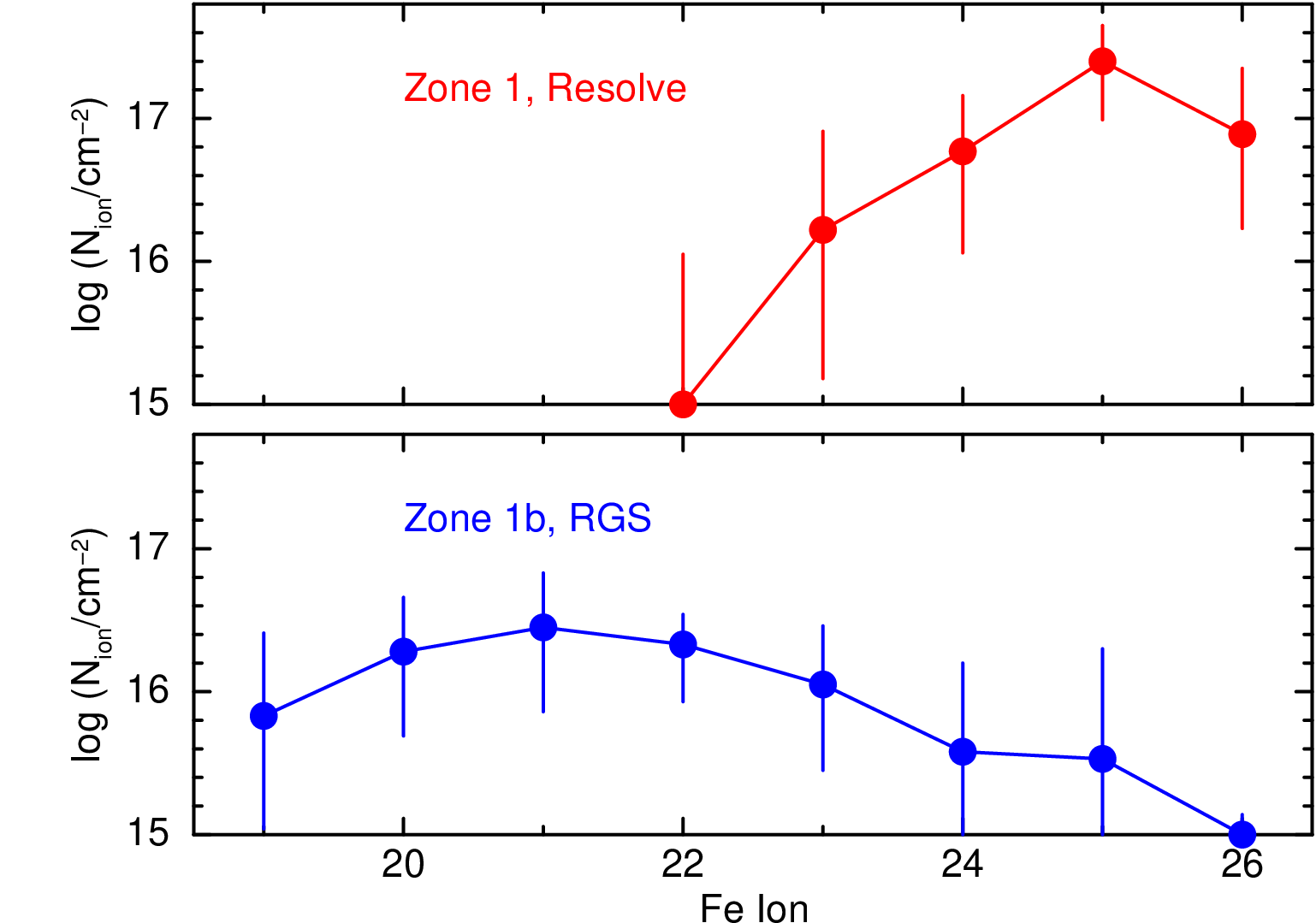}}
\rotatebox{-90}{\includegraphics[height=8.5cm]{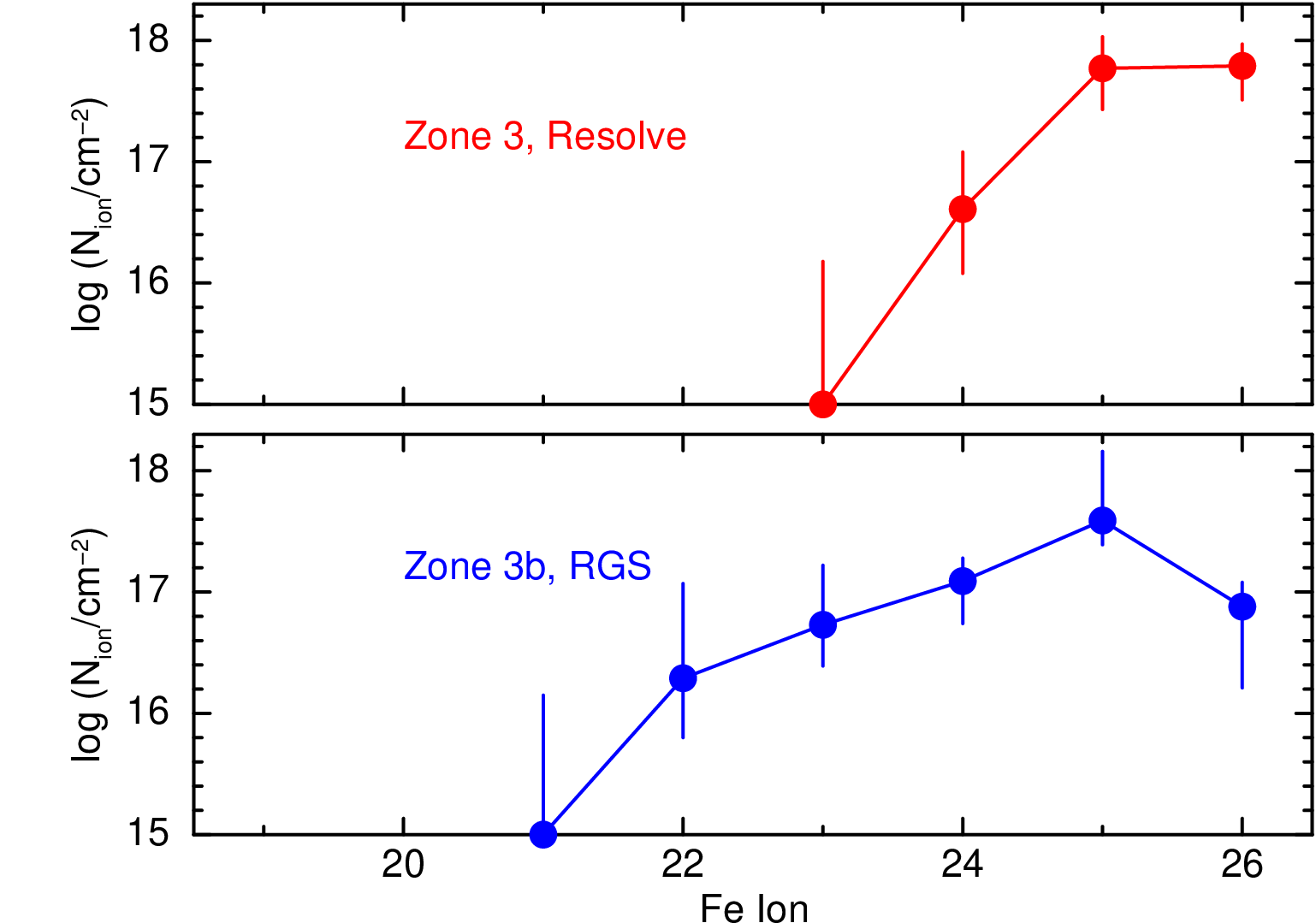}}
\end{center}
\caption{Ionic column densities for zone~1 (left) and zone~3 (right), where the RGS columns are denoted in blue. The distribution of the charges states of iron skew to lower values in the RGS, in order to reproduce the iron L-shell absorption observed at soft X-rays. The differences in ionization may suggest variations in either the density or the distance of the zones from the black hole. Note Fe\,\textsc{xxvi} corresponds to H-like Fe.}
\vspace{0.6cm}
\label{fig:columns}
\end{figure*}

The statistical significances of the Resolve zones 1--6 was investigated in paper I, where Monte Carlo simulations were performed to account for the look elsewhere effect by searching over multiple energy/velocity bins; 
e.g. see Figure~5 in paper I. While the lower velocity zones\,1-3 in Resolve were found to be highly significant, for the faster zones, only zone~6 was of marginal significance ($>68$\% significance), while zones 4 and 5 were not significant. However, in the RGS and Resolve joint fit, the improvement in fit statistic for these faster zones 4 and 6 was noticeably higher; e.g, $\Delta C$ values of $-21.6$ and $-21.3$ for zones 4 and 6 (Table~4), versus $\Delta C=-10.0$ and $-12.5$ respectively in paper I. From comparing to the $\Delta C$ versus probability distribution calculated in Figure~5 in paper I, the significances of zones 4 and 6 increases to $>99$\% confidence. This is most likely due to the combination of the higher Fe abundance, which better models the depth of the lines and the precise broad band definition of the continuum. 
Only zone~5 remains formally not significant according to the Monte Carlo distribution of $\Delta C$ values, where $\Delta C=-9.8$. It is marginally significant, at 94.5\%, according to the AIC test (see Table~4). 

Figure~7 shows a visual comparison between the velocities and ionizations of the Resolve and RGS zones reported in Table~4. 
Overall, the higher velocity zones tend to be found at higher ionizations. Figure~7 can be fitted with a power-law relation of the form $v_{\rm out} \propto \xi^b$. Here, a positive relation 
is found with $b=0.4\pm0.1$.
In terms of the individual absorbers, zones 1/1b and 3/3b are kinematically consistent 
between the soft X-ray and Fe K bands, where the RGS  counterparts skew to lower ionizations (see below). 
The RGS--fast zone is intermediate in outflow velocity between Resolve zones 4 and 5, but appears to be closer in ionization (and column density) to zone~4. 
There is no corresponding soft X-ray counterparts of either Resolve zone 2 ($v/c=-0.102$) 
or the highest velocity zones 5 and 6. Nonetheless, their presence is consistent with the RGS spectrum, as they are included in the joint RGS and Resolve fit. Their ionizations are simply too high to produce detectable soft X-ray absorption, especially for zones 5--6 which arise from mainly H-like Fe at $\log\xi=6$; see Table~B1 for a list of ionic columns per zone.

We re-examined the physical characteristics of the absorption components using
photoionisation models generated with \textsc{cloudy} (v23.01; \cite{Chat23}).  Solar elemental abundances  \citep{Asplund21} were adopted, 
with the exception of iron, for which we assumed an abundance of $3\times$ Solar, based on our spectral fitting with
\textsc{xstar}. For the input continuum, we used the SED described in Paper I (see Section 3).
The ionic column densities were calculated for each zone, these values are listed in Appendix~B.

Figure~8 shows the comparison of the ionic column densities for Fe\,\textsc{xix-xxvi} for zones 1/1b and 3/3b, which cover both the soft and Fe K bands at different ionizations (Table~4).  
For the Resolve zones 1 and 3, their ionic column densities peak at Fe\,\textsc{xxv} (i.e. He-like Fe) and due to their higher ionizations, then fall off rapidly in ionic column at lower charges states, especially for Fe\,\textsc{xxiii} and below. 
Thus the Fe K absorbers do not cover a sufficiently wide range in ionic species to account for the range of Fe L-shell absorption observed in the RGS spectrum. In contrast, their soft X-ray counterparts, zones 1b and 3b produce a lower ionization tail on the distribution of ionic columns which do account for the observed soft X-ray absorption. This implies there is some dispersion in the ionization associated to these velocity zones, possibly relating to the variations in the gas density. 
 
\subsection{The Absorption Measure Distribution} \label{sec:amd}
 
The combined Resolve and RGS results in Table~4 can be used to determine how the column density varies with absorber ionization. 
In particular, the Absorption Measure Distribution or AMD \citep{Holczer07,Behar09} quantifies the distribution of the line of sight column density versus $\log\xi$, where the AMD can be expressed as:-

\begin{equation}
{\rm AMD} = \frac {d N_{\rm H}}{d(\log\xi)} \propto \xi^{a}
\end{equation}

\noindent which is simply the column density measured per unit $\Delta \log\xi$ increment. The integral over the AMD is the total column density of all the zones. We bin the AMD in increments of $\Delta\log\xi=0.5$ over the range from $\log\xi=3-5$. Thus zone 1b (RGS) only constitutes the $\log\xi=3.5-4.0$ bin, the $\log\xi=4.0-4.5$ bin consists of the sum of the zones 1, 3b (RGS) and RGS--fast zones and $\log\xi=4.5-5.0$ 
consists of the Resolve zones 2--4. The exception to this are zones 5 and 6, which due to their larger uncertainties in Table~4 form a single bin from $\log\xi=5.5-6.5$. As there is no intrinsic absorption at ionizations below $\log\xi<3.5$ in the 2024 spectrum, we calculated an upper limit to the column density between $\log\xi=3.0-3.5$ for comparison to the other AMD points. 
In calculating the AMD, all of the measured $N_{\rm H}$ values were corrected for special relativistic aberration along the line of sight, which effectively increases the columns by a factor of $\frac{1-\beta}{1+\beta}$ \citep{Luminari20}, which is due to the de-boosting of the ionizing continuum as seen by the outflowing absorber. 
The corrected values ($N_{\rm H, corr}$) are reported in Table~4.

\begin{figure*}
\begin{center}
\rotatebox{-90}{\includegraphics[width=6cm]{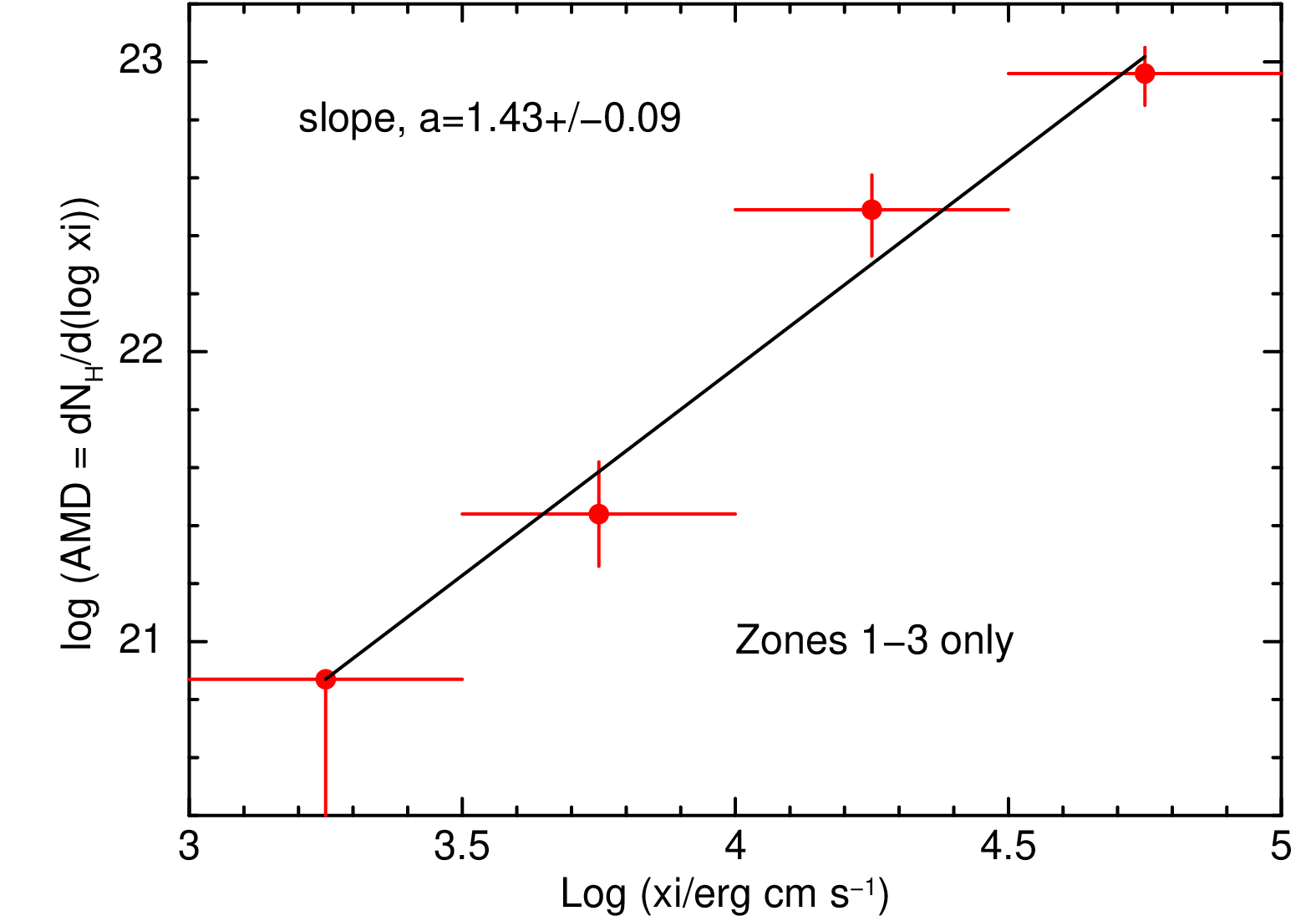}}
\rotatebox{-90}{\includegraphics[width=6cm]{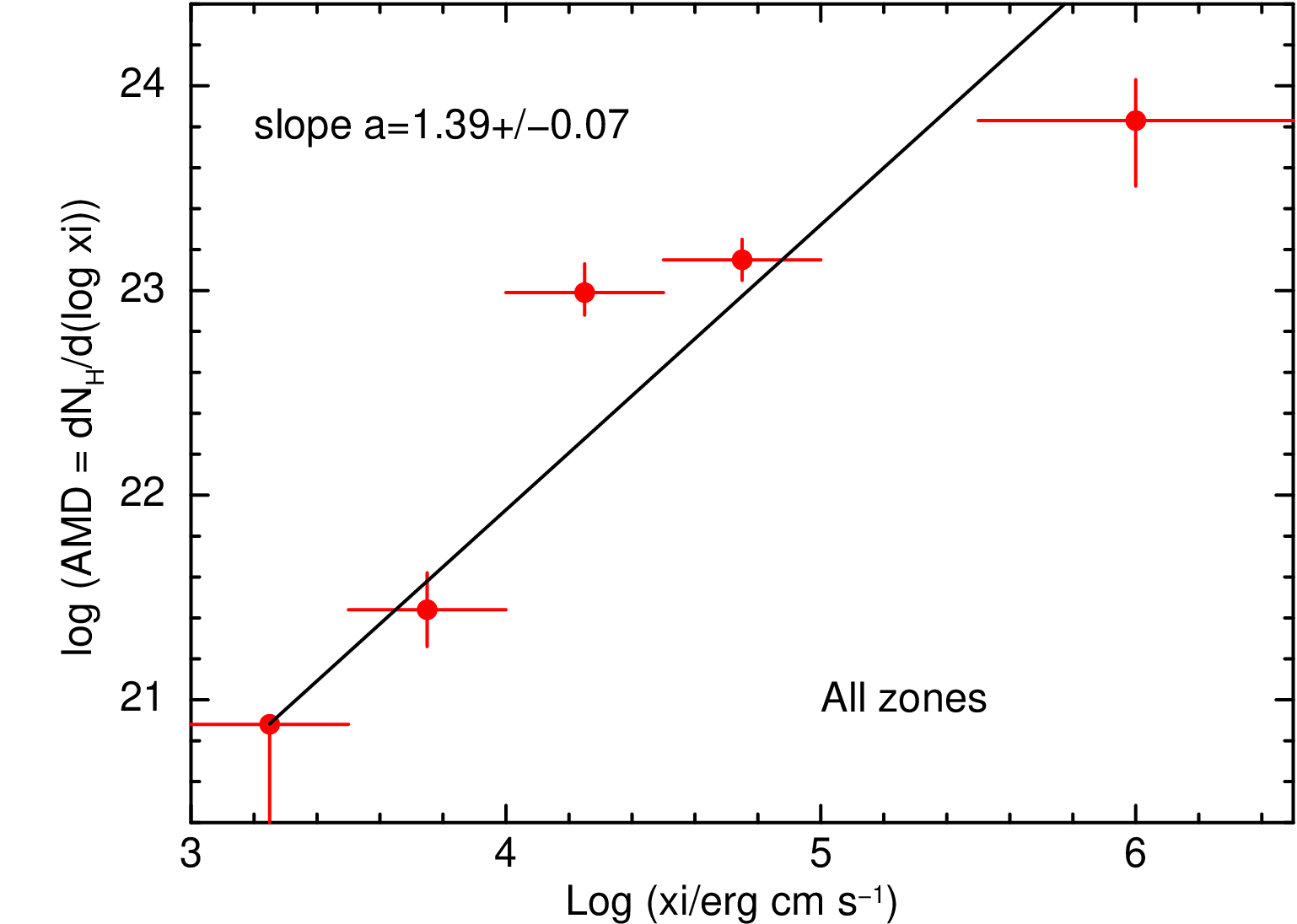}}
\end{center}
\caption{The Absorption Measure Distribution (AMD) of \pg, plotting the change in column density as a function of the ionization, binned into $\Delta(\log\xi=0.5)$ increments. 
The left panel is limited to the lower velocity zones 1--3, while the right panel show all of the velocity zones in Table~4 in the RGS and Resolve spectra. 
The column increases with ionization, with a slope of $a\approx1.4$ and translates to a radial density profile varying as $n(r)\propto r^{-5/3}$. 
Alternatively, density variations may occur from the size-scale of the absorbing clumps.}
\label{fig:amd}
\end{figure*} 

Figure~9 shows the results of the AMD fits, where the fitted slope ($a$) of the AMD is obtained over a log--log scale as per equation~2. Two cases were considered; (i) where the AMD is comprised of only zones 1--3 
(including the RGS counterparts 1b and 3b) over the narrow velocity range of $0.074-0.12c$ and (ii) over all the Resolve plus RGS zones, covering the full range in velocity and ionization as per Table~4. 
In both cases, the slopes of the AMD were found to be consistent, with $a=1.43\pm0.09$ (zones 1--3) and $a=1.39\pm0.07$ (all zones). 
Considering all the zones, there may be some evidence that the AMD flattens at the very highest ionizations out to $\log\xi=6$.  
This may be due a physical limitation on the maximum wind column density of $N_{\rm H}<10^{24}$\,cm$^{-2}$, such that the wind does not become completely Compton thick at high ionization, 
but this is tentative due to the poor constraints on zones 5 and 6. 
Regardless, the AMD displays a steep slope of $a>1$, which is significantly steeper than the typical values measured in Seyfert 1 warm absorbers 
\citep{Behar09}. 

\section{Discussion} \label{sec:discussion}

\pg\ provides a further example of a clumpy ultra fast outflow revealed by \xrism, where up to six discrete narrow velocity components have been measured in the Resolve spectrum at Fe K (paper I).  
Presently, the other two known cases with \xrism\ are PDS\,456 \citep{xrism25,Xu25} and IRAS~05189$-$2524 \citep{Noda25}. IRAS~05189$-$2524 is a type 2 QSO which is highly absorbed below 2\,keV, while 
PDS\,456 was caught during a prolonged period of enhanced obscuration at soft X-rays during its 2024 \xrism\ campaign. Thus any soft X-ray wind components could not be measured at high resolution with RGS for either of these AGN. 
The current 2024 \pg\ campaign has provided the first opportunity to simultaneously measure both the Fe K and soft X-ray counterparts of the multi-velocity UFO at high resolution, where \pg\ was captured during a historically bright period. 
At least two of the Fe K velocity zones were revealed to have soft X-ray counterparts, zone 1 at $v/c=-0.074c$ and zone 3 at $v/c=-0.12c$, while a further faster RGS zone was also measured at $v/c=-0.33c$. 
The soft X-ray absorbers are somewhat less ionized than their corresponding Resolve counterparts, as can be seen in Figures 7 and 8. 

\subsection{The Structure of the Ultra Fast Outflow}

The combined spectral fit of RGS and Resolve (Table~4) in the soft X-ray and Fe K bands reveal multiple outflowing zones, ranging over at least two orders of magnitude in both column density and ionization parameter. 
This makes it possible to probe the structure of the UFO in \pg, through the AMD (Section~5.1). 
The AMD slope, which parameterizes the distribution of column density versus ionization in the wind, is found to be unusually steep in \pg, with $a=1.4$, compared to the slower Seyfert warm absorbers where typically $a<<1$. 

\subsubsection{A large scale wind}

We first consider the case where the wind is in the form of a continuous, large scale outflow, where the radial
density profile can be expressed as:-

\begin{equation}
n(r) \propto r^{-\alpha}.
\end{equation}

\noindent Following the discussion in \citet{Behar09}, e.g. see their Section~3.1, the AMD slope $a$ relates to the profile index, $\alpha$ as:-

\begin{equation}
\alpha = \frac{1+2a}{1+a} \pm \frac{\Delta a}{1 + a^{2}}
\end{equation}

\noindent where $\Delta a$ is the fitted error on the AMD index. 

The AMD slope for \pg\ from Section~5.1 is $a=1.43\pm0.09$ and thus $\alpha=1.67\pm0.02$ or $n(r) \propto r^{-5/3}$. 
An index of $\alpha=1$, corresponding to $a=0$ and a flat AMD is strongly ruled out. This is also the case for a constant ionization wind from a radial outflow, 
where $\alpha=2$ and $a$ tends towards $\infty$; i.e. a singular peaked distribution of $N_{\rm H}$ vs $\log\xi$. The measured index is closer to the value of $\alpha=1.5$, 
which might be expected from magnetically driven winds \citep{BP82,CL94,Fukumura10}. The AMD slope is also consistent with 
the NGC\,4151 \xrism\ observations, where a highly structured wind with a wide range of outflow velocities can be fitted by a radial profile of $\alpha\approx1.5$ \citep{Xiang25}. 

As the ionization parameter is $\xi = L/n(r)r^2$, then the wind ionization will vary versus radius as:-
\begin{equation}
\xi(r) \propto r^{\alpha -2}.
\end{equation}

\noindent Thus for $\alpha=5/3$ in \pg, the ionization scales as $\xi(r) \propto r^{-1/3}$ in a smooth radial profile. 

The above dependence between $\xi$ and $r$ in equation~5 can be combined with an outflow velocity law scaling as $v \propto r^{-1/2}$, i.e. in proportion to the Keplerian disk velocity, 
as might be expected in magneto-centrifugal winds. This yields a relation between outflow velocity $v$ and $\xi$:-

\begin{equation}
v \propto \xi^{1/2(2-\alpha)}.
\end{equation}

\noindent From the velocity versus ionization correlation in Figure~7, we have $v \propto \xi^{0.4\pm0.1}$. From equation~6, this translates into a radial density slope of $\alpha=0.75\pm0.25$, which 
is substantially lower than the steep value of $\alpha=1.67\pm0.02$ value inferred from the AMD fit and appears to be ruled out by the data (e.g. Figure~9).  

An intuitive comparison comes from comparing the physical ranges of the ionization and velocity values with radius. 
As the wind ionization decreases with radius as $r^{-1/3}$, the two orders of magnitude observed range in $\xi$ implies a six orders range in $r$. 
On the other hand, if $v \propto r^{-1/2}$, then for a range of outflow velocities of just under one order magnitude ($0.07-0.40c$), the wind should cover less than two orders in radius. 
Thus the inferred velocity and ionization ranges for a smooth, large scale wind are incompatible. Physically, the observed range in ionization may instead be enhanced by the wind being 
intrinsically clumpy. This is considered below in terms of the AMD relation.

\subsubsection{A clumpy outflow}

A more realistic scenario is a clumpy wind, where zones with the same or similar velocity (e.g. zones 1--3), but different ionizations, can be explained by localized density variations. 
Hydrodynamical simulations of disk winds for AGN in the high Eddington regime \citep{Takeuchi13} predict such wind structures and can potentially reproduce the properties 
of the clumpy wind first revealed by \xrism\ in PDS\,456  \citep{Hu25}. 
Following the calculations of \citet{Behar09} for a localized, but clumpy wind, where the gas exists at a similar radius $r$, 
then the density $n$ is expressed in terms of the clump size ($\Delta r$) as:-

\begin{equation}
n \propto (\Delta r)^{-\alpha}.
\end{equation}

\noindent While in this scenario, $\alpha$ relates to the AMD slope $a$ as:-

\begin{equation}
\alpha = \frac{1}{1+a} \pm \frac{\Delta a}{1 + a^{2}}.
\end{equation} 

\noindent Furthermore, as $\xi \propto n^{-1}$ (for a constant distance) and as $N_{\rm H} = n \Delta r$, then the equivalent relations are:-
\begin{equation}
\xi \propto (\Delta r)^{\alpha}~{\rm and}~  
N_{\rm H} \propto (\Delta r)^{1 - \alpha}.  
\end{equation}

\noindent For \pg\ with $a=1.43\pm0.09$, then $\alpha=0.41\pm0.02$. Hence, $n \propto (\Delta r)^{-2/5}$, $N_{\rm H} \propto (\Delta r)^{3/5}$ and $\xi \propto (\Delta r)^{2/5}$. 

Smaller co-spatial clumps embedded within the wind will have higher densities with lower ionizations and column densities. This matches the behavior seen towards absorber zones 1/1b and 3/3b, 
where the soft X-ray zones have somewhat lower ionizations and columns and which can produce the extended distributions of ionic species plotted in Figure~8. 

Most of the absorbers fall within a range of $\Delta (\log\xi)=1$ and thus as $\xi \propto (\Delta r)^{2/5}$, their ionization distribution could be accounted for by a two orders of magnitude range in 
clump size. The exception would be the very highly ionized zones 5 and 6, which could be homogeneous.   
However as the absorbers would be at a similar distance, this does not explain the range of velocities observed (Figure~7). Instead a hybrid scenario may be more 
plausible whereby the density (and ionization) depends upon both $R$ and $\Delta R$, where faster winds are launched from closer in.  
Overall the wind could be both large-scale and clumpy, launched over a range of radii, where
inhomogeneities account for the lower ionization gas for both the fast and slower zones.
Observationally, this is similar to the entrained UFOs \citep{Serafinelli19}, which have been suggested to explain previous examples of lower ionization, but fast soft X-ray winds.

We can estimate the properties of the soft X-ray absorbers, if the absorber size-scale is similar to the X-ray source size; i.e. $\Delta r \approx R_{\rm corona} \approx 10^{14}$\,cm. Here, a gravitational radius in 
\pg\ is $R_{\rm g}  =10^{13}$\,cm. If the absorber is much smaller than this, then it will not absorb enough continuum flux, much bigger and the density fluctuations would not be observed. 
For zone~3b, with $N_{\rm H}\approx10^{22}$\,cm$^{-2}$, then the density is $n\approx 10^{8}$\,cm$^{-3}$. From its ionization ($\log\xi=4.1$) and 1--1000\,Rydberg luminosity ($L_{\rm ion}=3\times10^{45}$\,erg\,s$^{-1}$, paper I), then its distance is $R=5\times10^{16}$\,cm (or a few $\times 10^{3}\,{\rm R}_{\rm g}$). This is consistent with the location of the higher ionization soft X-ray absorber deduced from the 2014 \xmm\ campaign, as seen from the absorber variability \citep{Reeves18}. 
For the slower, less ionized, zone~1b soft absorber, the clumps would either be placed slightly further out, or be somewhat more dense and compact if at the same distance as zone~3b. 

\begin{figure}
\begin{center}
\rotatebox{-90}{\includegraphics[width=6cm]{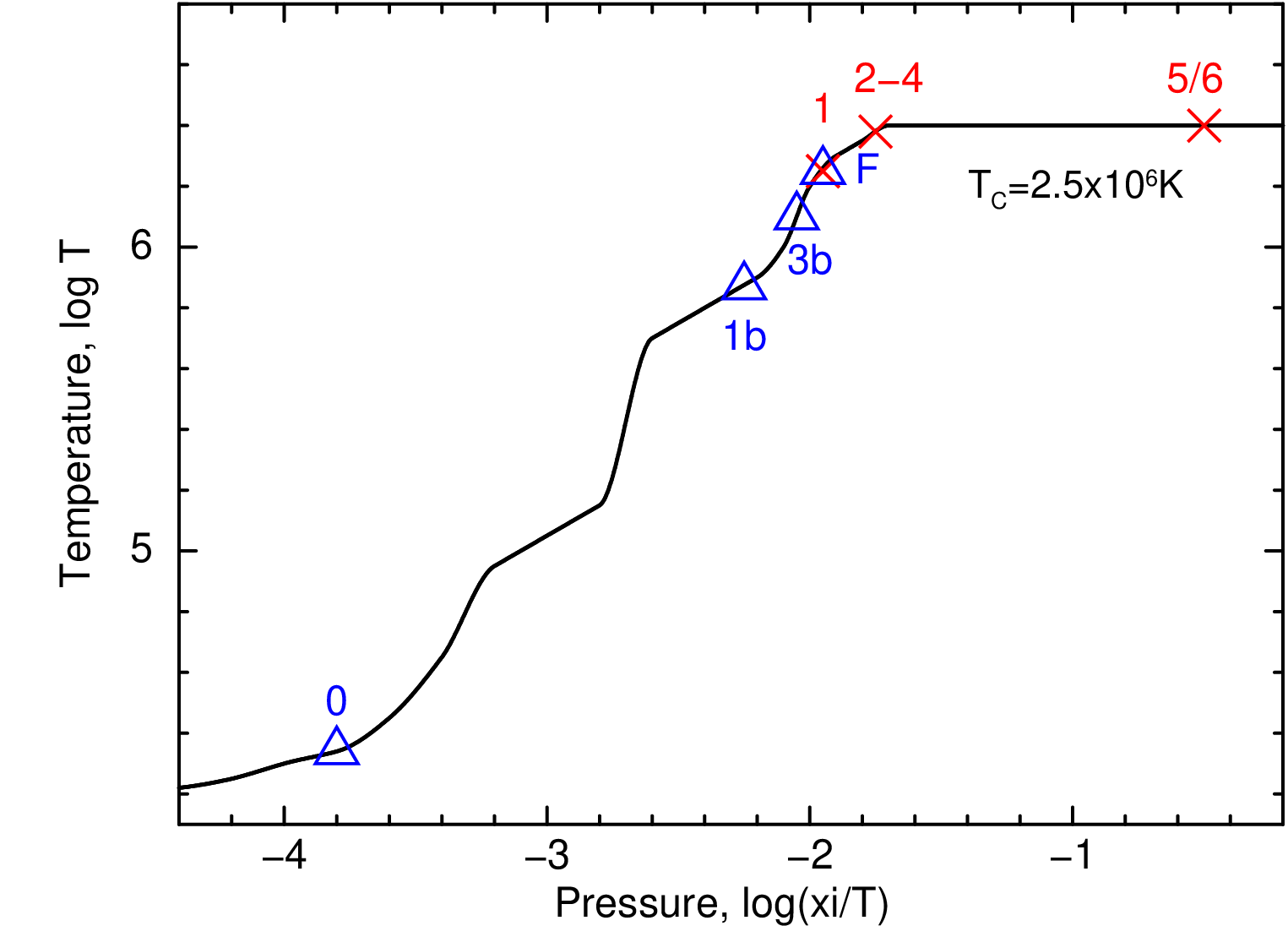}}
\end{center}
\caption{The S-curve for the absorbers, of temperature versus pressure, computed from the SED of \pg. There are no unstable regions, as a result of the steep SED. RGS zones are marked as blue triangles, Resolve zones as red crosses, with the highest ionization components falling on the Compton branch of the curve in the upper right. Here, the Compton temperature is $T_{\rm C}=2.5\times10^6$\,K.}
\label{fig:scurve}
\end{figure} 

As noted above, hydrodynamical simulations of near (or super) Eddington sources can predict clumpy winds. One question is whether this is the case for the sub-Eddington AGNs? 
While there are now three observed cases of structured, clumpy UFOs in high accretion rate AGN (PDS\,456, IRAS~05189$-$2524 and \pg), the situation may be less clear in 
lower luminosity Seyfert 1s. The variability of the UFO troughs in NGC 3783, for instance, may be more connected to the presence of a variable disk/corona and subsequent magnetic reconnection events and ejecta \citep{Gu25}.  
Other UFOs in lower luminosity AGN as observed by \xrism\ may not display the same range of velocities as per the high Eddington rate sources, for example in MCG$-$6$-$30$-$15 \citep{Brenneman25}, where the weak 
UFOs can have simpler profiles. The Resolve spectra of many Seyfert 1s are also complicated by the presence of low velocity warm absorbers on larger radial scales (e.g. NGC\,4151, \citealt{Xiang25}), 
which are absent in \pg. Further high resolution examples, spanning the luminosity range, are required to test how the properties of UFOs vary with accretion rate. 

\subsection{The Physical Properties of the Ultra Fast Outflow}

To understand the physical properties of the gas, we calculated the S-curve (Stability curve) for \pg, adopting the SED obtained in paper I. The S-curve plots the temperature against pressure \citep{Krolik81}, 
where the radiation to gas pressure ratio ($\Xi$) is defined as:-

\begin{equation}
\Xi = \frac{\xi}{T} \frac{1}{4\pi c k} \approx 2\times10^{4} \frac{\xi}{T}
\end{equation}
 
\noindent and $k=1.38\times10^{-16}$\,ergs\,K$^{-1}$ is the Boltzmann constant. 

\begin{table*}[htbp]
\centering
\caption{Mass Outflow Rates for Soft X-ray Zones.}
\label{tab:rates}
\begin{tabular}{lccccccc}
\hline
Zone & v/c & $\Psi$ & $n/10^8\,{\rm cm}^{-3}$ & $R/10^{16}\,{\rm cm}$ & $\Delta R/R$ & $\dot{M}\,/{\rm M}_{\odot}\,{\rm yr}^{-1}$ & $L_{\rm k} / 10^{44}~{\rm ergs}~{\rm s}^{-1}$\\
\hline
1b & $-0.074$  & 0.74  & 0.14 & 19 & $5\times10^{-4}$ & 0.1 & 0.2 \\
3b & $-0.012$  & 0.62  & 1.0 & 3.8 & $3\times10^{-3}$ & 0.3 & 1.3 \\
RGS--fast & $-0.331$  & 0.25  & 3.4 & 1.1 & $9\times10^{-3}$ & 0.9 & 30 \\
Total & & & & & & 1.3 & 32 \\
\hline
\end{tabular}
\end{table*}

Figure~10 plots the temperature against pressure ratio (here $\xi/T$), where the individual absorbers from Table~4 are superimposed upon the curve. 
As the gas is highly ionized, most of the absorbers occupy the upper right portion of the curve, where the Fe K band Resolve absorbers lie on the flat, Compton branch of the curve. 
The soft RGS band absorbers lie on the positive gradient portion of the curve just below this, suggesting they can undergo some heating and/or cooling in equilibrium with the continuum. 
However, all the absorbers on the S-curve are thermally stable, as none occur in a region with a negative gradient.
This suggests the UFOs in \pg\ should remain stable over time, with a high duty cycle, which is consistent with the presence of the soft X-ray UFOs across multiple epochs 
of \xmm\ observations \citep{Pounds03,Reeves18}. The observed variability of the soft X-ray UFOs in \pg\ \citep{Reeves18} may instead be connected to the clumpy nature of the gas.

The shape of the S-curve, with no unstable regions, is a result of the SED of \pg, as is shown in Figure~2 in paper I. \pg\ has a high Eddington ratio, of close to 1 \citep{Lobban16} and that tends to result in a softer SED shape, with strong UV to soft X-ray bumps, weak hard X-ray emission and a steep photon index \citep{Jin12,Mitchell23}. Even at hard X-rays, the photon index in \pg\ is steeper than $\Gamma=2$ and thus at every portion of its SED above the ionization potential of H\,\textsc{i}, the $\nu F_{\nu}$ flux (energy flux per logarithmic interval) declines with energy. 
As a result of the steep SED, the Compton temperature is very low, with $T=2.5\times10^{6}$\,K and prevents the wind from becoming fully ionized. 
Overall, other high Eddington rate AGN, with similar, steep SEDs, may also predict similar S-curves, which are conducive to the stability of the UFOs. 
One example is in PDS\,456, which also accretes near to Eddington and has a steep SED. Here the derived S-curve is similar to the one in \pg\ \citep{Xu25} and the UFO components are generally stable.

\subsection{The Soft X-ray Mass Outflow Rate}

The mass outflow rate can be calculated from $\dot{M} = 4\pi b f_{\rm cov} \mu m_{\rm p} v_{\rm out} nR^2$, where $b=\Omega/4\pi$ is the geometrical covering fraction, $f_{\rm vol}\approx \Delta R/R$ is the volume 
filling factor, $\mu m_{\rm p}$ is the average atomic mass, $v_{\rm out}$ is the outflow velocity and $nR^2 = L^{\prime}_{\rm ion} / \xi$ from the definition of the ionization parameter. 
Here,  $L^\prime_\mathrm{ion}$ is the irradiating luminosity after the correction for the Doppler de-boosting factor ($\Psi$) due to the motion of the gas away from the X-ray source, where $L^\prime_\mathrm{ion} = \Psi L_{\rm{ion}}$. 
From Equation~2 in \citet{Luminari20}, $\Psi = \gamma^{-4}(1+\beta)^{-4}$ for radially outflowing gas, where $\gamma$ is the Lorentz factor and $\beta=v_{\rm out}/c$. 

As per paper I, the $\dot{M}$ value can be expressed in terms of the typical observable parameters in \pg\ as:-

\begin{equation}
\dot{M} = \mu f_{\text{cov}} \left(\frac{f_{\text{vol}}}{0.01}\right) \left(\frac{v_{\text{out}}}{0.1c}\right) \left(\frac{L^\prime_{\text{ion}}/\xi}{10^{41}\,\text{cgs}} \right)M_{\odot}~\text{yr}^{-1}.
\label{eq:mdot_normalized}
\end{equation}

\noindent For the $\dot{M}$ estimate, we adopt $\mu=1.4$ for the average atomic mass, $L_{\rm ion}=3\times10^{45}$\,ergs\,s$^{-1}$ and $b=0.5$ for the covering factor. The latter value was used in paper I for the Fe K outflows, which 
is similar to the frequency of detections of UFOs in nearby AGN \citep{Tombesi10,Gofford13}. Otherwise, the main uncertainty is the filling factor $\Delta R/R$. 
A clump size of $\Delta R = 10^{14}$\,cm (or $10R_{\rm g}$ in \pg) is adopted as a conservative choice, as the clump sizes should not be too much smaller than the X-ray source extent to produce sufficiently deep 
absorption lines. The density is calculated as $n=N_{\rm H}/\Delta R$ and subsequently the absorber distance as $R^2 = L^{\prime}_{\rm ion} / n\xi$. This yields an 
estimate for $\Delta R/R$ for each zone and thus $\dot{M}$, which are listed in Table~5.

The soft X-ray outflow has an estimated total outflow rate of the order $\approx 1\,{\rm M}_{\odot}$\,yr$^{-1}$, with the fastest zone (RGS--fast) having the largest contribution. 
This is consistent with the Fe K outflow estimate in paper I for zones 1--3. The overall mass outflow rate appears similar to the Eddington rate of $2\,{\rm M}_{\odot}$\,yr$^{-1}$ for a black hole mass of $10^8\,{\rm M}_{\odot}$ in \pg.
The clumpy nature of the soft X-ray gas may restrict $\dot{M}$ such that it does not exceed the Eddington rate, despite its lower ionization compared to the Fe K zones. 
The mechanical power can be calculated as $L_{\rm k} = (\gamma - 1) \dot{M} c^2$ for each zone and it is dominated by the fastest zone (see Table~5), of about 30\% of Eddington. 
In contrast, the two slower zones (1b and 3b) contribute 1\% or less of the Eddington luminosity. 

\subsection{The Wind Driving Mechanism}

One possibility is the fast winds are driven by electron scattering in an Eddington limited regime, e.g. see \cite{KP03}. 
To examine this, we compare the radiation force due to electron scattering ($F_{\rm rad}$) to the observed momentum thrust of the wind components ($\dot{p}_{\rm out}$). 
From paper 1, the mass outflow rate in \pg\ is calculated to be $\dot{M}_{\rm out}\approx 1$\,M$_{\odot}$\,yr$^{-1}$ for the slower ($v/c \approx 0.1$) zones 1--3. 
This translates into a outward momentum thrust of $\dot{p}_{\rm out} = 2\times10^{35}$\,dynes. 

The radiation force due to electron scattering is then:-

\begin{equation}
F_{\rm rad} = \frac{L}{c} \sigma_{\rm T} N_{\rm H} f_{\rm cov} = \frac{L}{c} \tau f_{\rm cov}
\end{equation}

\noindent where $\sigma_{\rm T}$ is the Thomson cross-section, $N_{\rm H}$ is the column density, $L=10^{46}$\,erg\,s$^{-1}$ is the bolometric luminosity of \pg\ and 
$\tau$ is the optical depth to electron scattering.  Here the factor $f_{\rm cov} = \Omega/4\pi$ relates to the solid angle subtended by the absorbers to the X-ray source and 
thus the fraction of the luminosity $L$ they can intercept.  

The absorber zones in Table~4 have $N_{\rm H}\approx10^{22}$\,cm$^{-2}$ typically and thus are Compton thin with $\tau\approx10^{-2}$. 
Values of $f_{\rm cov}<1$ will dilute the radiation incident upon the absorbers, unless the gas is spherically distributed. 
Hence the resultant radiation force may be two orders of magnitude lower than the observed wind thrust and electron scattering appears to be an unlikely mechanism to accelerate most of the observed zones. The exception to this could be the very highly zones 5 and 6, where the Compton depths can be closer to $\tau\sim1$.

Alternatively, magneto-hydrodynamical (MHD) models could reproduce the broad (ionization, velocity) properties of UFOs \citep{Fukumura10,Kraemer18}, as specifically applied to the earlier \xmm\ spectra of \pg\ \citep{Fukumura15}. . 
However, steady-state MHD models tend to predict broad absorption profiles from a smooth wind configuration \citep{Fukumura22}, as opposed to the narrow profiles observed by \xrism. 
Future investigations are required to test whether MHD mechanisms can produce structured, clumpy winds, potentially via time dependent magnetic fields or thermal instabilities.  Variable magnetic field configurations, concurrent with X-ray coronal flares, could produce fast clumps of gas, as recently observed following a strong flare in the \xrism\ observations of NGC\,3783 \citep{Gu25}. We note the \xrism\ observations of \pg\ also occurred post-flare, as seen by \swift\ in Figure~1.

For radiation to have an appreciable effect on the acceleration, significant line driving opacity is required to produce a notable force multiplier \citep{Castor75} and this will only occur if there is a substantive column of lower ionization gas. For zones~2--6, the force multipliers predicted by the \textsc{cloudy} models (see Section~5) are close to unity ($1.0-1.3$), as expected in highly ionized gas with $\log\xi=4-6$ \citep{Dannen19,Kraemer18}. The only exception is the least ionized zone~1b (Table~4) seen in RGS, where the predicted force multiplier is 2.7, consistent
with some radiative acceleration. 
However, past epochs of RGS observations have shown more enhanced lower ionization absorption in the soft X-ray band in \pg\ \citep{Pounds03,Reeves18}. 
A mini obscuration event was observed during the multi--orbit \xmm\ campaign in 2014 (Figure~1), where the low flux spectrum shows more substantive low ionization absorption, 
including a broad, blueshifted Fe M-shell UTA trough between 15--16\,\AA. 
Here, the column density of the low ionization UFO component (with $\log\xi=1.8$) increased by more than an order 
of magnitude to $N_{\rm H}=5\times10^{21}$\,cm$^{-2}$ over a duration of a few days \citep{Reeves18}. 

Such gas could provide the substantial UV and soft X-ray opacity, as required for line driving, which is not present in the high flux 2024 observation. 
However it could be located closer to the accretion disk plane where it may be shielded from the central X-ray source \citep{PK04}. The gas can then accelerate up 
to the escape velocity, becoming highly ionized once in our sightline to the X-ray source and reaching its coasting speed \citep{Nomura16,Nomura20}.  
One possible caveat could be the inclination angle, which may need to be closer to equatorial in this scenario, whereas more polar inclinations may be preferred in observations 
of \pg\ \citep{Zoghbi15,Danehkar18}.
Nonetheless, occasionally the lower ionization gas may be lifted into our sightline before becoming more ionized, resulting in the soft X-ray opacity variations observed in 2014.  
Future work will quantitatively explore the long-term variability of the soft X-ray wind in \pg\ and the impact of the lower ionization gas between epochs.  

\section{Conclusions} \label{sec:conclusions}

We have presented an in-depth analysis of the high resolution soft X-ray spectrum of PG\,1211+143 with \xmm\ RGS in December 2024, which occurred simultaneously with the \xrism\ observations presented in paper I. 
This revealed a rich soft X-ray spectrum, with soft X-ray absorption lines originating primarily from highly ionized O (H-like O\,\textsc{viii}) and a range of lines from L-shell Fe (Fe\,\textsc{xix-xxiv}). 
The soft X-ray absorption features can be modeled with three distinct velocity components. The two lower velocity RGS absorbers ($v/c=-0.074$ and $v/c=-0.120$) are kinematically consistent with their 
corresponding components (zones 1 and 3) observed in Resolve through their Fe K band absorption profiles. A third faster zone is present in the RGS, with $v/c=-0.33$, which is intermediate in velocity 
between zones 4 and 5 observed in Resolve. 
The observations imply that the outflow in \pg\ is not a simple homogeneous wind, but is clumpy and structured. 
The data may pose a challenge to current wind models, where steady-state MHD prescriptions need to account for the clumpy wind structures and radiative acceleration requires significant columns of lower ionization gas for line driving to occur.  
Further simultaneous \xrism\ and RGS observations will reveal whether
the clumpy properties are inherent to most UFOs.


\section{Acknowledgements}

We thank Norbert Schartel for arranging the \xmm\ TOO observations of \pg\ and Fiona Harrison and Karl Forster for arranging the \nustar\ TOO. 
We also thank Dr Kim Page for her assistance in providing the \swift\ UVOT pointings for the 2014 campaign. 

JR and VB acknowledge support through NASA grants 80NSSC25K7845 and 80NSSC22K0474. 
Part of this work was supported by JSPS KAK609
ENHI Grant Number JP21K13958 (MM), JP21K13963, JP24K00638 (KH), JP19K21884, JP20H01947 
JP20KK0071, JP23K20239, JP24K00672, JP25H00660 (HN), JP20K14525 (MN), JP24K17104 (SO),
JP20H01946 (YU), JP23K13154 (SY), STFC
through grant ST/T000244/1 (CD), Yamada Science Foundation (MM), Exploratory Research Grant for Young Scientists, Hirosaki University (MN),
and the Kagoshima University postdoctoral research program (KU-DREAM, AT). 
This work has been partially supported by the ASI-INAF program I/004/11/6. 
Based on observations obtained with \xmm, an ESA science mission with instruments and contributions directly funded by ESA Member States and NASA.

\begin{contribution}
JNR was responsible for writing and submitting the manuscript. JNR, VB and MM have led the data reduction, analysis and discussion. SBK worked on the photoionization modeling. 
JNR and MM are the PIs of the \xrism\ observation.
All other authors contributed to the discussion and review of the manuscript. 
\end{contribution}

\facility{XRISM, XMM-Newton, NuSTAR, Swift}

\software{HEASoft/FTOOLS, XMMSAS}

\appendix

\restartappendixnumbering

\section{Scans of RGS spectrum in redshift space}

As was stated in Section~4.4, scans of the \textsc{xstar} absorbers were run on the RGS spectrum in redshift space, to ensure the best-fit absorber parameters were found.

This was performed by varying the absorber redshift (as a proxy for outflow velocity) over the range from $z=+0.1$ to $z=-0.3$, which from equation~1 covers the outflow velocity range from slightly redshifted 
($v/c=+0.02c$) to highly blue-shifted ($-0.41c$) in the rest frame of \pg. The latter value corresponds to the maximum outflow velocity measured in the {\it Resolve} spectrum (zone 6, see paper I). 
Increments of $\Delta z= 1\times10^{-3}$ were used for the scans and the absorber was fitted for $N_{\rm H}$ and $\log\xi$ at each $z$ increment. The continuum was also allowed to vary. 
The search initially focused on higher ionization gas, restricted to $\log\xi>3$. The most significant (in $\Delta C$) zone was then successively added, the spectrum refitted and the scans were repeated 
until no absorption components were revealed at $>99$\% confidence according to the AIC statistic. Having modeled all the higher ionization absorption, then the scans were rerun for lower ionization gas with $\log\xi=0-3$. 

The results are shown in Figure~A1 below. 
Panels (a--d) in Figure~A1 show the scans performed in the high ionization range. At the first absorber scan (panel a), three pronounced minima were found in redshift space, with the most significant occurring at $z=-0.042$, leading to an improvement in fit statistic of $\Delta C=-29.5$. Subsequent scans confirmed two further significant velocity zones (panels b, c), with $z=+3\times10^{-3}$ and $z=-0.234$.  Upon the fourth scan (panel d), no further absorption was required. 
For the low ionization absorption, the first scan (panel e) revealed a significant absorber at exactly $z=0$ and 
no significant absorbers were found in any subsequent scans (panel f).  

\begin{figure*}
\begin{center}
\rotatebox{-90}{\includegraphics[height=8.5cm]{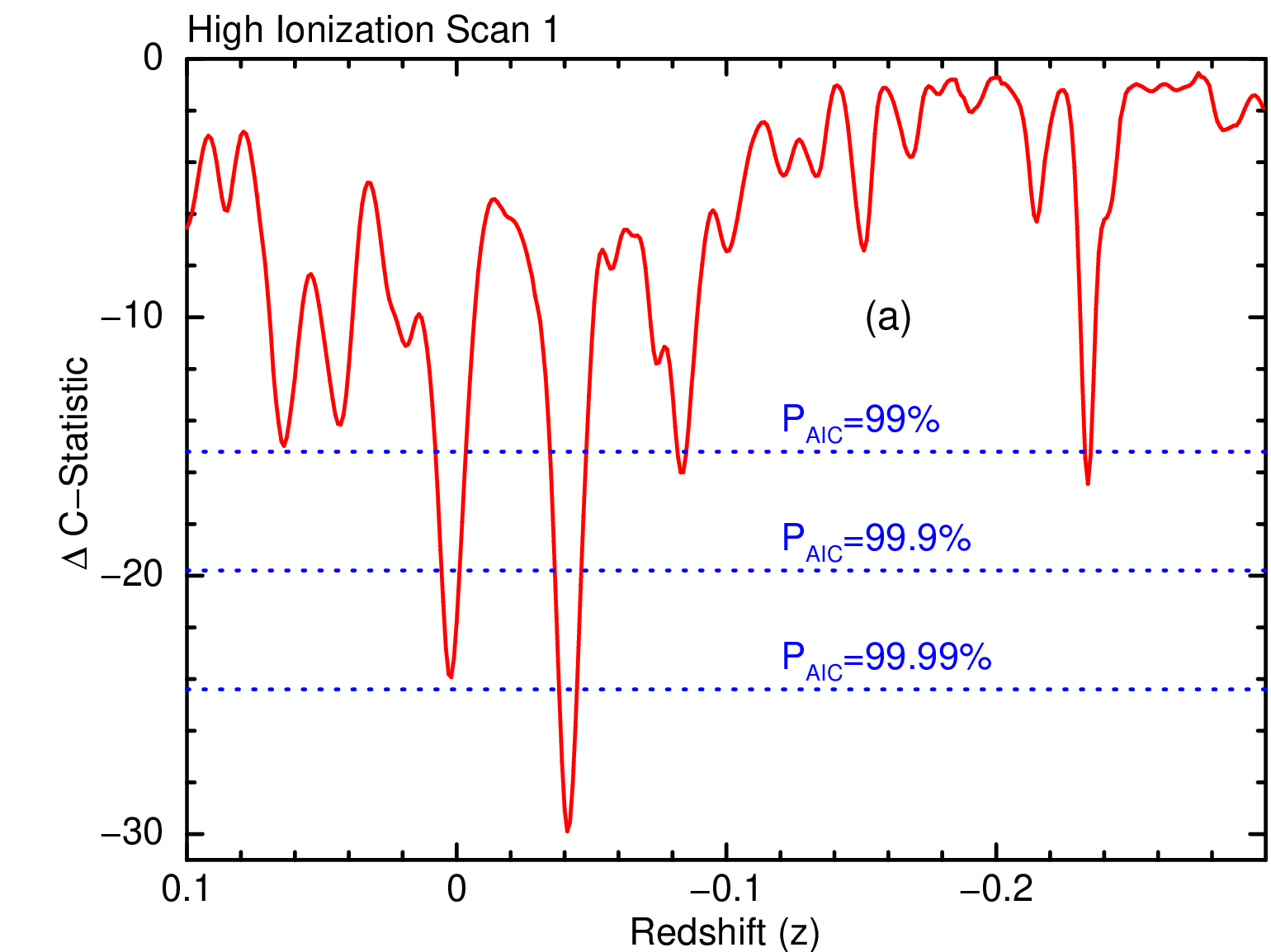}}
\rotatebox{-90}{\includegraphics[height=8.5cm]{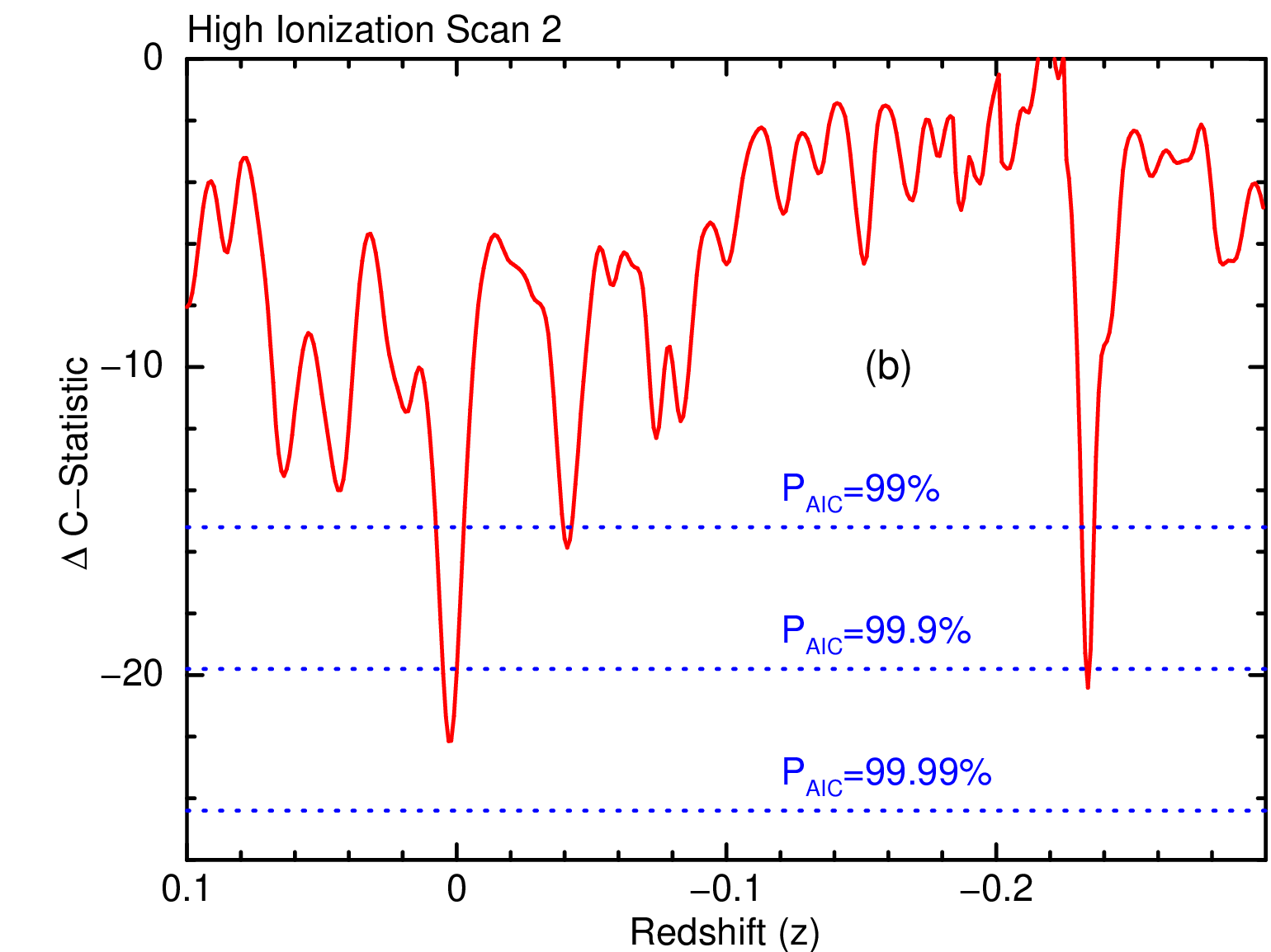}}
\rotatebox{-90}{\includegraphics[height=8.5cm]{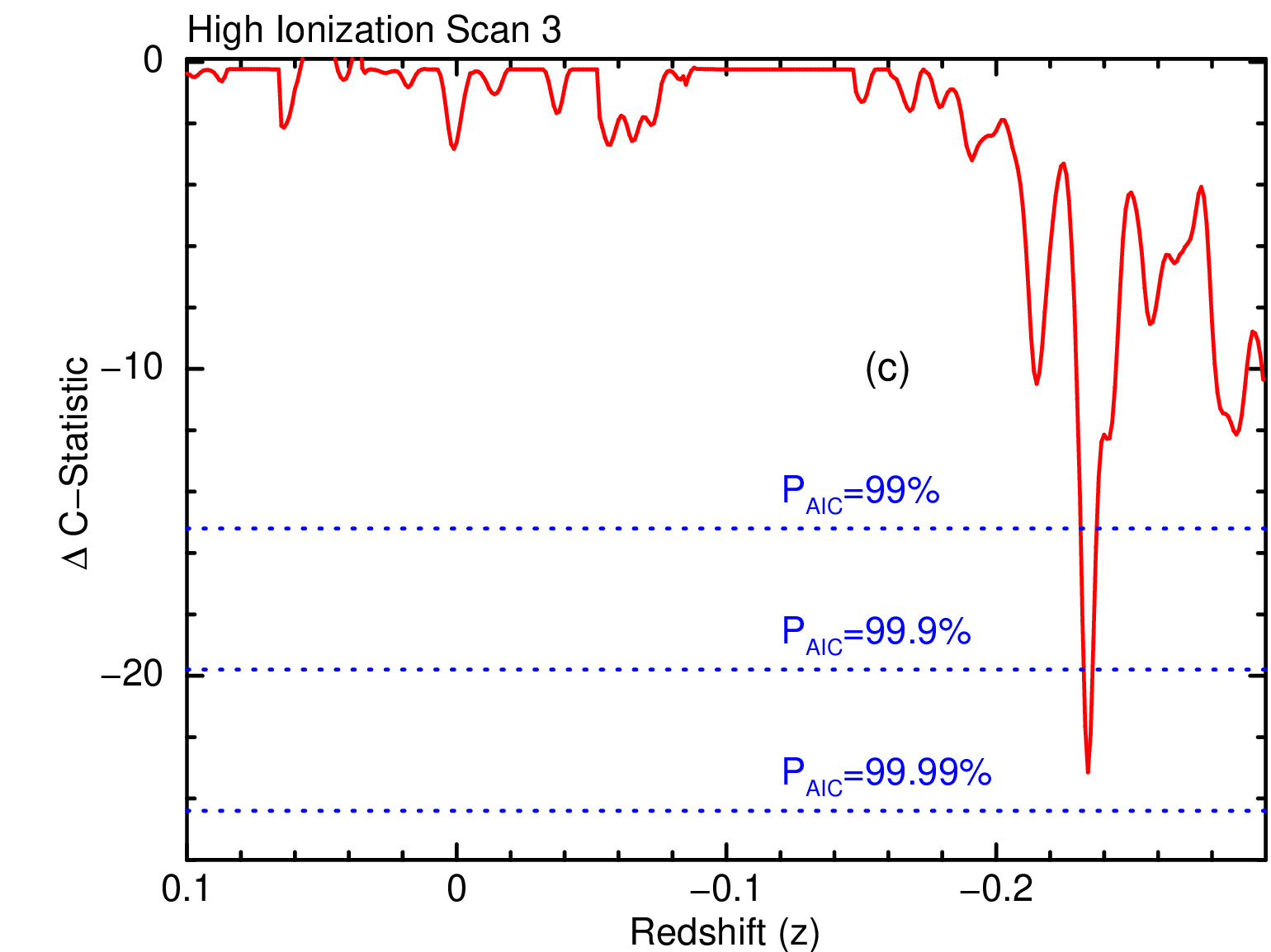}}
\rotatebox{-90}{\includegraphics[height=8.5cm]{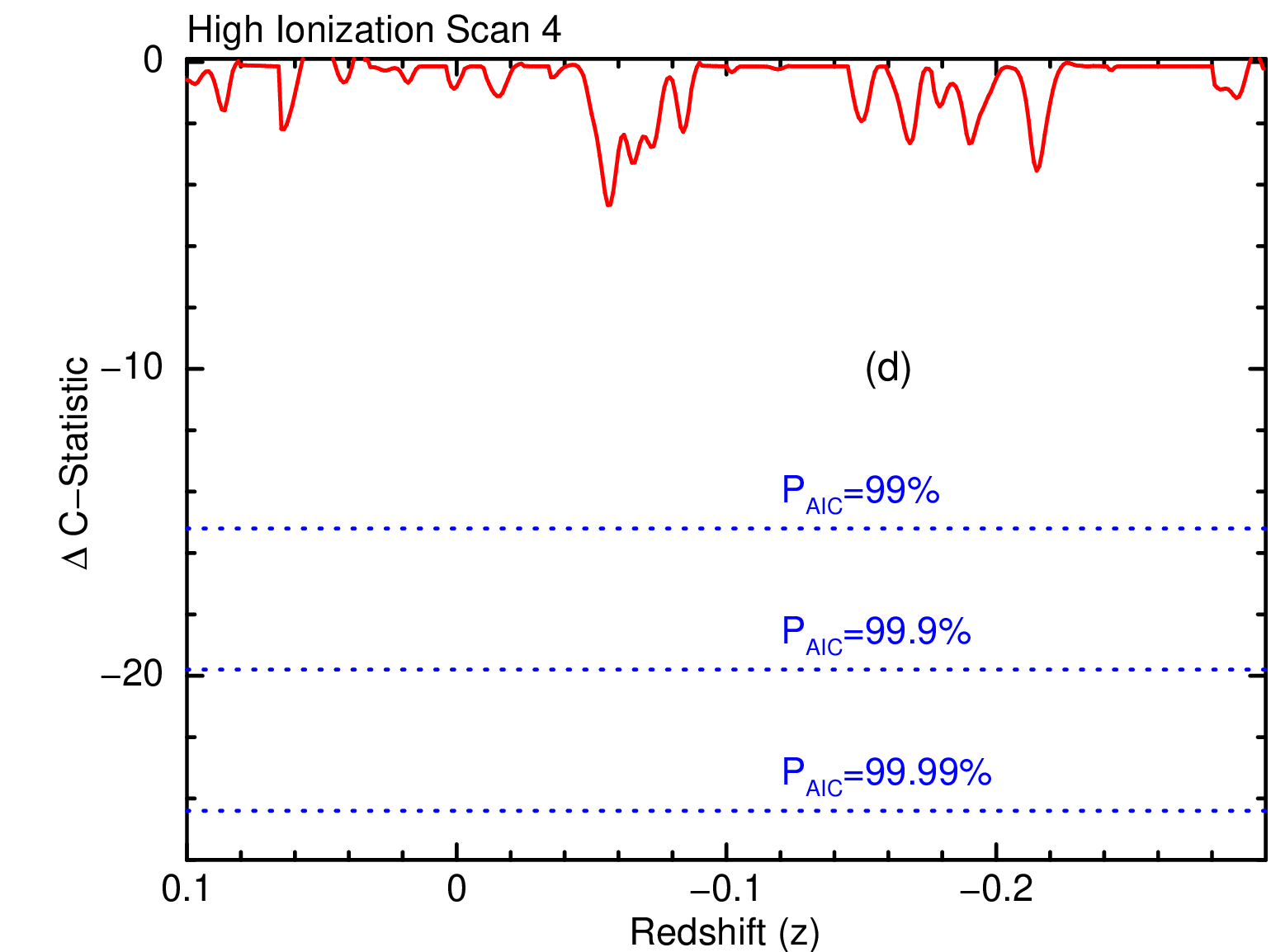}}
\rotatebox{-90}{\includegraphics[height=8.5cm]{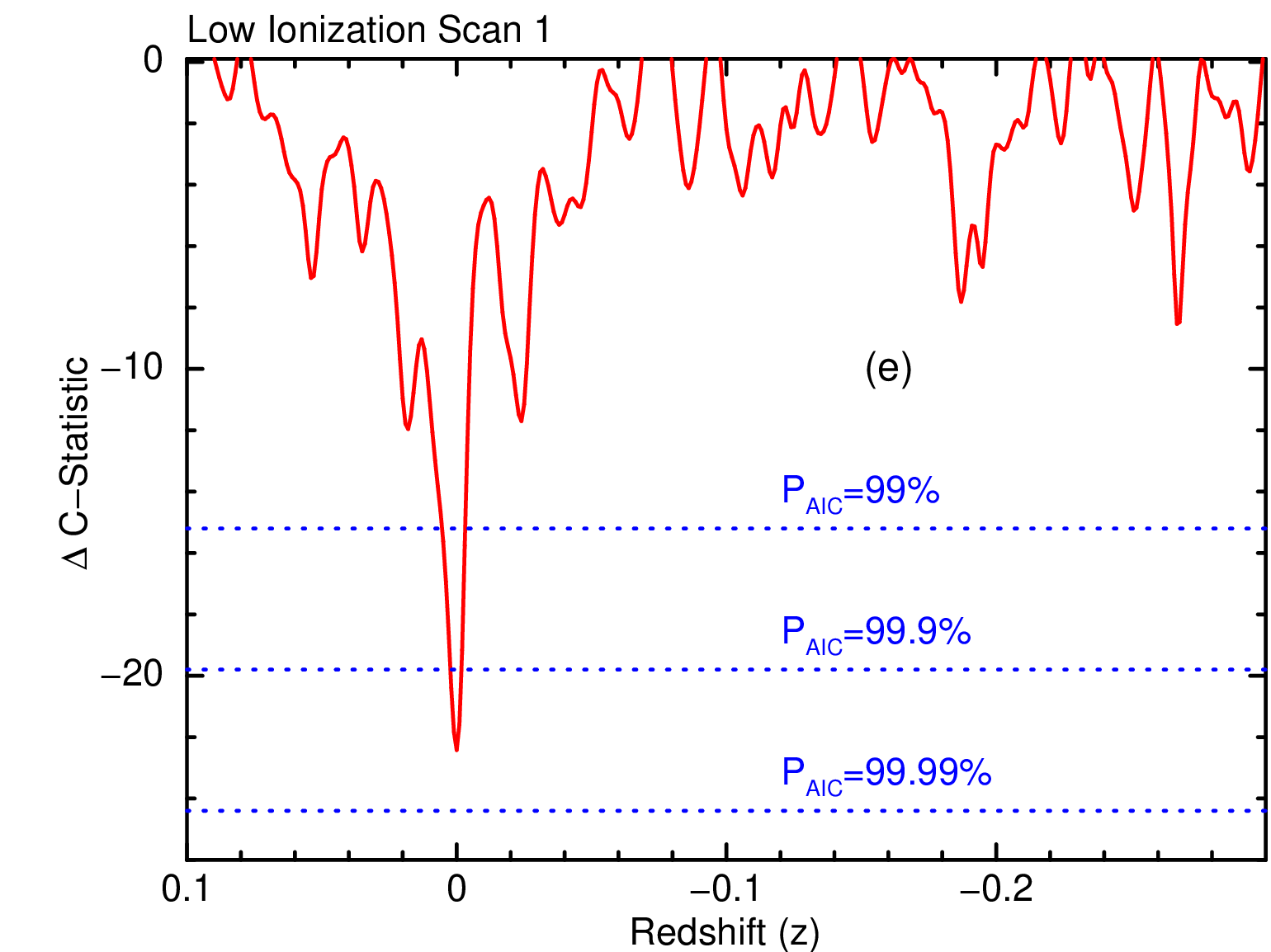}}
\rotatebox{-90}{\includegraphics[height=8.5cm]{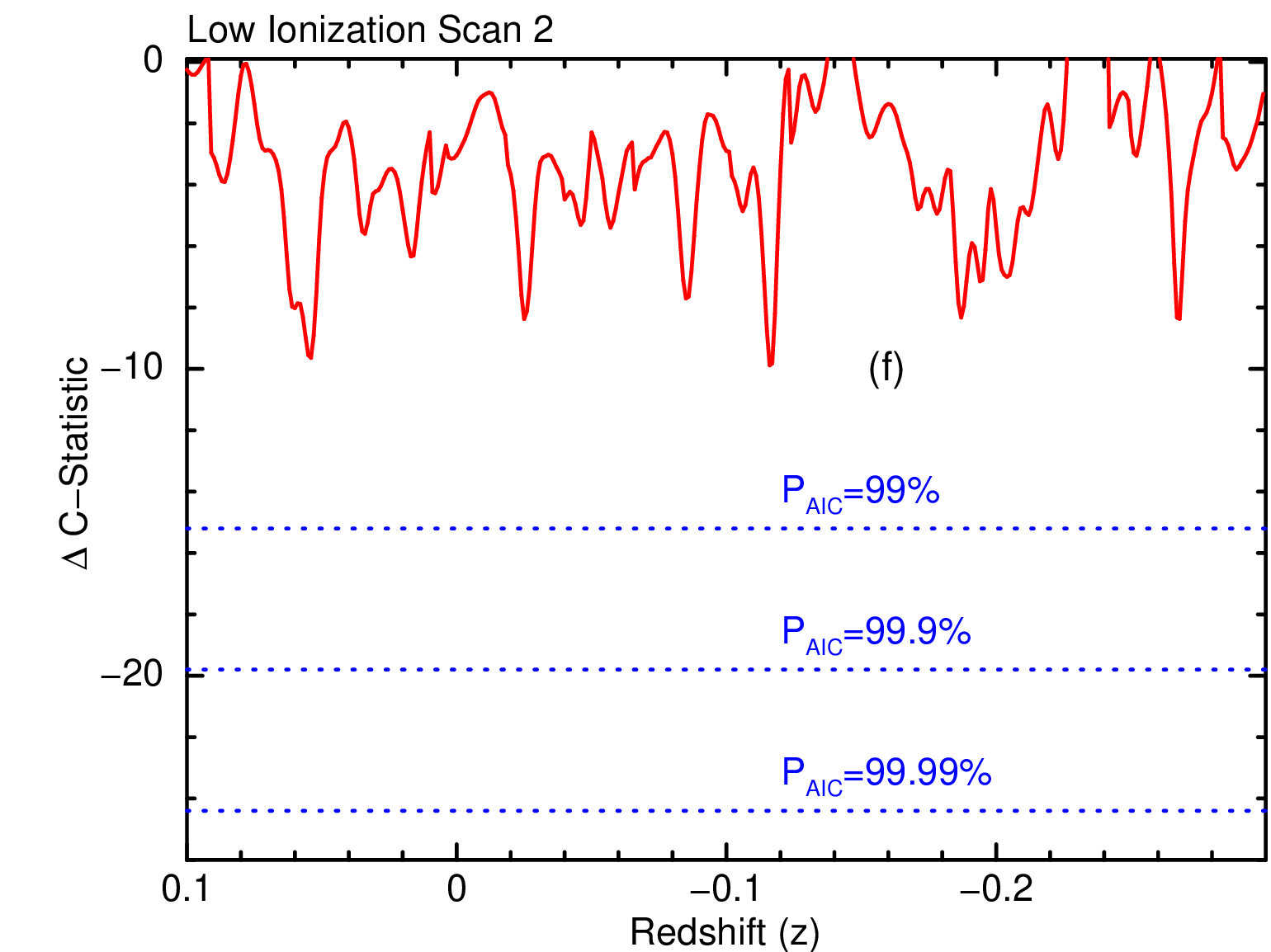}}
\end{center}
\caption{Results of the scans for the addition of the absorbers to the RGS spectrum, showing the improvement in fit statistic $\Delta C$ against redshift. Panels a--d show the scans for adding three successive high ionizations absorbers, where the fourth scan (panel d) then showed no further statistical improvement. Panels e--f show the scans for low ionization absorbers, where only one absorber at $z=0$ is required (panel e) and no further zones were then required (panel f).} 
\label{fig:appendix}
\end{figure*}

\section{Ionic Column Densities}

The individual ionic column densities for all the absorption zones are listed in Table~A1. The logarithmic values are given, where the units are cm$^{-2}$. Columns are quoted for Fe\,\textsc{xix-xxvi}, O\,\textsc{viii} and 
Ne\,\textsc{x} which constitute the absorption from all the outflowing zones. The zones are the same as those defined in Table~4. Values or upper limits for columns below $\log(N_{\rm ion})<15$ are not quoted and are 
below the detection limits.

\newpage

\begin{deluxetable*}{lcccccccccc}
\tabletypesize{\scriptsize}
\tablecaption{Distribution of Ionic Columns per zone}
\tablewidth{0pt}
\tablehead{
\colhead{} & \colhead{Fe~\textsc{xxvi}} & \colhead{Fe~\textsc{xxv}} & \colhead{Fe~\textsc{xxiv}} & \colhead{Fe~\textsc{xxiii}} & \colhead{Fe~\textsc{xxii}} & \colhead{Fe~\textsc{xxi}} & \colhead{Fe~\textsc{xx}} & 
\colhead{Fe~\textsc{xix}} & \colhead{O~\textsc{viii}} & \colhead{Ne~\textsc{x}}}
\startdata						
Zone~1 & $16.89^{+0.46}_{-0.66}$ & $17.40^{+0.25}_{-0.41}$ & $16.77^{+0.39}_{-0.71}$ & $16.22^{+0.69}_{-1.04}$ & $<16.05$ & & & & $15.71^{+0.47}_{-0.66}$ & $15.48^{+0.46}_{-0.65}$\\
Zone~1b & $<15.14$ & $15.53^{+0.77}_{-0.93}$ & $15.68^{+0.62}_{-0.78}$ & $16.05^{+0.41}_{-0.60}$ & $16.33^{+0.21}_{-0.40}$ & $16.45^{+0.17}_{-0.38}$ & $16.28^{+0.38}_{-0.59}$ & $15.83^{+0.58}_{-0.79}$ 
& $15.92^{+0.38}_{-0.39}$ & $16.03^{+0.37}_{-0.38}$\\
Zone~2$^a$ & $17.56^{+0.20}_{-0.30}$ & $17.54^{+0.20}_{-0.30}$ & $16.38^{+0.20}_{-0.30}$ & $<15.47$ & & & & & $15.48^{+0.20}_{-0.30}$	 & $15.63^{+0.20}_{-0.30}$\\
Zone~3 & $17.79^{+0.18}_{-0.28}$ & $17.77^{+0.26}_{-0.24}$ & $16.61^{+0.47}_{-0.53}$ & $<16.18$ & $<15.05$ & & & & $15.72^{+0.35}_{-0.38}$ & $15.86^{+0.35}_{-0.42}$\\
Zone~3b & $16.88^{+0.20}_{-0.67}$ & $17.56^{+0.57}_{-0.20}$ & $17.09^{+0.19}_{-0.35}$ & $16.73	^{+0.49}_{-0.54}$  & $16.29^{+0.78}_{-0.49}$ & $<16.15$ & $<15.27$ & & $16.07^{+0.41}_{-0.39}$ & 
$16.20^{+0.40}_{-0.39}$\\
Zone~4$^a$ & $17.76^{+0.21}_{-0.30}$ & $17.74^{+0.21}_{-0.30}$ & $16.58^{+0.21}_{-0.30}$ & $15.47^{+0.22}_{-0.30}$ & & & & & $15.69^{+0.21}_{-0.30}$ & $15.83^{+0.21}_{-0.30}$\\
RGS--fast	& $17.45^{+0.22}_{-0.32}$ & $17.95^{+0.11}_{-0.20}$ & $17.32^{+0.21}_{-0.31}$ & $16.76^{+0.24}_{-0.44}$ & $16.13^{+0.49}_{-0.57}$ & $<15.97$ & $<15.05$ & & $16.25^{+0.24}_{-0.31}$ 
& $16.38^{+0.23}_{-0.31}$\\
Zone~5 & $17.99^{+0.26}_{-0.46}$ & $16.65^{+0.27}_{-0.47}$ & & & & & & & $<15.40$ & $<15.60$\\
Zone~6 & $18.32^{+0.39}_{-0.47}$ & $17.00^{+0.69}_{-0.56}$	& $<15.50$ & & & & & & $<15.99$ & $<16.08$
\enddata
\tablenotetext{a}{For zones 2 and 4, the ionization parameter was fixed at the value found for zone 3, as per Table~4. Otherwise, if allowed to vary, they are consistent in ionization.}
\label{tab:ionic}
\end{deluxetable*}


\begin{thebibliography}

\bibitem[H.~Akaike(1974)]{Akaike74} Akaike, H.\ 1974, IEEE Transactions on Automatic Control, 19, 716. doi:10.1109/TAC.1974.1100705
\bibitem[\protect\citeauthoryear{K.~A.~Arnaud}{1996}]{Arnaud96}Arnaud K. A., 1996, in Jacoby G. H., Barnes J., eds, ASP Conf. Ser. Vol. 101, Astronomical Data Analysis Software and Systems V. Astron. Soc. Pac., San Francisco, p. 17
\bibitem[M.~Asplund et al.(2021)]{Asplund21} Asplund, M., Amarsi, A.~M., \& Grevesse, N.\ 2021, \aap, 653, A141. doi:10.1051/0004-6361/202140445
\bibitem[E.~Behar(2009)]{Behar09} Behar, E.\ 2009, \apj, 703, 2, 1346. doi:10.1088/0004-637X/703/2/1346
\bibitem[E.~Behar et al.(2001)]{Behar01} Behar, E., Sako, M., \& Kahn, S.~M.\ 2001, \apj, 563, 2, 497. doi:10.1086/323966
\bibitem[R.~D.~Blandford \& D.~G.~Payne(1982)]{BP82} Blandford, R.~D. \& Payne, D.~G.\ 1982, \mnras, 199, 883. doi:10.1093/mnras/199.4.883
\bibitem[T.~A.~Boroson \& R.~F.~Green(1992)]{BorosonGreen92} Boroson, T.~A. \& Green, R.~F.\ 1992, \apjs, 80, 109. doi:10.1086/191661
\bibitem[L.~W.~Brenneman et al.(2025)]{Brenneman25} Brenneman, L.~W., Wilkins, D.~R., Ogorza{\l}ek, A., et al.\ 2025, \apj, 995, 2, 200. doi:10.3847/1538-4357/ae1225
\bibitem[D.~N.~Burrows et al.(2005)]{Burrows05} Burrows, D.~N., Hill, J.~E., Nousek, J.~A., et al.\ 2005, \ssr, 120, 3-4, 165. doi:10.1007/s11214-005-5097-2
\bibitem[W.~Cash(1979)]{Cash79} Cash, W.\ 1979, \apj, 228, 939. doi:10.1086/156922
\bibitem[J.~I.~Castor et al.(1975)]{Castor75} Castor, J.~I., Abbott, D.~C., \& Klein, R.~I.\ 1975, \apj, 195, 157. doi:10.1086/153315
\bibitem[M.~Chatzikos et al.(2023)]{Chat23} Chatzikos, M., Bianchi, S., Camilloni, F., et al.\ 2023, \rmxaa, 59, 327. doi:10.22201/ia.01851101p.2023.59.02.12
\bibitem[G.~Chartas et al.(2003)]{Chartas02} Chartas, G., Brandt, W.~N., \& Gallagher, S.~C.\ 2003, \apj, 595, 1, 85. doi:10.1086/377299
\bibitem[G.~Chartas et al.(2002)]{Chartas03} Chartas, G., Brandt, W.~N., Gallagher, S.~C., et al.\ 2002, \apj, 579, 1, 169. doi:10.1086/342744
\bibitem[J.~Contopoulos \& R.~V.~E.~Lovelace(1994)]{CL94} Contopoulos, J. \& Lovelace, R.~V.~E.\ 1994, \apj, 429, 139. doi:10.1086/174307
\bibitem[A.~Danehkar et al.(2018)]{Danehkar18} Danehkar, A., Nowak, M.~A., Lee, J.~C., et al.\ 2018, \apj, 853, 2, 165. doi:10.3847/1538-4357/aaa427
\bibitem[R.~C.~Dannen et al.(2019)]{Dannen19} Dannen, R.~C., Proga, D., Kallman, T.~R., et al.\ 2019, \apj, 882, 2, 99. doi:10.3847/1538-4357/ab340b
\bibitem[J.~W.~den Herder et al.(2001)]{denHerder01} den Herder, J.~W., Brinkman, A.~C., Kahn, S.~M., et al.\ 2001, \aap, 365, L7. doi:10.1051/0004-6361:20000058
\bibitem[L.~Ferrarese \& D.~Merritt(2000)]{FM00} Ferrarese, L. \& Merritt, D.\ 2000, \apjl, 539, 1, L9. doi:10.1086/312838
\bibitem[K.~Fukumura et al.(2022)]{Fukumura22} Fukumura, K., Dadina, M., Matzeu, G., et al.\ 2022, \apj, 940, 1, 6. doi:10.3847/1538-4357/ac9388
\bibitem[K.~Fukumura et al.(2015)]{Fukumura15} Fukumura, K., Tombesi, F., Kazanas, D., et al.\ 2015, \apj, 805, 1, 17. doi:10.1088/0004-637X/805/1/17
\bibitem[K.~Fukumura et al.(2010)]{Fukumura10} Fukumura, K., Kazanas, D., Contopoulos, I., et al.\ 2010, \apjl, 723, 2, L228. doi:10.1088/2041-8205/723/2/L228
\bibitem[J.~Garc{\'\i}a et al.(2014)]{Garcia14} Garc{\'\i}a, J., Dauser, T., Lohfink, A., et al.\ 2014, \apj, 782, 2, 76. doi:10.1088/0004-637X/782/2/76
\bibitem[K.~Gebhardt et al.(2000)]{Gebhardt00} Gebhardt, K., Bender, R., Bower, G., et al.\ 2000, \apjl, 539, 1, L13. doi:10.1086/312840
\bibitem[V.~E.~Gianolli et al.(2024)]{Gianolli24} Gianolli, V.~E., Bianchi, S., Petrucci, P.-O., et al.\ 2024, \aap, 687, A235. doi:10.1051/0004-6361/202348908
\bibitem[J.~Gofford et al.(2015)]{Gofford15} Gofford, J., Reeves, J.~N., McLaughlin, D.~E., et al.\ 2015, \mnras, 451, 4, 4169. doi:10.1093/mnras/stv1207
\bibitem[J.~Gofford et al.(2013)]{Gofford13} Gofford, J., Reeves, J.~N., Tombesi, F., et al.\ 2013, \mnras, 430, 1, 60. doi:10.1093/mnras/sts481
\bibitem[L.~Gu et al.(2025)]{Gu25} Gu, L., Fukumura, K., Kaastra, J., et al.\ 2025, \aap, 704, A146. doi:10.1051/0004-6361/202557189
\bibitem[T.~Holczer et al.(2007)]{Holczer07} Holczer, T., Behar, E., \& Kaspi, S.\ 2007, \apj, 663, 2, 799. doi:10.1086/518416
\bibitem[P.~F.~Hopkins et al.(2016)]{Hopkins16} Hopkins, P.~F., Torrey, P., Faucher-Gigu{\`e}re, C.-A., et al.\ 2016, \mnras, 458, 1, 816. doi:10.1093/mnras/stw289
\bibitem[H.~Hu et al.(2025)]{Hu25} Hu, H., Asahina, Y., Yoshioka, S., et al.\ 2025, , arXiv:2510.17696. doi:10.48550/arXiv.2510.17696
\bibitem[C.~Jin et al.(2012)]{Jin12} Jin, C., Ward, M., \& Done, C.\ 2012, \mnras, 425, 2, 907. doi:10.1111/j.1365-2966.2012.21272.x
\bibitem[P.~M.~W.~Kalberla et al.(2005)]{Kalberla05} Kalberla, P.~M.~W., Burton, W.~B., Hartmann, D., et al.\ 2005, \aap, 440, 2, 775. doi:10.1051/0004-6361:20041864
\bibitem[T.~R.~Kallman et al.(2004)]{Kallman04} Kallman, T.~R., Palmeri, P., Bautista, M.~A., et al.\ 2004, \apjs, 155, 2, 675. doi:10.1086/424039
\bibitem[S.~Kaspi et al.(2000)]{Kaspi00} Kaspi, S., Smith, P.~S., Netzer, H., et al.\ 2000, \apj, 533, 2, 631. doi:10.1086/308704
\bibitem[A.~R.~King(2010)]{King10} King, A.~R.\ 2010, \mnras, 402, 3, 1516. doi:10.1111/j.1365-2966.2009.16013.x
\bibitem[A.~R.~King(2003)]{King03} King, A.\ 2003, \apjl, 596, 1, L27. doi:10.1086/379143
\bibitem[A.~R.~King \& K.~A.~Pounds(2003)]{KP03} King, A.~R. \& Pounds, K.~A.\ 2003, \mnras, 345, 2, 657. doi:10.1046/j.1365-8711.2003.06980.x
\bibitem[S.~B.~Kraemer et al.(2018)]{Kraemer18} Kraemer, S.~B., Tombesi, F., \& Bottorff, M.~C.\ 2018, \apj, 852, 1, 35. doi:10.3847/1538-4357/aa9ce0
\bibitem[G.~A.~Kriss et al.(2018)]{Kriss18} Kriss, G.~A., Lee, J.~C., Danehkar, A., et al.\ 2018, \apj, 853, 2, 166. doi:10.3847/1538-4357/aaa42b
\bibitem[J.~H.~Krolik et al.(1981)]{Krolik81} Krolik, J.~H., McKee, C.~F., \& Tarter, C.~B.\ 1981, \apj, 249, 422. doi:10.1086/159303
\bibitem[M.~Laurenti et al.(2025)]{Laurenti25} Laurenti, M., Tombesi, F., Cond{\`o}, P., et al.\ 2025, \aap, accepted, arXiv:2512.06077. 
\bibitem[A.~P.~Lobban et al.(2016)]{Lobban16} Lobban, A.~P., Vaughan, S., Pounds, K., et al.\ 2016, \mnras, 457, 1, 38. doi:10.1093/mnras/stv2896
\bibitem[A.~Luminari et al.(2020)]{Luminari20} Luminari, A., Tombesi, F., Piconcelli, E., et al.\ 2020, \aap, 633, A55. doi:10.1051/0004-6361/201936797
\bibitem[P.~Marziani et al.(1996)]{Marziani96} Marziani, P., Sulentic, J.~W., Dultzin-Hacyan, D., et al.\ 1996, \apjs, 104, 37. doi:10.1086/192291
\bibitem[G.~A.~Matzeu et al.(2023)]{Matzeu23} Matzeu, G.~A., Brusa, M., Lanzuisi, G., et al.\ 2023, \aap, 670, A182. doi:10.1051/0004-6361/202245036
\bibitem[G.~A.~Matzeu et al.(2017)]{Matzeu17} Matzeu, G.~A., Reeves, J.~N., Braito, V., et al.\ 2017, \mnras, 472, 1, L15. doi:10.1093/mnrasl/slx129
\bibitem[M.~Mehdipour et al.(2025)]{Mehdipour25} Mehdipour, M., Kaastra, J.~S., Eckart, M.~E., et al.\ 2025, \aap, 699, A228. doi:10.1051/0004-6361/202555623
\bibitem[J.~A.~J~Mitchell et al.(2023)]{Mitchell23} Mitchell, J.~A.~J., Done, C., Ward, M.~J., et al.\ 2023, \mnras, 524, 2, 1796. doi:10.1093/mnras/stad1830
\bibitem[M.~Mizumoto et al.(2026)]{Mizumoto25} Mizumoto, M., Reeves, J.~N., Braito, V., et al.\ 2026, \apj, 997, 2, 219. doi:10.3847/1538-4357/ae2853
\bibitem[E.~Nardini et al.(2015)]{Nardini15} Nardini, E., Reeves, J.~N., Gofford, J., et al.\ 2015, Science, 347, 6224, 860. doi:10.1126/science.1259202
\bibitem[H.~Noda et al.(2025)]{Noda25} Noda, H., Yamada, S., Ogawa, S., et al.\ 2025, \apjl, 993, 2, L53. doi:10.3847/2041-8213/ae14e8
\bibitem[M.~Nomura et al.(2020)]{Nomura20} Nomura, M., Ohsuga, K., \& Done, C.\ 2020, \mnras, 494, 3, 3616. doi:10.1093/mnras/staa948
\bibitem[M.~Nomura et al.(2016)]{Nomura16} Nomura, M., Ohsuga, K., Takahashi, H.~R., et al.\ 2016, \pasj, 68, 1, 16. doi:10.1093/pasj/psv124
\bibitem[B.~M.~Peterson et al.(2004)]{Peterson04} Peterson, B.~M., Ferrarese, L., Gilbert, K.~M., et al.\ 2004, \apj, 613, 2, 682. doi:10.1086/423269
\bibitem[P.~O.~Petrucci et al.(2020)]{Petrucci20} Petrucci, P.-O., Gronkiewicz, D., Rozanska, A., et al.\ 2020, \aap, 634, A85. doi:10.1051/0004-6361/201937011
\bibitem[P.~O.~Petrucci et al.(2018)]{Petrucci18} Petrucci, P.-O., Ursini, F., De Rosa, A., et al.\ 2018, \aap, 611, A59. doi:10.1051/0004-6361/201731580
\bibitem[K.~A.~Pounds \& K.~Page(2025)]{PP25} Pounds, K. \& Page, K.\ 2025, \mnras, 540, 3, 2530. doi:10.1093/mnras/staf637
\bibitem[K.~A.~Pounds et al.(2016a)]{Pounds16a} Pounds, K., Lobban, A., Reeves, J., et al.\ 2016a, \mnras, 457, 3, 2951. doi:10.1093/mnras/stw165
\bibitem[K.~A.~Pounds et al.(2016b)]{Pounds16b} Pounds, K.~A., Lobban, A., Reeves, J.~N., et al.\ 2016b, \mnras, 459, 4, 4389. doi:10.1093/mnras/stw933
\bibitem[K.~A.~Pounds \& J.~N.~Reeves(2009)]{PR09} Pounds, K.~A. \& Reeves, J.~N.\ 2009, \mnras, 397, 1, 249. doi:10.1111/j.1365-2966.2009.14971.x
\bibitem[K.~A.~Pounds et al.(2003)]{Pounds03} Pounds, K.~A., Reeves, J.~N., King, A.~R., et al.\ 2003, \mnras, 345, 3, 705. doi:10.1046/j.1365-8711.2003.07006.x
\bibitem[D.~Proga \& T.~R.~Kallman(2004)]{PK04} Proga, D. \& Kallman, T.~R.\ 2004, \apj, 616, 2, 688. doi:10.1086/425117
\bibitem[J.~N.~Reeves et al.(2018)]{Reeves18} Reeves, J.~N., Lobban, A., \& Pounds, K.~A.\ 2018, \apj, 854, 1, 28. doi:10.3847/1538-4357/aaa776
\bibitem[J.~N.~Reeves et al.(2009)]{Reeves09} Reeves, J.~N., O'Brien, P.~T., Braito, V., et al.\ 2009, \apj, 701, 1, 493. doi:10.1088/0004-637X/701/1/493
\bibitem[J.~N.~Reeves et al.(2003)]{Reeves03} Reeves, J.~N., O'Brien, P.~T., \& Ward, M.~J.\ 2003, \apjl, 593, 2, L65. doi:10.1086/378218
\bibitem[P.~W.~A.~Roming et al.(2005)]{Roming05} Roming, P.~W.~A., Kennedy, T.~E., Mason, K.~O., et al.\ 2005, \ssr, 120, 3-4, 95. doi:10.1007/s11214-005-5095-4
\bibitem[A.~R{\'o}{\.z}a{\'n}ska et al.(2015)]{Rozanska15} R{\'o}{\.z}a{\'n}ska, A., Malzac, J., Belmont, R., et al.\ 2015, \aap, 580, A77. doi:10.1051/0004-6361/201526288
\bibitem[M.~Sako et al.(2001)]{Sako01} Sako, M., Kahn, S.~M., Behar, E., et al.\ 2001, \aap, 365, L168. doi:10.1051/0004-6361:20000081
\bibitem[R.~Serafinelli et al.(2019)]{Serafinelli19} Serafinelli, R., Tombesi, F., Vagnetti, F., et al.\ 2019, \aap, 627, A121. doi:10.1051/0004-6361/201935275
\bibitem[S.~A.~Sim et al.(2010)]{Sim10} Sim, S.~A., Miller, L., Long, K.~S., et al.\ 2010, \mnras, 404, 3, 1369. doi:10.1111/j.1365-2966.2010.16396.x
\bibitem[S.~A.~Sim et al.(2008)]{Sim08} Sim, S.~A., Long, K.~S., Miller, L., et al.\ 2008, \mnras, 388, 2, 611. doi:10.1111/j.1365-2966.2008.13466.x
\bibitem[L.~Str{\"u}der et al.(2001)]{Struder01} Str{\"u}der, L., Briel, U., Dennerl, K., et al.\ 2001, \aap, 365, L18. doi:10.1051/0004-6361:20000066
\bibitem[S.~Takeuchi et al.(2013)]{Takeuchi13} Takeuchi, S., Ohsuga, K., \& Mineshige, S.\ 2013, \pasj, 65, 88. doi:10.1093/pasj/65.4.88
\bibitem[M.~Tashiro et al.(2025)]{Tashiro25} Tashiro, M., Kelley, R., Watanabe, S., et al.\ 2025, \pasj, 77, S1. doi:10.1093/pasj/psaf023
\bibitem[L.~Titarchuk(1994)]{Titarchuk94} Titarchuk, L.\ 1994, \apj, 434, 570. doi:10.1086/174760
\bibitem[F.~Tombesi et al.(2013)]{Tombesi13} Tombesi, F., Cappi, M., Reeves, J.~N., et al.\ 2013, \mnras, 430, 2, 1102. doi:10.1093/mnras/sts692
\bibitem[F.~Tombesi et al.(2010)]{Tombesi10} Tombesi, F., Cappi, M., Reeves, J.~N., et al.\ 2010, \aap, 521, A57. doi:10.1051/0004-6361/200913440
\bibitem[M.~J.~L.~Turner et al.(2001)]{Turner01} Turner, M.~J.~L., Abbey, A., Arnaud, M., et al.\ 2001, \aap, 365, L27. doi:10.1051/0004-6361:20000087
\bibitem[D.~A.~Verner et al.(1996)]{Verner96} Verner, D.~A., Ferland, G.~J., Korista, K.~T., et al.\ 1996, \apj, 465, 487. doi:10.1086/177435
\bibitem[A.~Y.~Wagner et al.(2013)]{Wagner13} Wagner, A.~Y., Umemura, M., \& Bicknell, G.~V.\ 2013, \apjl, 763, 1, L18. doi:10.1088/2041-8205/763/1/L18
\bibitem[J.~Wilms et al.(2000)]{Wilms00} Wilms, J., Allen, A., \& McCray, R.\ 2000, \apj, 542, 2, 914. doi:10.1086/317016
\bibitem[X.~Xiang et al.(2025)]{Xiang25} Xiang, X., Miller, J.~M., Behar, E., et al.\ 2025, \apjl, 988, 2, L54. doi:10.3847/2041-8213/adee9b
\bibitem[XRISM Collaboration et al.(2025)]{xrism25} Xrism Collaboration, Audard, M., Awaki, H., et al.\ 2025, \nat, 641, 8065, 1132. doi:10.1038/s41586-025-08968-2
\bibitem[XRISM Collaboration et al.(2025b)]{xrism25b} Xrism Collaboration, Audard, M., Awaki, H., et al.\ 2025, \aap, 702, A147. doi:10.1051/0004-6361/202556000
\bibitem[Y.~Xu et al.(2025)]{Xu25} Xu, Y., Gallo, L.~C., Hagino, K., et al.\ 2025, \pasj, 77, S223. doi:10.1093/pasj/psaf070
\bibitem[S.~Yamada et al.(2024)]{Yamada24} Yamada, S., Kawamuro, T., Mizumoto, M., et al.\ 2024, \apjs, 274, 1, 8. doi:10.3847/1538-4365/ad5961
\bibitem[A.~Zoghbi et al.(2015)]{Zoghbi15} Zoghbi, A., Miller, J.~M., Walton, D.~J., et al.\ 2015, \apjl, 799, 2, L24. doi:10.1088/2041-8205/799/2/L24


\end{thebibliography}
\end{document}